\documentclass[useAMS,usenatbib]{mn2e}

\usepackage{longtable}
\usepackage{multirow}
\usepackage{graphicx,amssymb,amsfonts,amsmath}
\usepackage{subfigure}
\usepackage{natbib}
\usepackage{lscape}

\newcommand{\Log}{{\rm log}~}
\newcommand{\NH}{N_{\rm H}}

\newcommand{\Ldisc}{L_{\rm opt-UV}}
\newcommand{\Lir}{L_{\rm ir}}

\newcommand{\ebvq}{E(B-V)_{\rm qso}}

\newcommand{\aopt}{\alpha_{\rm OPT}}
\newcommand{\anir}{\alpha_{\rm NIR}}
\newcommand{\Lirmu}{L_{12\mu \rm m}}
\newcommand{\size}{s_{12\mu \rm m}}
\newcommand{\rinf}{r_{\rm infl}}
\newcommand{\vrec}{v_{\rm kick}}
\newcommand{\rinfk}{r_{\rm infl,recoil}}

\newcommand{\Ldiscr}{L_{\rm opt-UV,red}}
\newcommand{\Ldiscder}{L_{\rm opt-UV,dered}}
\newcommand{\Lhost}{L_{\rm host}}
\newcommand{\msun}{M_\odot}

\DeclareRobustCommand{\ion}[2]{%
\relax\ifmmode
\ifx\testbx\f@series
{\mathbf{#1\,\mathsc{#2}}}\else
{\mathrm{#1\,\mathsc{#2}}}\fi
\else\textup{#1\,{\mdseries\textsc{#2}}}%
\fi}

\newcommand{\rev}[1]{{ #1}}

\title[BH binary candidates: II. spectral energy distribution atlas]{The nature of massive black hole binary candidates: II. Spectral energy distribution atlas}
\author[E. Lusso et al.]{E.~Lusso$^{1}$\thanks{lusso@mpia.de}, R.~Decarli$^{1}$, M.~Dotti$^{2,3}$,  C.~Montuori$^{4,5}$, David~W.~Hogg$^{1,6}$, P.~Tsalmantza$^{7}$, \newauthor M.~Fumagalli$^{8,9,10}$, and J.~X.~Prochaska$^{11}$\\
$^{1}$Max Planck Institut f\"{u}r Astronomie, K\"{o}nigstuhl 17, D-69117, Heidelberg, Germany.\\
$^{2}$Dipartimento di Fisica G. Occhialini, Universit\`{a} degli Studi di Milano Bicocca, Piazza della Scienza 3, 20126 Milano, Italy.\\
$^{3}$INFN, Sezione Milano-Bicocca, Piazza della Scienza 3, 20126 Milano, Italy.\\
$^{4}$Technion, Department of Physics, IL-32000, Haifa, Israel.\\
$^{5}$Dipartimento di Scienza e Alta Tecnologia, Universit\`{a} dell' Insubria, via Valleggio 11, 22100 Como, Italy.\\
$^{6}$Center for Cosmology and Particle Physics, Department of Physics, New York University, 4 Washington Place, New York, NY 10003, USA.\\
$^{7}$Klausenpfad 22, Heidelberg, D-691221, Germany.\\
$^{8}$Carnegie Observatories, 813 Santa Barbara Street, Pasadena, CA 91101, USA.\\
$^{9}$Department of Astrophysics, Princeton University, Princeton, NJ 08544-1001, USA.\\
$^{10}$Hubble Fellow.\\
$^{11}$Department of Astronomy and Astrophysics, UCO/Lick Observatory, University of California, 1156 High Street, Santa Cruz, CA 95064, USA.
}
\begin{document}
\date{Draft, \today}

\pagerange{\pageref{firstpage}--\pageref{lastpage}} \pubyear{2002}

\maketitle

\label{firstpage}

\begin{abstract}
Recoiling supermassive black holes (SMBHs) are considered one plausible physical mechanism to explain high velocity shifts between narrow and broad emission lines sometimes observed in quasar spectra.
If the sphere of influence of the recoiling SMBH is such that only the accretion disc is bound, the dusty torus would be left behind, hence the SED should then present distinctive features (i.e. a mid-infrared deficit).
Here we present results from fitting the Spectral Energy Distributions (SEDs) of 32 Type-1 AGN with high velocity shifts between broad and narrow lines.
The aim is to find peculiar properties in the multi-wavelength SEDs of such objects by comparing their physical parameters (torus and disc luminosity, intrinsic reddening, and size of the 12$\mu$m emitter) with those estimated from a control sample of $\sim1000$ \emph{typical} quasars selected from the Sloan Digital Sky Survey in the same redshift range.
We find that all sources, with the possible exception of J1154+0134,
analysed here present a significant amount of 12~$\mu$m emission. This is in
contrast with a scenario of a SMBH displaced from the center of the galaxy, as
expected for an undergoing recoil event.

\end{abstract}


\section{Introduction} 
\label{Introduction}
It is now widely accepted that galaxies host supermassive black holes (SMBHs)
in their centre with masses of the order of $10^{6-9}M_\odot$
\citep{1964ApJ...140..796S,1969Natur.223..690L}.  Locally, the SMBH mass
correlates with the mass (\citealt{1998AJ....115.2285M,2003ApJ...589L..21M}),
with the velocity dispersion
(\citealt{2000ApJ...539L...9F,2002ApJ...574..740T}), and with the luminosity
of the host galaxy bulge (\citealt{1995ARA&A..33..581K}).  The existence of
these correlations implies that the growth of the SMBH is tightly linked with
the galaxy evolution, playing a crucial role in the star-formation history of
the galaxy itself.  \par From the theoretical point of view, according to the
hierarchical galaxy formation model, a main channel of galaxy growth is
through mergers, that enhance star formation and trigger active galactic
nuclei (AGN).  Given that SMBH ubiquitously populate the center of galaxies,
it is expected that black hole binaries (BHB) will be formed in the course of
a merger event.  \par If the binary hypothesis is correct, BHB systems should
leave characteristic features in the spectrum of the system such as a shift of
the peak of their broad emission lines (BLs associated to the central BHs)
with respect the narrow ones (NLs associated to the host galaxy;
\citealt{1980Natur.287..307B,1983LIACo..24..473G}). For example, objects
showing {\it double-peaked emission} lines (DPEs) in their spectra might be
interpreted as possible BHB candidates. Those systems present displaced
broad-line peaks (one blueshifted and one redshifted compared with the narrow
line redshift) as a result of the orbital motion of the ionized gas
gravitationally bound to the BH pairs.  To date, long multiepoch
  campaigns have ruled out the BHB interpretation for most of the DPEs studied
  (e.g., \citealt{1994ApJS...90....1E}). However, AGN showing single peaked
  shifted BLs are still considered plausible BHB candidates, as they could be
  related to close binaries in which only one MBH is active
  (e.g. \citealt{2009ApJ...697..288B, 2009MNRAS.398L..73D}) and monitoring
  campaigns are ongoing to prove their nature
  (e.g. \citealt{2013MNRAS.433.1492D} hereafter Paper I,
  \citealt{2012ApJS..201...23E,2013arXiv1312.6694L}; \rev{\citealt{2013ApJ...777...44J}}).  \par Other alternative
scenarios that might explain the BLs displacement are: {\it 1)} the projection
of two unrelated quasars (QSOs) viewed, by chance, along similar sight lines,
{\it 2)} a recoil induced by the anisotropic emission of gravitational waves
in the final merge of SMBHs pairs (\citealt{PhysRev.128.2471}).  The first
interpretation relies on the probability of finding two unassociated quasars
aligned with the line of sight. This turns out to be unlikely in several cases
(see for example \citealt{2009Natur.458...53B,2009MNRAS.397..458D}).
The latter is a prediction of general relativity, confirmed by numerical
simulations, which imply that, after BHB coalescence, the resulting BH can
recoil at velocities up to several thousand km s$^{-1}$ due to anisotropic
gravitational wave emission (e.g. \citealt{2007ApJ...668.1140B,
  2007ApJ...659L...5C, 2007PhRvL..98w1102C, 2012PhRvD..85h4015L, 2013PhRvD..87h4027L}).
Observationally, the recoiling SMBH and its surrounding gas would give rise to spectra where BLs are shifted from NLs of the host galaxy (e.g., \citealt{2008ApJ...678L..81K}; \rev{\citealt{2009ApJ...702L..82C}}; \citealt{2012AdAst2012E..14K,2012ApJ...752...49C}).
\par
\par From a multi-wavelength perspective we expect the spectral energy
distribution (SED) of DPEs or BHB candidates to have features similar to a
``typical" QSO SED: "infrared bump" at $\sim10$ $\mu$m, and an upturn in the
optical-UV, the so-called "big-blue bump" (BBB,
\citealt{1989ApJ...347...29S,1994ApJS...95....1E,2006ApJS..166..470R,2011ApJS..196....2S,2012ApJ...759....6E}). 
The BBB is thought to be representative of the emission from the accretion
disc around the SMBH, while the infrared bump is due to the presence of dust
(from sub-parsec to hundreds parsec scale) which re-radiates a fraction of the
optical-UV disc photons at infrared wavelengths.  \par The SED of a recoiling
BH significantly displaced from the center of the host galaxy might have
instead different features.  For example, if the sphere of influence of the
recoiling BH on the gas were such that only the accretion disc is bound, all
the larger scale structures (like a dusty torus) would be left behind, hence
the SED should present the BBB only
(\citealt{2010ApJ...724L..59H}, H10 hereafter; \citealt{2011ApJ...729..125G}).
\par
In this paper we will present the broad-band SEDs, from $\sim$1500\AA{} to
$\sim$20~$\mu$m, of a sample of 32 quasars identified by
\citet{2011ApJ...738...20T}. This sample has been selected from the Sloan
Digital Sky Survey (SDSS; \citealt{2000AJ....120.1579Y}) spectroscopic
database based on large velocity shifts ($>$1000 km s$^{-1}$) between NLs and
BLs, and it spans the redshift range $0.136\leq z\leq 0.713$. In Paper I we presented their spectral properties (fluxes, line luminosities, widths, broad line profiles and their evolution).  Source are subsequently divided in four classes on the basis of the shape of the line profile: {\it 1)} fairly bell-shaped, strongly shifted BLs identify good BHB candidates; {\it 2)} BLs with tentative evidence of double-horned profiles are classified as DPEs; {\it 3)} objects with lines showing a rather symmetric base, centered at the redshift of the NLs, but an asymmetric core, resulting in a shifted peak, are called ``Asymmetric"; {\it 4)} other more complex profiles, or lines with relatively small shifts, or lines with asymmetric wings but modest peak shift are labelled as ``Others" (see Fig.~2 in Paper I for examples of each class of objects).
The main aim of the present analysis is to test whether the multi-wavelength information can give further hints on the BHB scenarios described above.
In order to quantify the contribution of host-galaxies and AGN reddening, as well as the disc and infrared AGN luminosities ($\Ldisc$ and $\Lir$), we will make use of the SED-fitting code presented in \citet{2013ApJ...777...86L}.
\par
\par
We adopted a concordance $\Lambda-$cosmology with $H_{0}=70\, \rm{km \,s^{-1}\, Mpc^{-1}}$, $\Omega_{M}=0.3$, $\Omega_{\Lambda}=1-\Omega_{M}$.

\section{Recoiling scenario}
\label{Recoiling scenario}
The SMBH ejected by gravitational wave recoil might carry along its accretion disc and broad line region, but not the dusty torus (e.g. H10; \citealt{2011ApJ...729..125G}), whose emission is peaked around $10-20~\mu$m and is usually interpreted as hot-dust reprocessing of optical-UV radiation from the accretion disc. The temperatures involved are around $1000-1900$ K depending on the different compositions of the dust grains\footnote{The dust sublimation temperatures ranges around $1500-1900$ K for graphite and $1000-1400$ K for silicate grains (e.g., \citealt{1993ApJ...402..441L}).}, which are located at scales that are dependent on the nuclear luminosity. 
In principle, the recoiling SMBH is able to drag along its surrounding dust only if its sphere of influence is larger than the torus size.
\par
The SMBH sphere of influence is defined as the distance where the force on a test mass is dominated by the BH.
The standard definition of sphere of influence is (see \citealt{2005LRR.....8....8M}):
\begin{equation}
 \label{soi}
 \rinf = \frac{GM}{\sigma^2} \simeq 10.8\, {\rm pc} \left(\frac{M}{10^8 M_\odot}\right) \left(\frac{\sigma}{200 \rm \,km\, s^{-1}}\right)^{-2},
\end{equation}
where $M$ is the mass of the single BH formed via coalescence of two BHs, and $\sigma$ is the 1D velocity dispersion of the stars in the nucleus. For a recoiling BH, $\rinf$ needs to be scaled by the kick velocity $\vrec$ such as 
\begin{equation}
 \label{soirec}
 \rinfk \simeq 0.43\, {\rm pc} \left(\frac{M}{10^8 M_\odot}\right) \left(\frac{\vrec}{1000 \rm \,km\, s^{-1}}\right)^{-2}.
\end{equation}
For a Milky Way BH mass of $4\times10^{6}M_\odot$ and a kick velocity of 2000 km s$^{-1}$ (e.g., \citealt{2012PhRvD..85h4015L}) $\rinfk$ is extremely small ($\sim4.3\times10^{-3}$ pc). Given that our sample consists of quasars, a more appropriate BH mass is of the order of  $\sim10^{8}M_\odot$ (e.g., \citealt{2008ApJ...689L..89K}), which lead to a sphere of influence of $0.1$ pc at the same kick velocity. 
\par
\citet{2009A&A...502...67T} found that the characteristic size of the $12~\mu$m emitter scales approximately as the square root of the AGN luminosity 
\begin{equation}
 \label{sizelum}
 \size \cong 5.7\times10^{-22} \left( \frac{\Lirmu}{\rm erg\; s^{-1}}\right)^{0.5} {\rm pc}.
\end{equation}
This relation has been estimated making use of high resolution interferometric observations of nearby AGN with the MID-infrared Interferometric instrument (MIDI) at the Very Large Telescope Interferometer (VLTI; see also \citealt{2011A&A...531A..99T}).
The size of the $12~\mu$m emitter vary from $\sim0.6$ pc at $\Lirmu\sim10^{42}$ erg s$^{-1}$, to $\sim20$ pc at $\Lirmu\sim10^{45}$ erg s$^{-1}$. 
\par
For a BH of $\sim10^{8}M_\odot$ and a typical $12~\mu$m luminosity of $\sim10^{45}$ erg s$^{-1}$, the torus size is $\sim18$ pc, hence the BH has no chance to carry along a significant amount of dust during the kick. The SED of such object would have a negligible or no infrared bump, or no $1~\mu$m inflection.
The disappearance of the torus emission once the BH is outside the torus region is extremely rapid, and can be roughly quantified as follow
\begin{equation}
 \label{timescale}
 \tau_{\rm dis} \cong \frac{R}{c} \cong 32.6 \frac{R}{10~\rm pc}  \quad \rm yr,
\end{equation}
while the crossing time of a black hole within a typical torus size ($\sim10$pc) is
\begin{equation}
 \label{timescale}
 \tau = \frac{R}{v_{\rm BH}} \cong 9.8\times 10^4 \frac{R}{10~\rm pc} \left(\frac{v_{\rm BH}}{1000~{\rm km s^{-1}}}\right)^{-1} \rm yr.
\end{equation}
These numbers reinforce the idea that the mid-infrared emission, once the BH is kicked, should become negligible on very short timescales.
Sources with low mid-infrared emission compared to the optical-UV one are known as {\it hot-dust-poor} AGN (HDP, \citealt{2010ApJ...724L..59H}, but see also \citealt{2009MNRAS.397.1326R,2010Natur.464..380J}).
\par
In the following, we will investigate if the QSOs presented in Paper I (which show large velocity shifts between NLs and BLs) may be interpreted in the context of the recoiling scenario through the analysis of their broad-band SEDs.

\section{Photometry}
\label{Photometry}
A well sampled SED over a broad range of wavelength is mandatory in order to
properly disentangle the emission associated to stellar light from that due to
accretion and to constrain physical parameters from SED decomposition.  We
thus have collected the multi-wavelength photometry from mid-infrared to
optical-UV: $W1$, $W2$, $W3$, and $W4$ bands centered at wavelengths of 3.4,
4.6, 12, and 22 $\mu$m from the {\it Wide-field Infrared Survey Explorer}
(WISE; \citealt{2010AJ....140.1868W}); $J(1.25\mu {\rm m})$, $H(1.65\mu {\rm
  m})$, and $K(2.16\mu {\rm m})$ bands from the {\it Two Micron All Sky
  Survey} (2MASS; \citealt{2003tmc..book.....C}); optical bands from the SDSS
seventh data release (DR7; \citealt{2009ApJS..182..543A}); NUV(2306\AA) and
FUV(1551\AA) from the {\it Galaxy Evolution Explorer} (GALEX;
\citealt{2007ApJS..172..468Z}). The total number of employed bands is 14.
\par All QSOs are detected in WISE and SDSS, only 3 objects have not been
detected in the 2MASS bands (J0927+2943, J0932+0318, and J1012+2613).  \par
GALEX fluxes are collected from the online database and images have been
inspected visually.  We have found that 28 objects have a GALEX detection and
7 out of these 28 sources have two observations from different surveys:
all-sky imaging survey (AIS) and from the medium imaging survey (MIS).  Three
quasar (J0221+0101, J0829+2728, and J1440+3319) out of 7 show a difference
between AIS and MIS photometry higher than 0.35 magnitudes in both GALEX
bands, which is a factor $\sim 2-3$ difference in fluxes.  
For the SED fitting of J0221+0101 we use  
the flux measurements coming from the longest exposures (the MIS data). For  
J0829+2728 we consider the AIS fluxes because the values from the MIS were too faint to
be consistent with the SDSS data (probably because this source was in the low state).
For J1440+3319 we have found three observations from different surveys:
  AIS, nearby galaxy survey (NGS), and from a guest investigation (GI)
  program.  Fluxes for J1440+3319 vary by a factor of $\sim$2.7 in the NUV
  band from the AIS to the NGS survey ($F_\nu=23.10-62.25~\mu$Jy,
  respectively). We have taken as reference flux value for this object the one
  from the GI program, which is the deepest (see Appendix~\ref{Notes on the
    SED general properties: GALEX data} for a more detailed discussion about
  the three varying objects).  Overall, 21 objects\footnote{J0918+3156 is
  detected only in the NUV band, while J0946+0139 has just a FUV detection
  from MIS. The latter is clearly visible in the NUV deep image, but it has
  been erroneously centered. The presence of a nearby galaxy might be the
  cause of the misalignment. For this source we have decided to consider the
  data coming from the AIS survey.} have data coming from MIS, 6 from AIS, and
one from the GI program; while for 4 objects we do not have any detection in
both GALEX bands.
\begin{figure*}
     \begin{center}
        \subfigure{\label{fig:first}
            \includegraphics[width=0.4\textwidth]{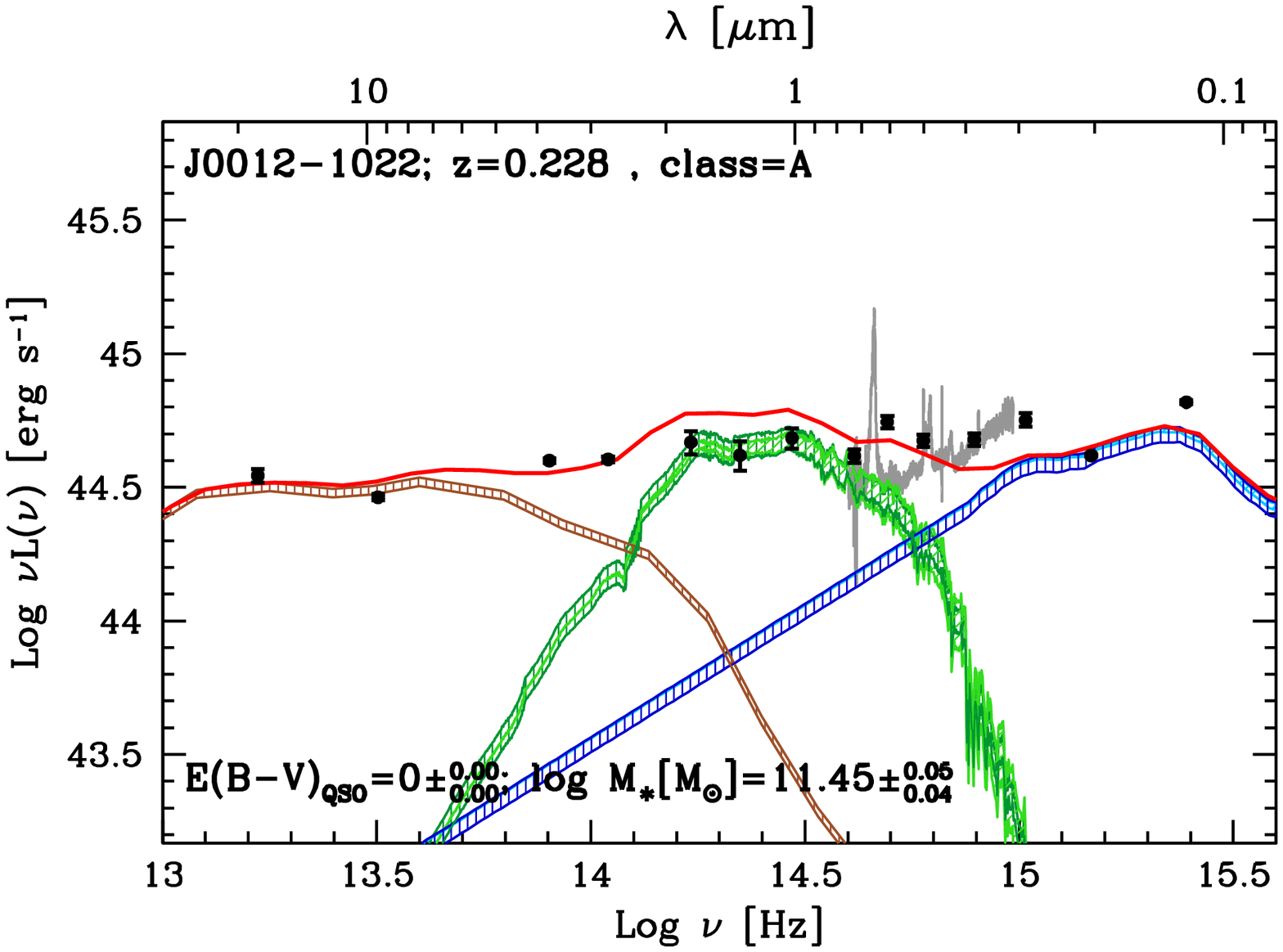}
        }
        \subfigure{\label{fig:second}
           \includegraphics[width=0.4\textwidth]{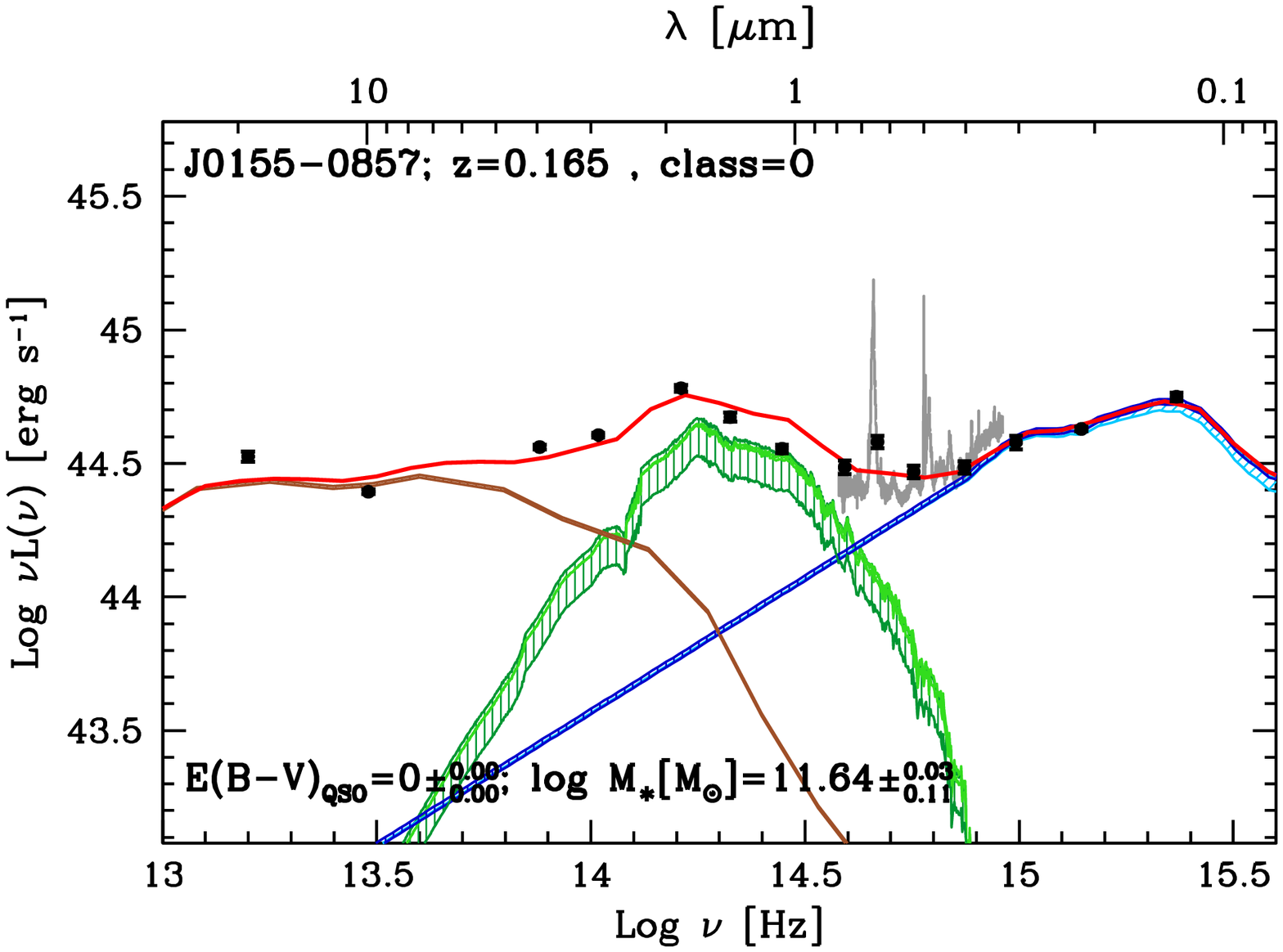}
        }\\
        \vspace{-2.2cm} 
        \subfigure{\label{fig:first}
            \includegraphics[width=0.4\textwidth]{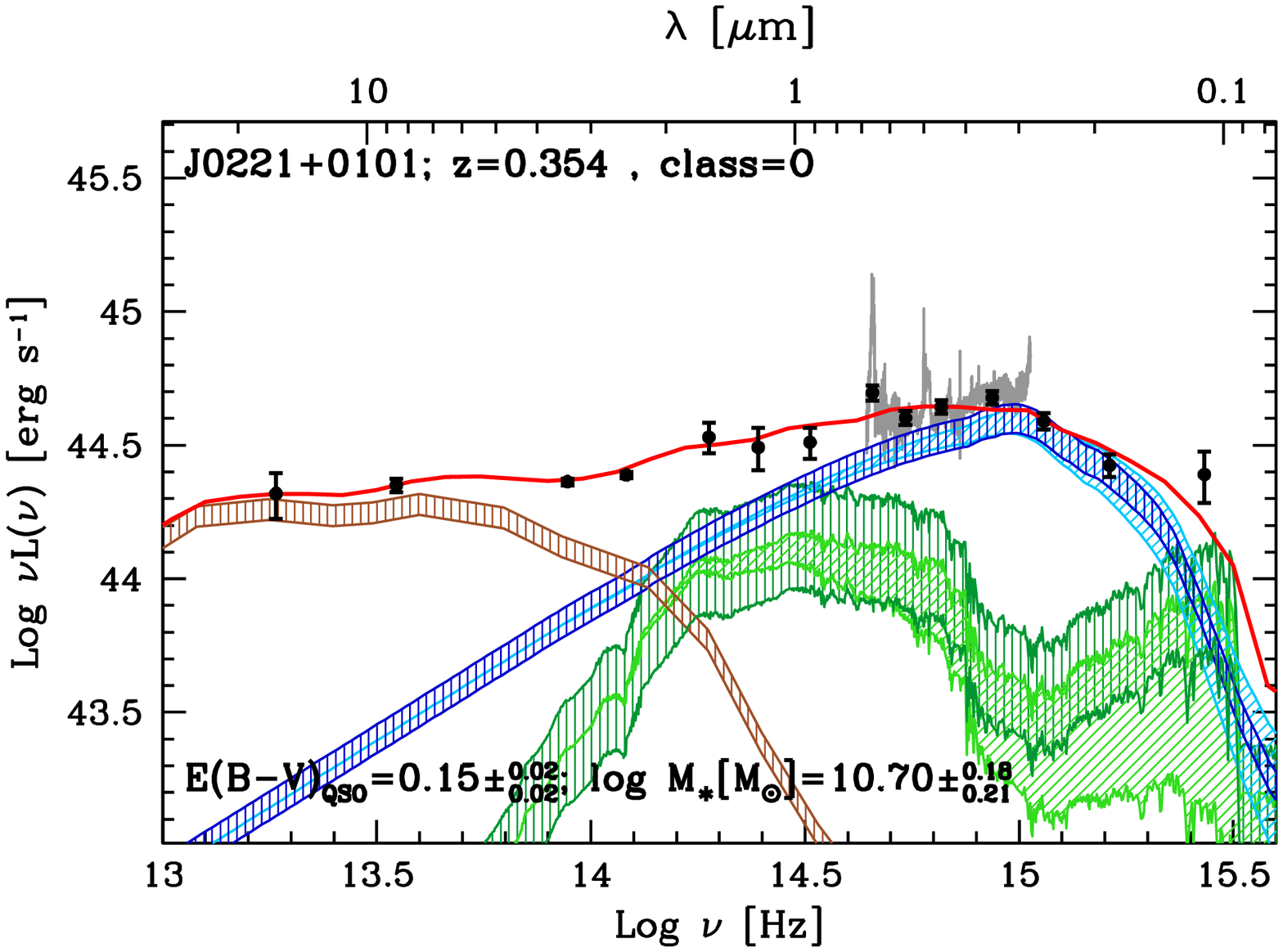}
        }
        \subfigure{\label{fig:second}
           \includegraphics[width=0.4\textwidth]{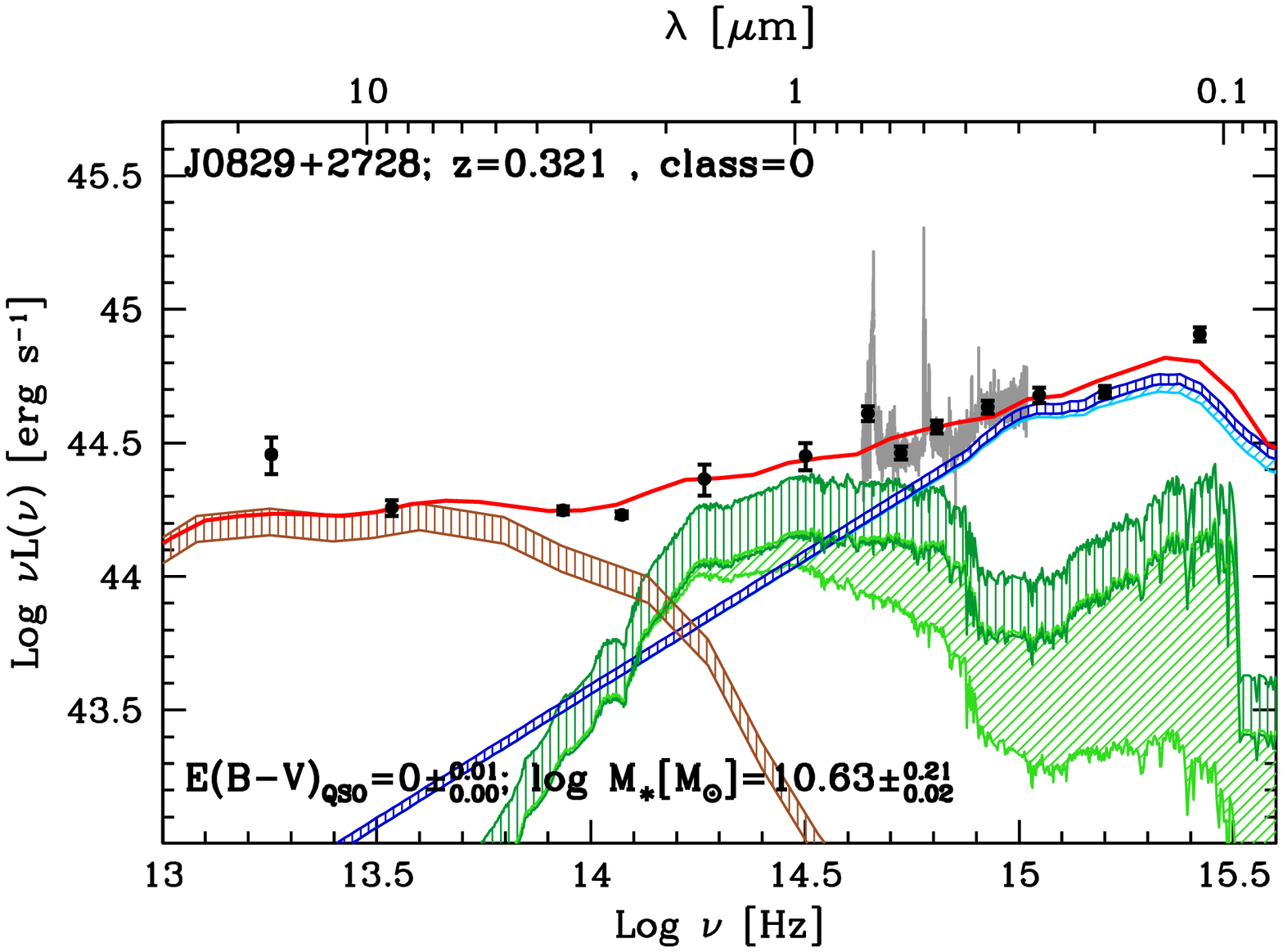}
        }
    \end{center}
    \caption{Rest-frame SED decompositions of the quasars sample. Each panel shows source's name, redshift, class, best-fit AGN reddening (estimated considering the Prevot et al. 1984 reddening law), and best-fit stellar mass with uncertainties. Object classification is based on the shape of Balmer lines: B- BHB candidates; D- DPEs; A- sources with Asymmetric line profiles; O- Others. Black circles are the observed photometry in the rest-frame (from infrared to optical-UV). 
The brown, dark green, and blue dashed regions correspond to the uncertainties (see \S~\ref{Model fitting}) on the normalization of the hot-dust from reprocessed AGN emission, galaxy, and BBB templates, respectively. The light green and cyan dashed regions encompass the uncertainty on the BBB and galaxy reddening, respectively.
The red line represents the best-fit SED. Grey line shows the optical spectrum from SDSS.}
   \label{bhbseds}
\end{figure*}
\begin{figure*}
\addtocounter{figure}{-1}
     \begin{center}
        \subfigure{\label{fig:first}
            \includegraphics[width=0.4\textwidth]{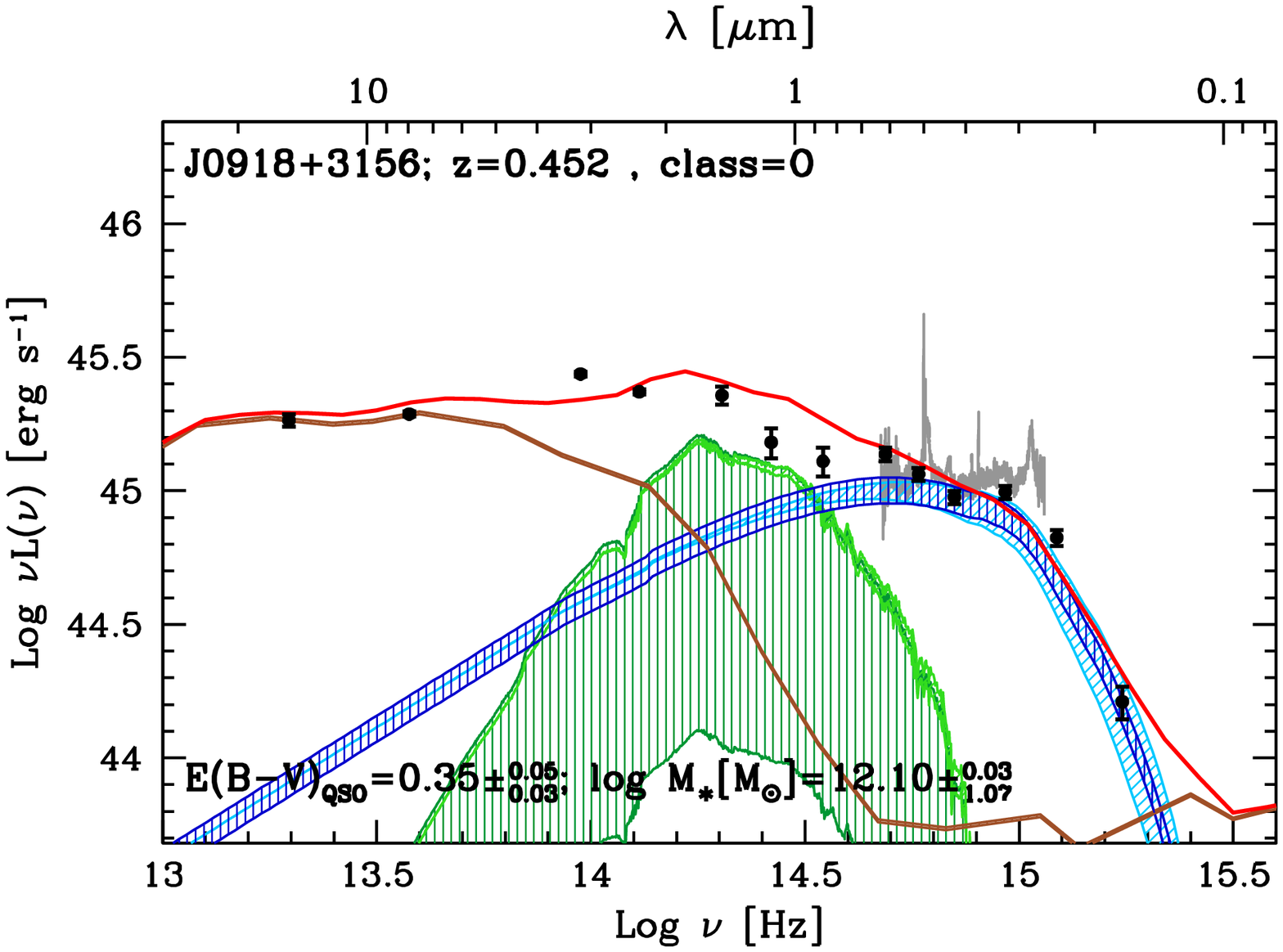}
        }
        \subfigure{\label{fig:second}
           \includegraphics[width=0.4\textwidth]{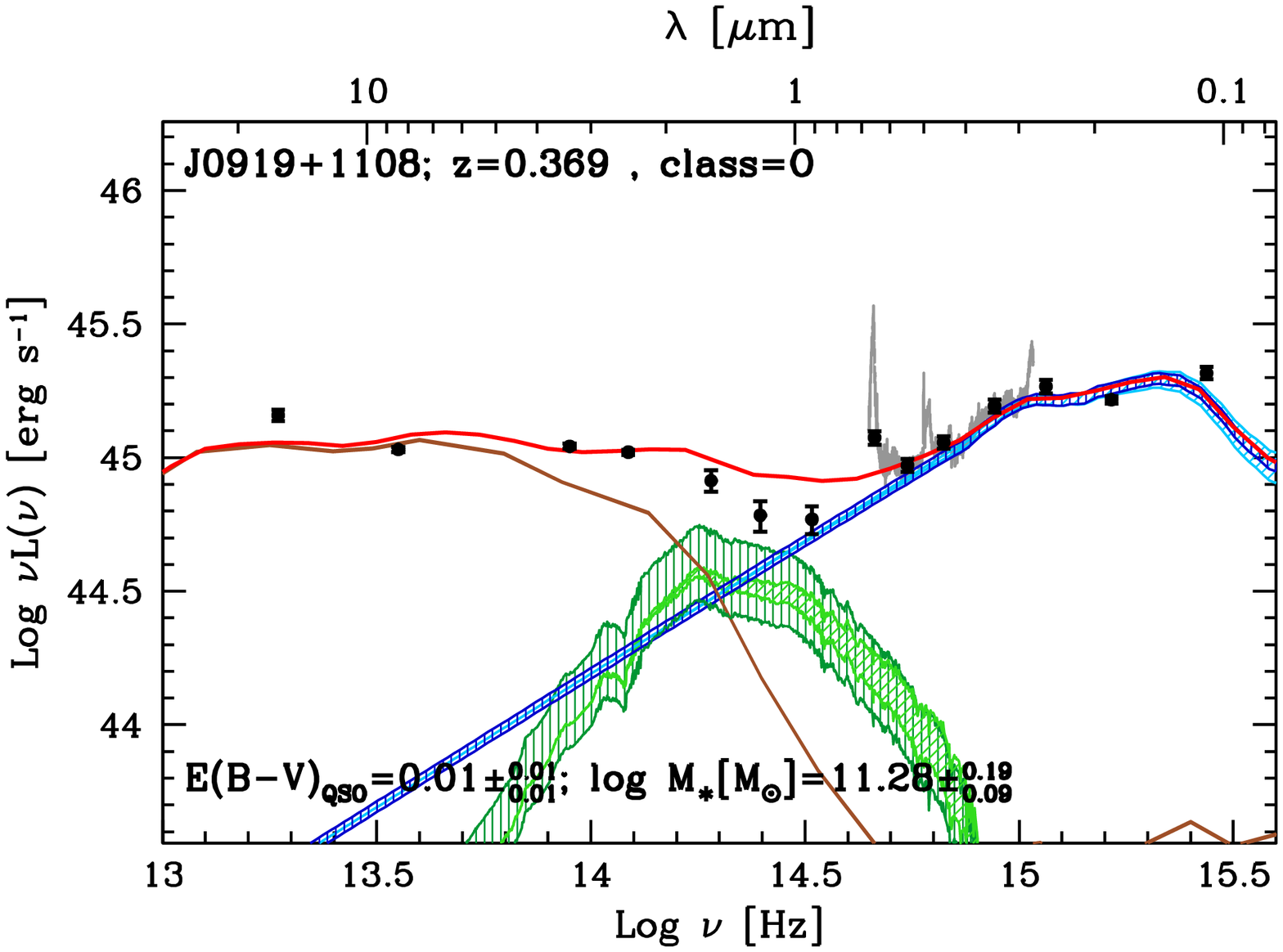}
        }\\   
	\vspace{-2.2cm} 
        \subfigure{
            \includegraphics[width=0.4\textwidth]{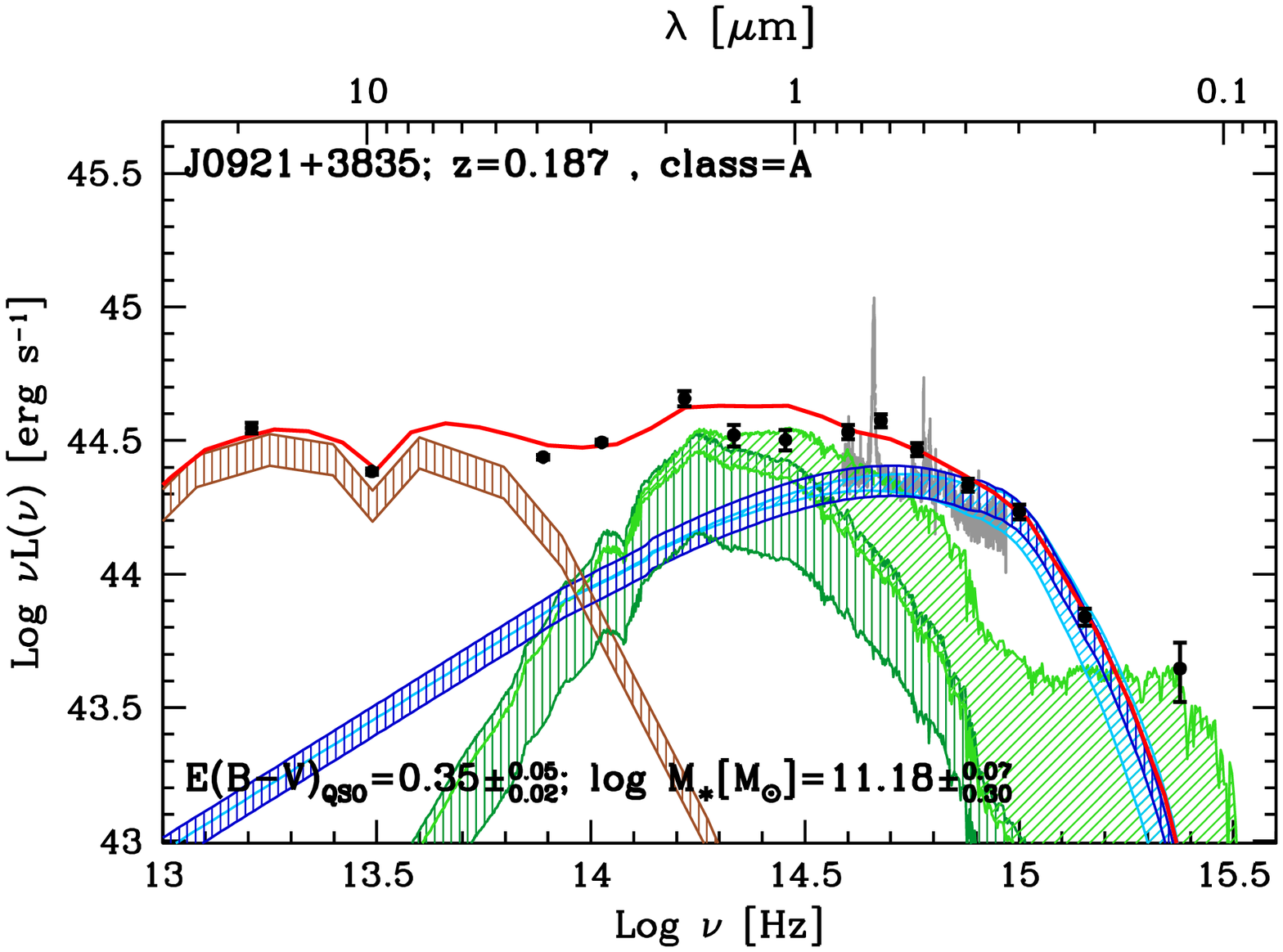}
        }
        \subfigure{
           \includegraphics[width=0.4\textwidth]{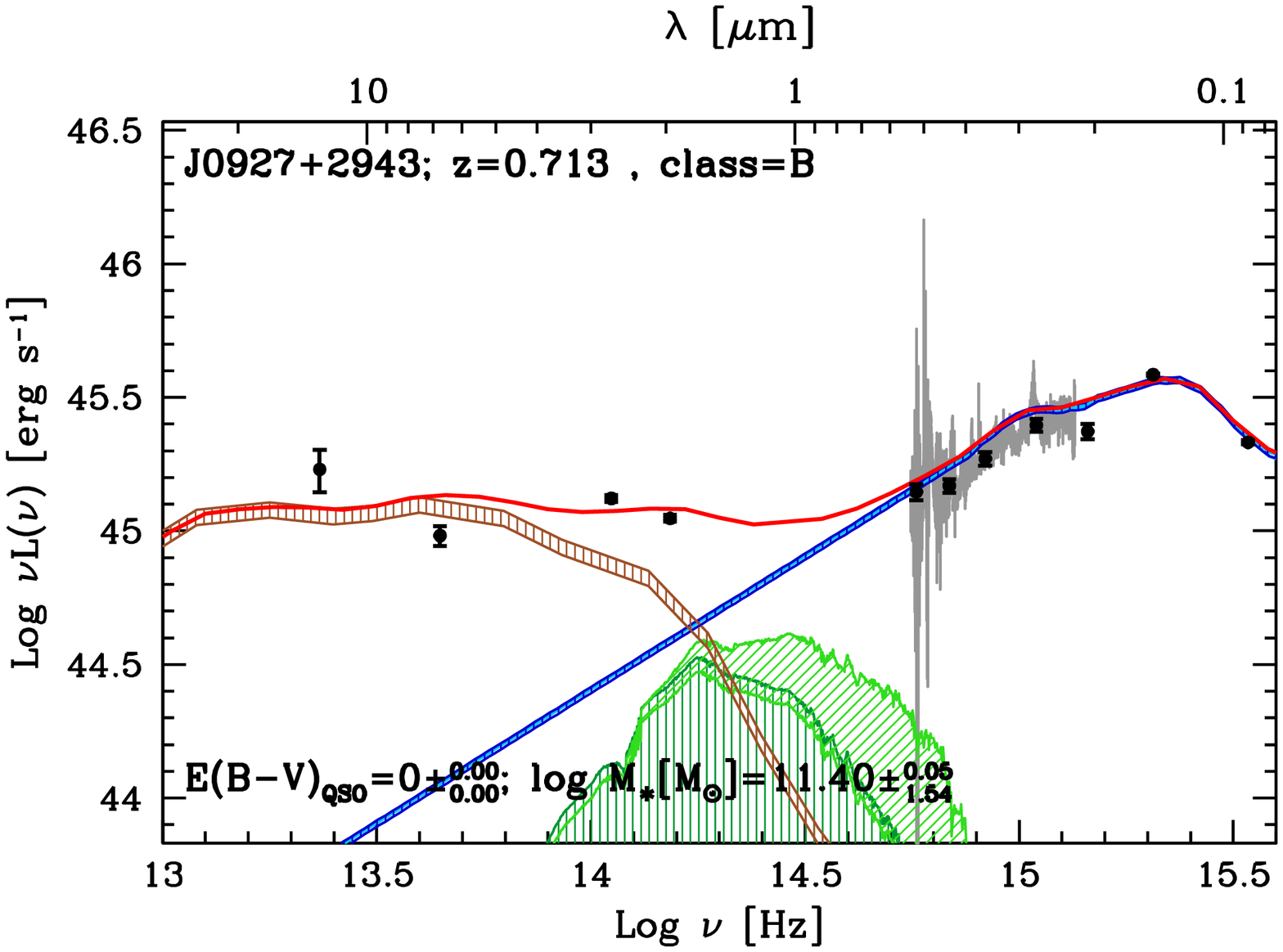}
        }\\ 
        \vspace{-2.2cm} 
        \subfigure{
            \includegraphics[width=0.4\textwidth]{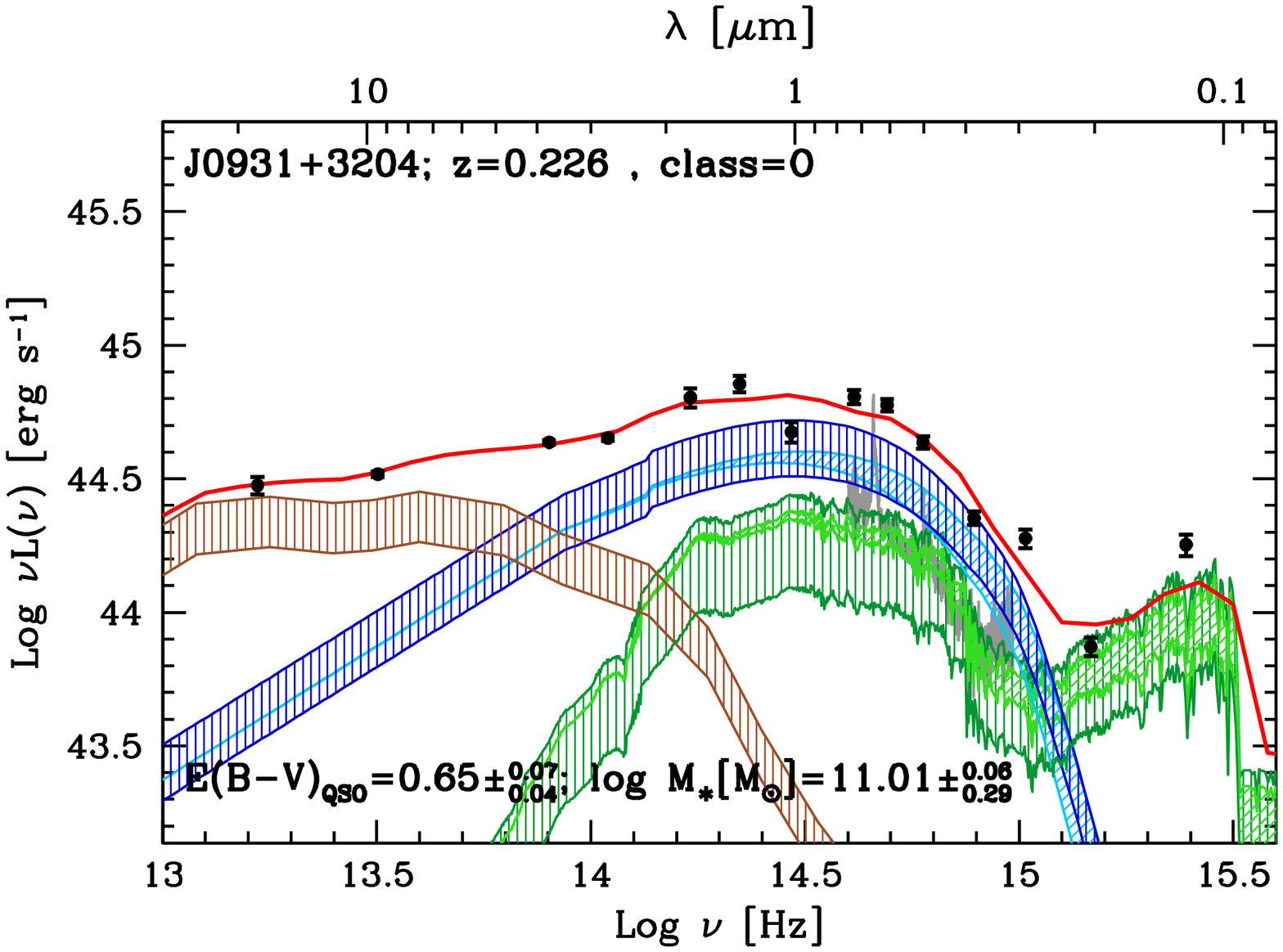}
        }
        \subfigure{
            \includegraphics[width=0.4\textwidth]{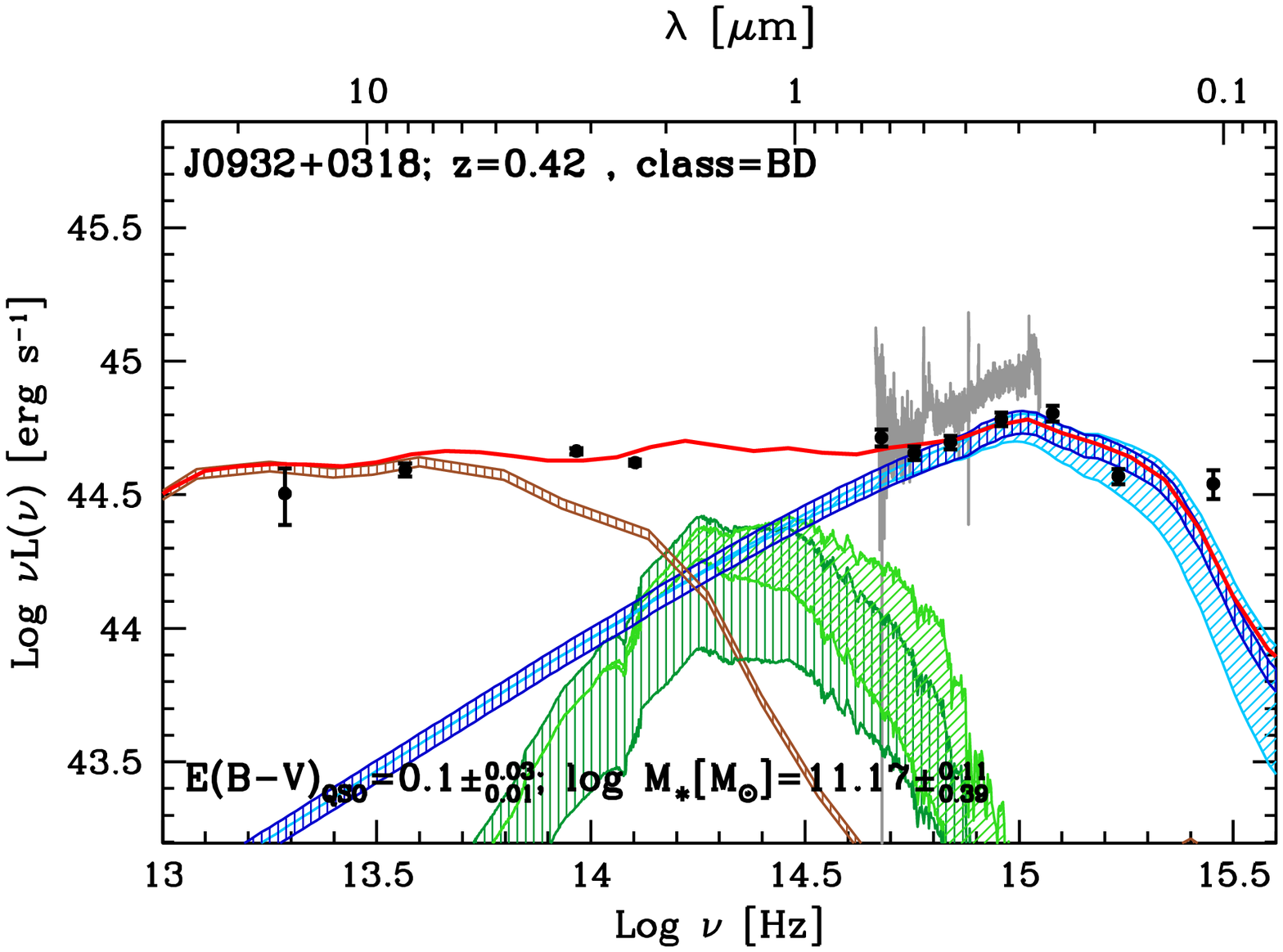}
        }\\ 
        \vspace{-2.2cm} 
        \subfigure{
            \includegraphics[width=0.4\textwidth]{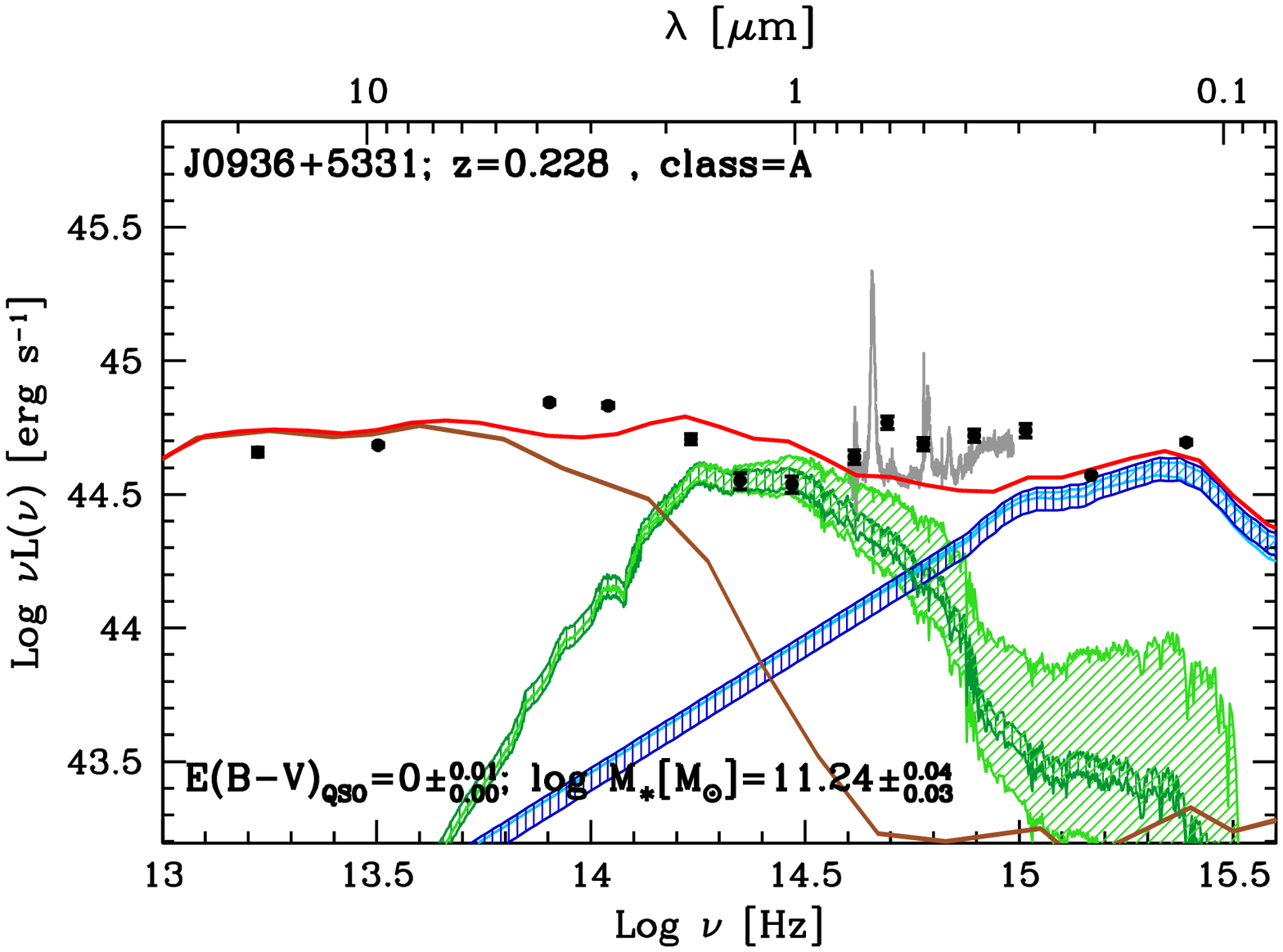}
        }
        \subfigure{
            \includegraphics[width=0.4\textwidth]{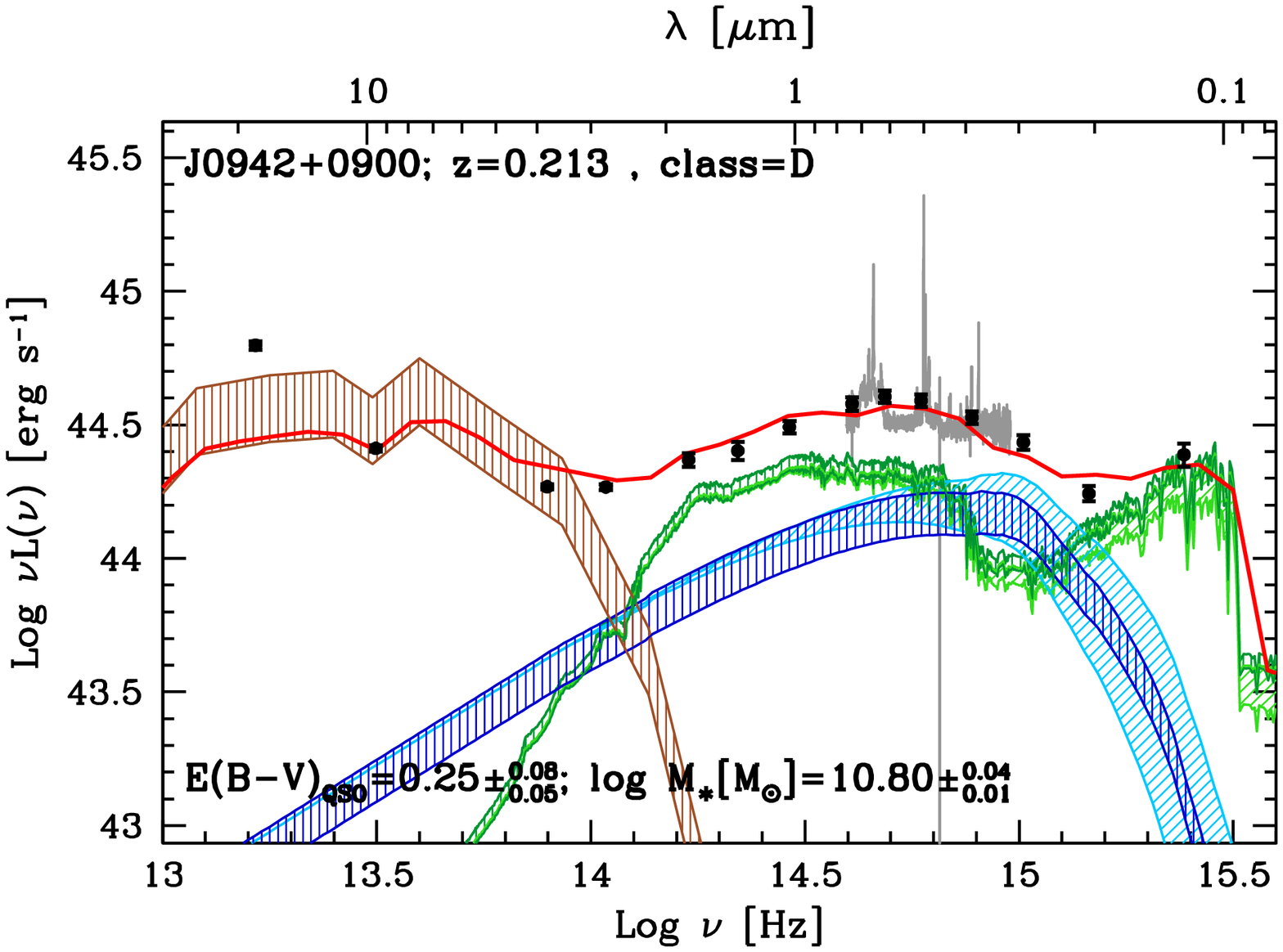}
        }
    \end{center}
    \caption[]{Continued}
\end{figure*}
\begin{figure*}
\addtocounter{figure}{-1}
     \begin{center}
        \subfigure{
            \includegraphics[width=0.4\textwidth]{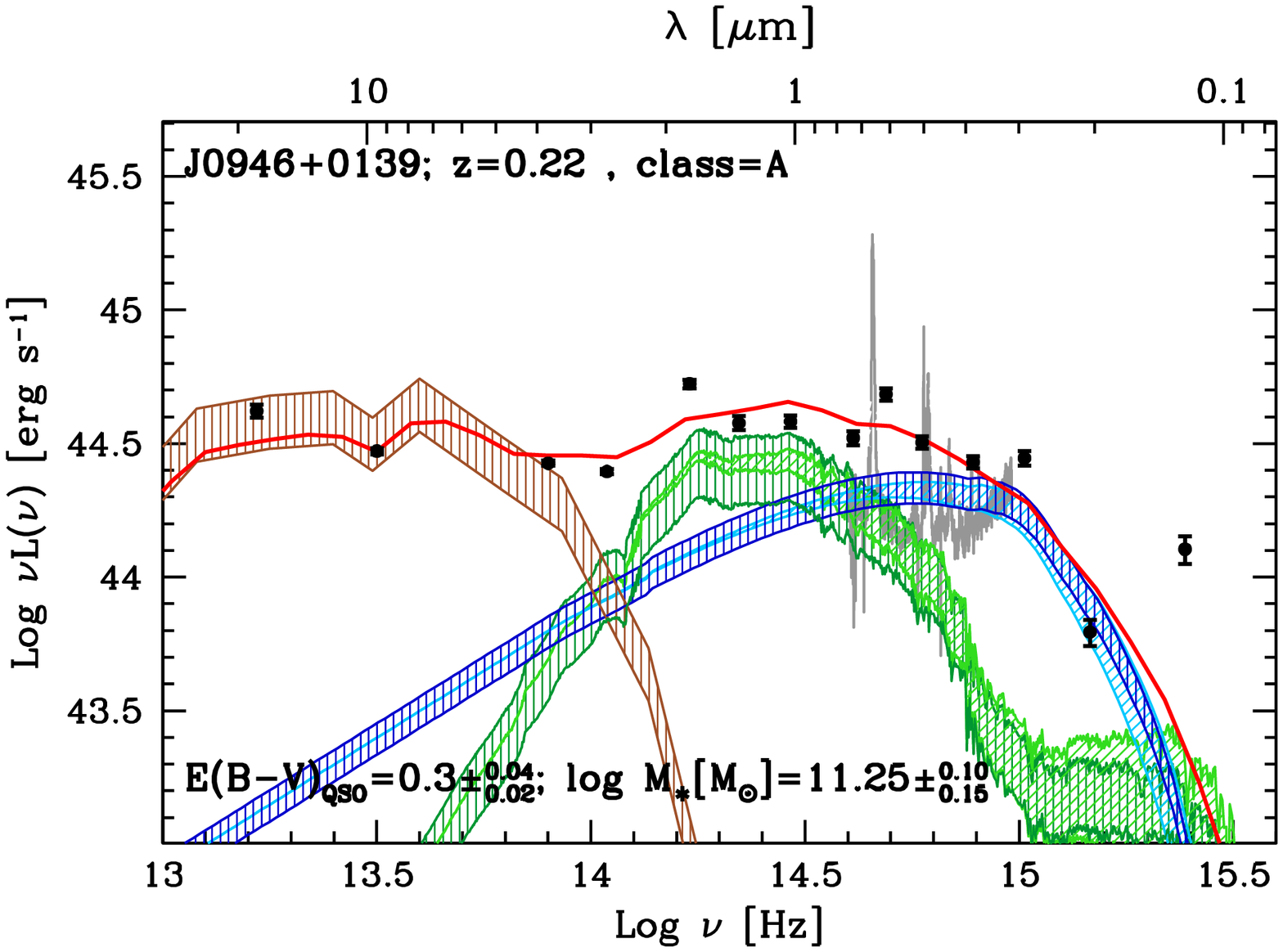}
        }
        \subfigure{
            \includegraphics[width=0.4\textwidth]{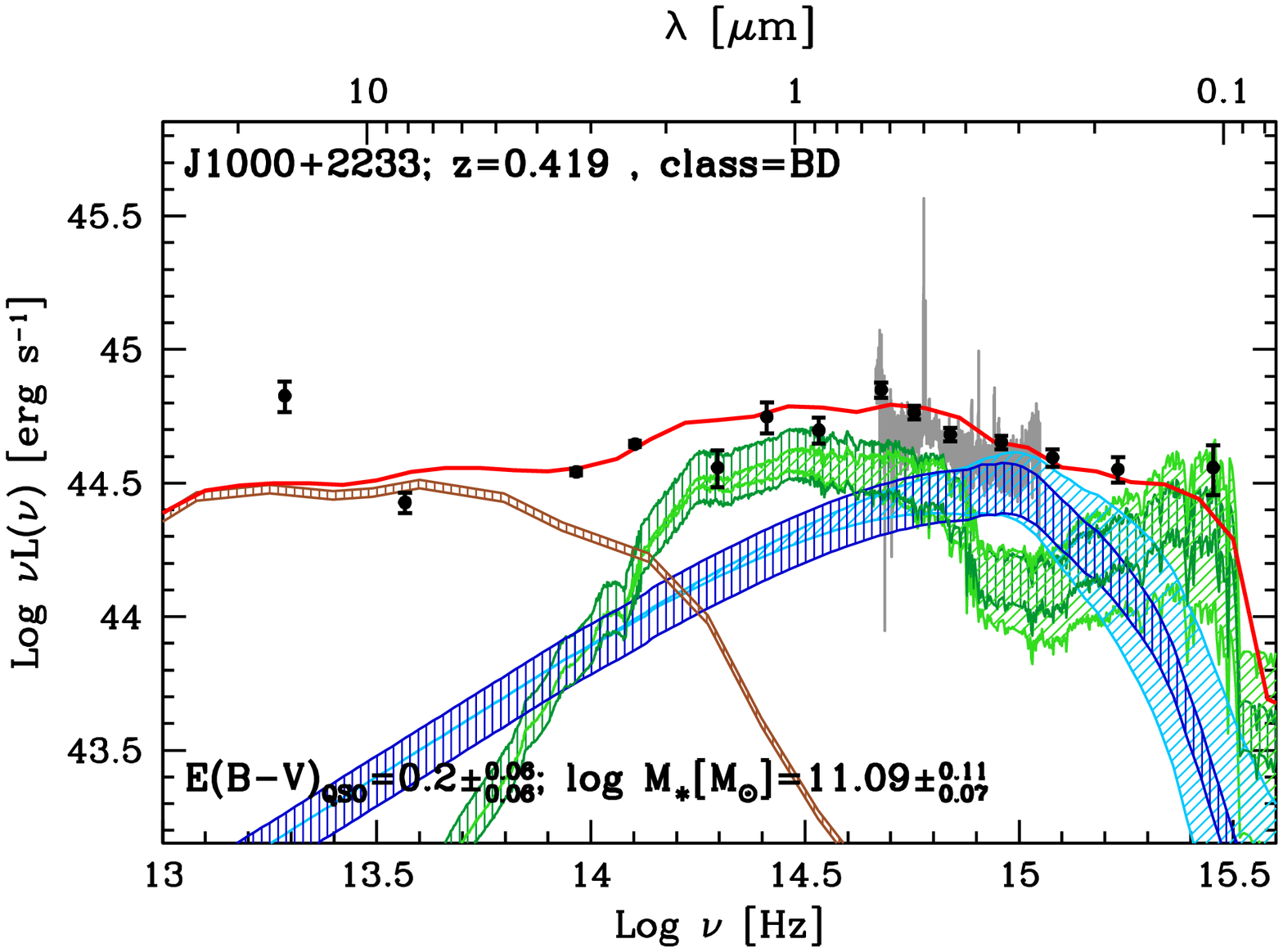}
        }\\
        \vspace{-2.2cm}     
        \subfigure{
            \includegraphics[width=0.4\textwidth]{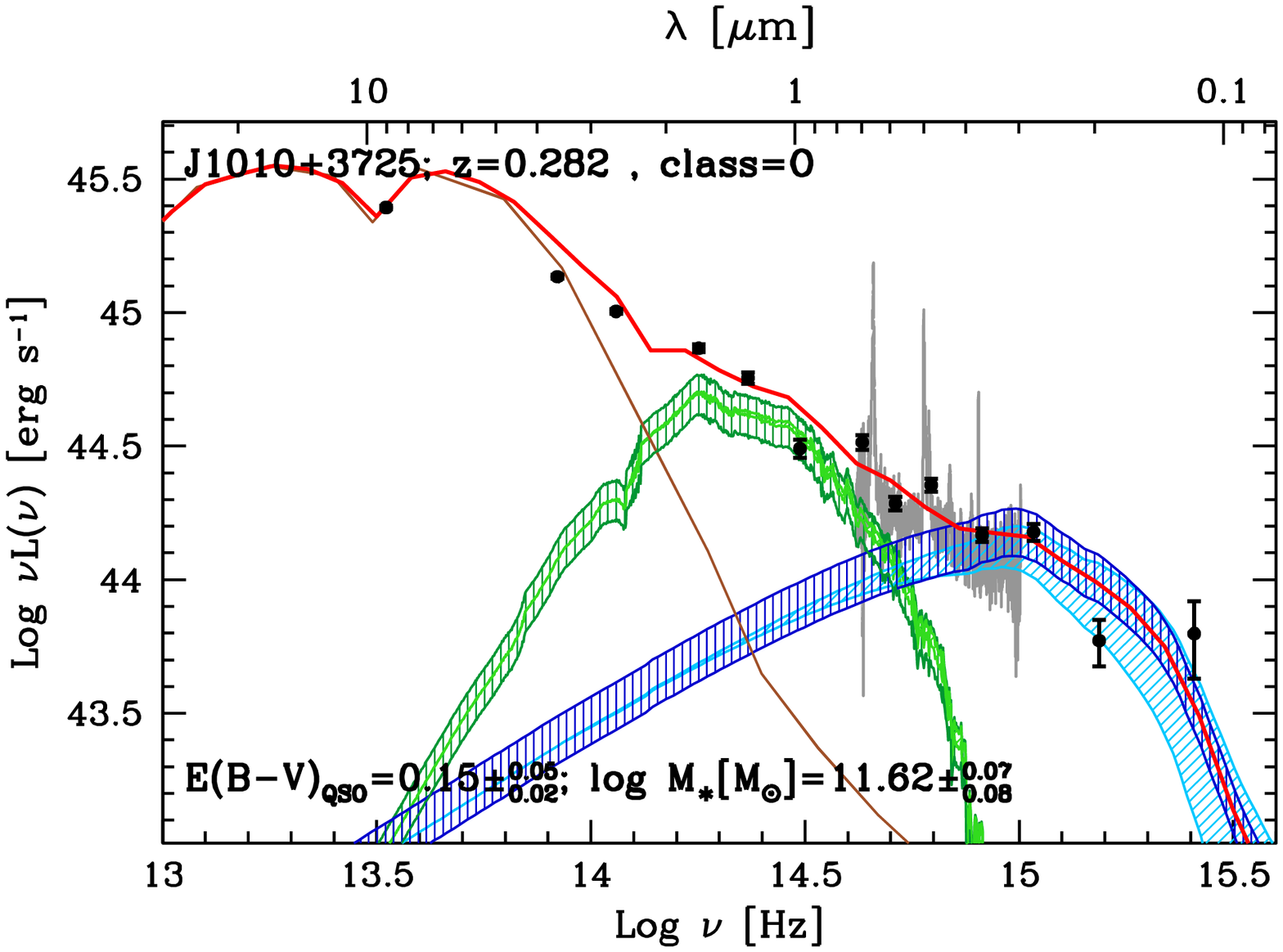}
        }
        \subfigure{
           \includegraphics[width=0.4\textwidth]{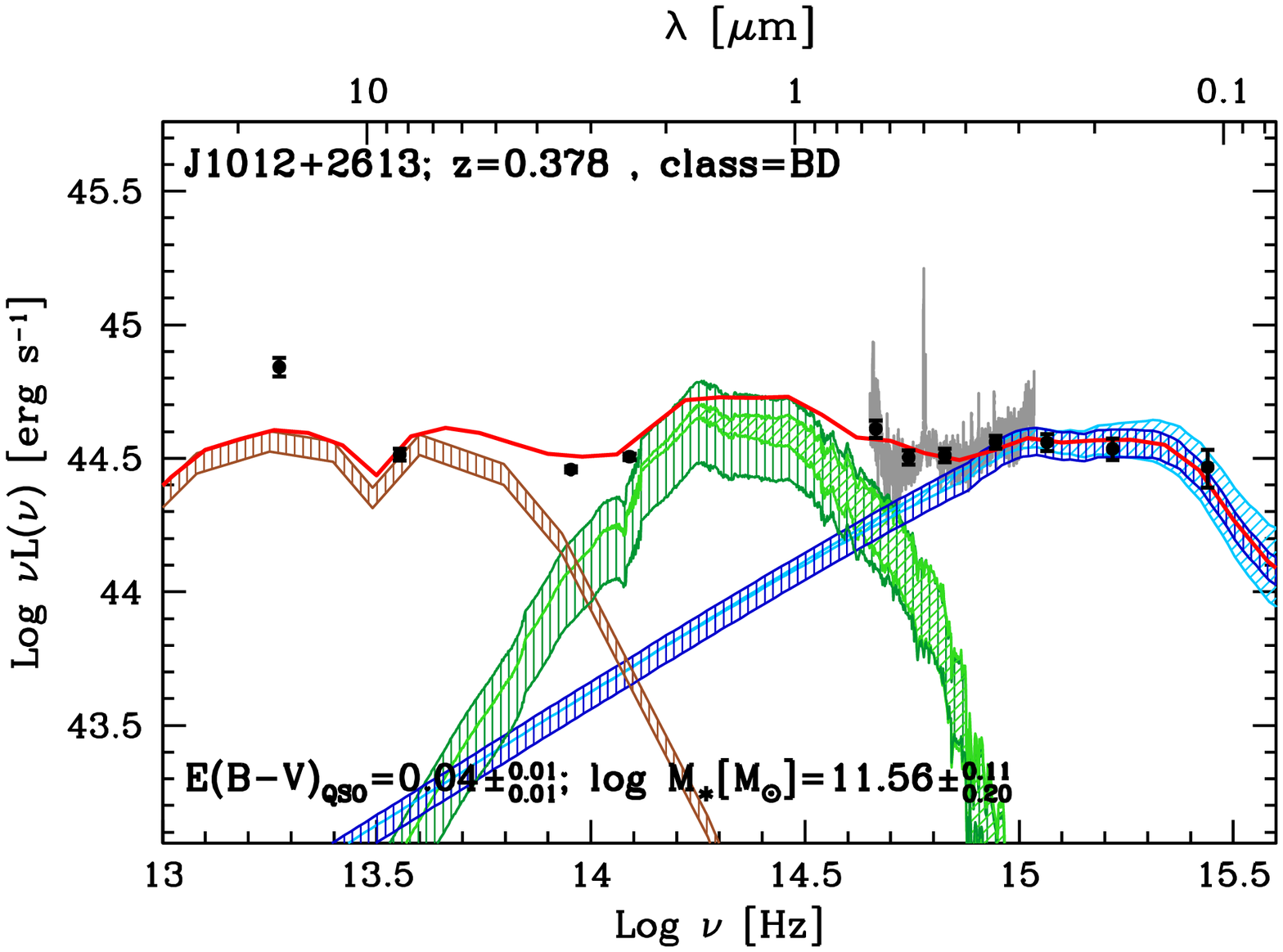}
        }\\ 
        \vspace{-2.2cm} 
        \subfigure{
            \includegraphics[width=0.4\textwidth]{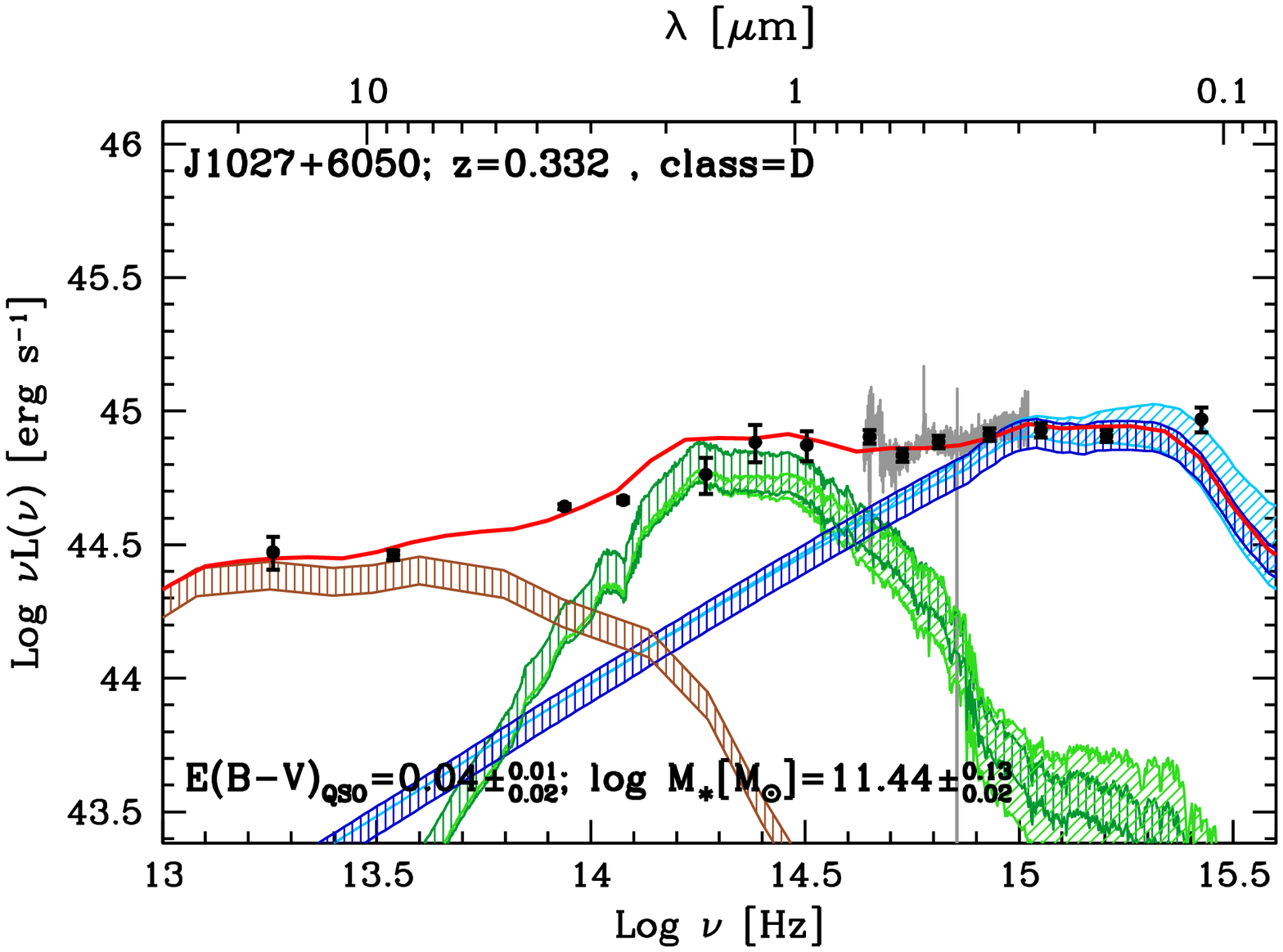}
        }
        \subfigure{
            \includegraphics[width=0.4\textwidth]{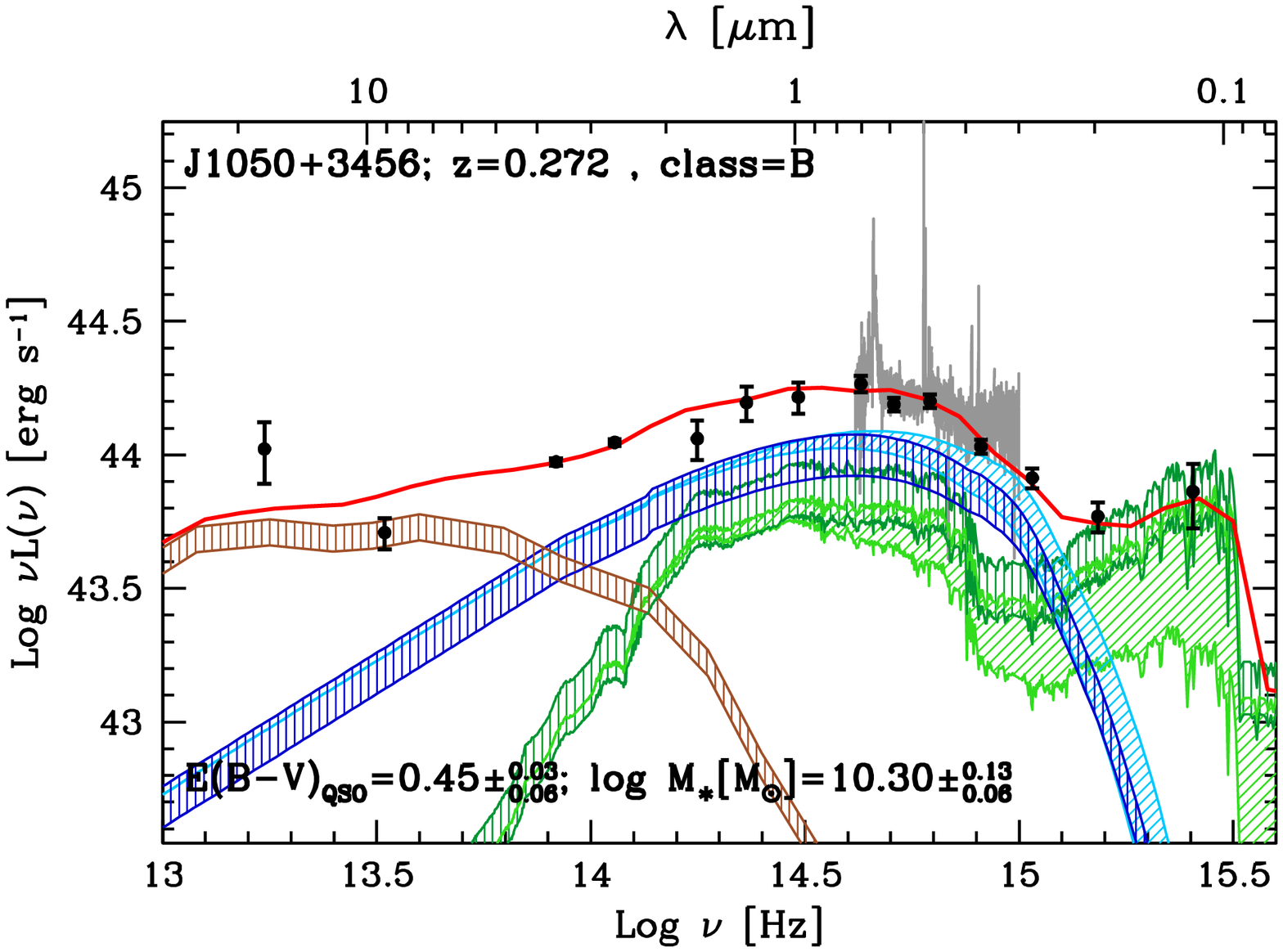}
        }\\ 
        \vspace{-2.2cm} 
        \subfigure{
            \includegraphics[width=0.4\textwidth]{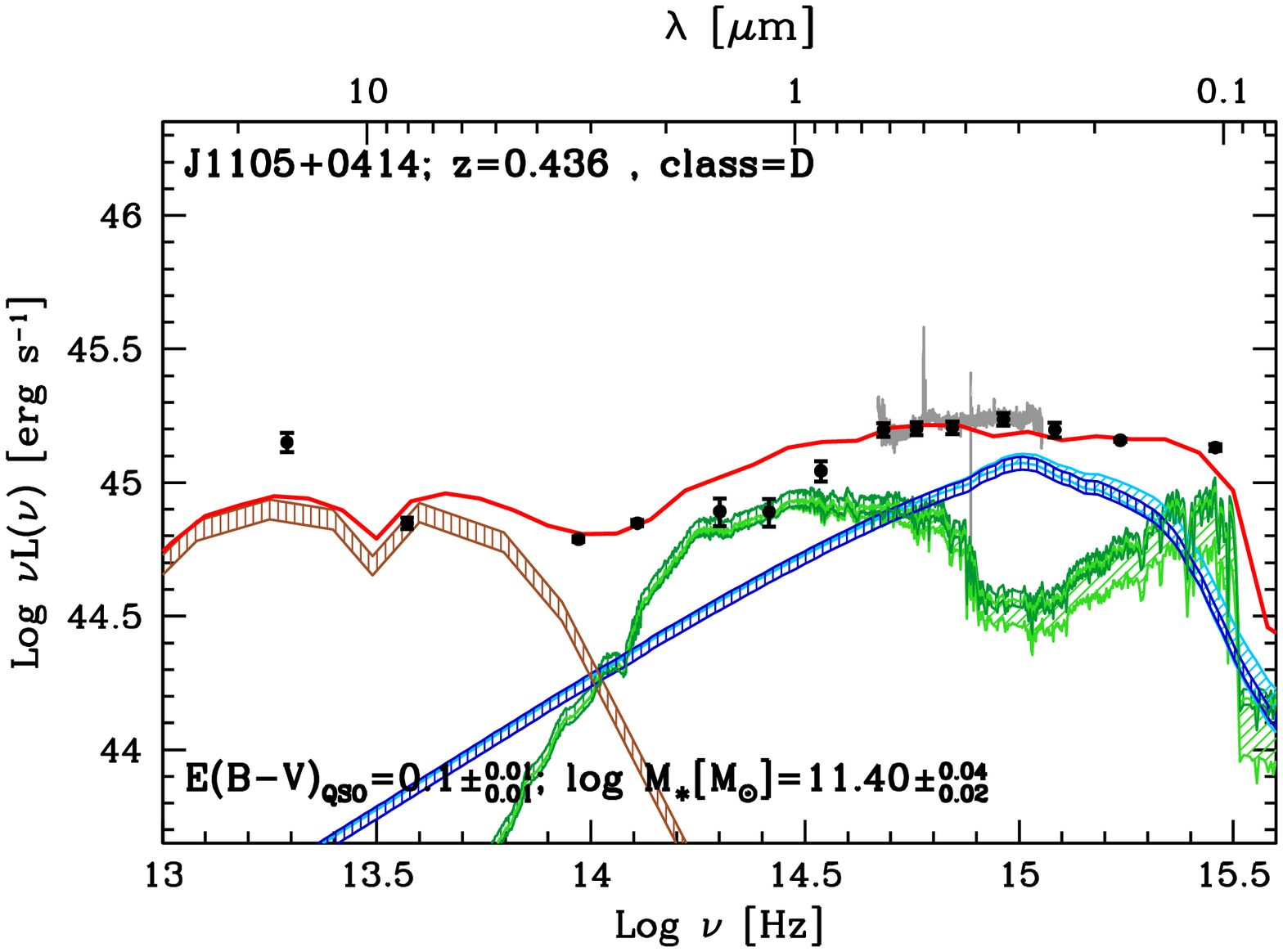}
        }
        \subfigure{
            \includegraphics[width=0.4\textwidth]{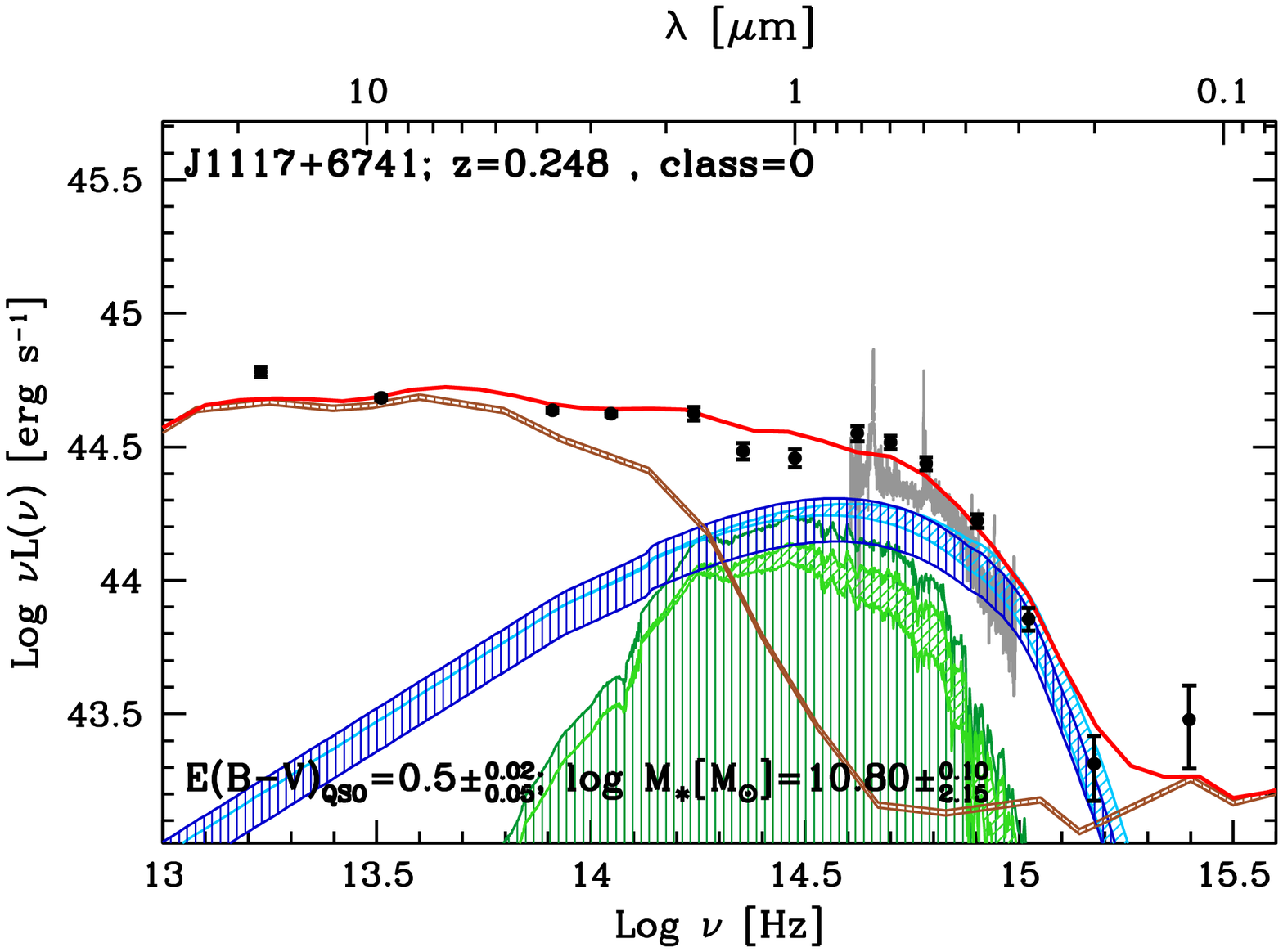}
        }\\
    \end{center}
    \caption[]{Continued}
\end{figure*}
\begin{figure*}
\addtocounter{figure}{-1}
     \begin{center}
        \subfigure{
            \includegraphics[width=0.4\textwidth]{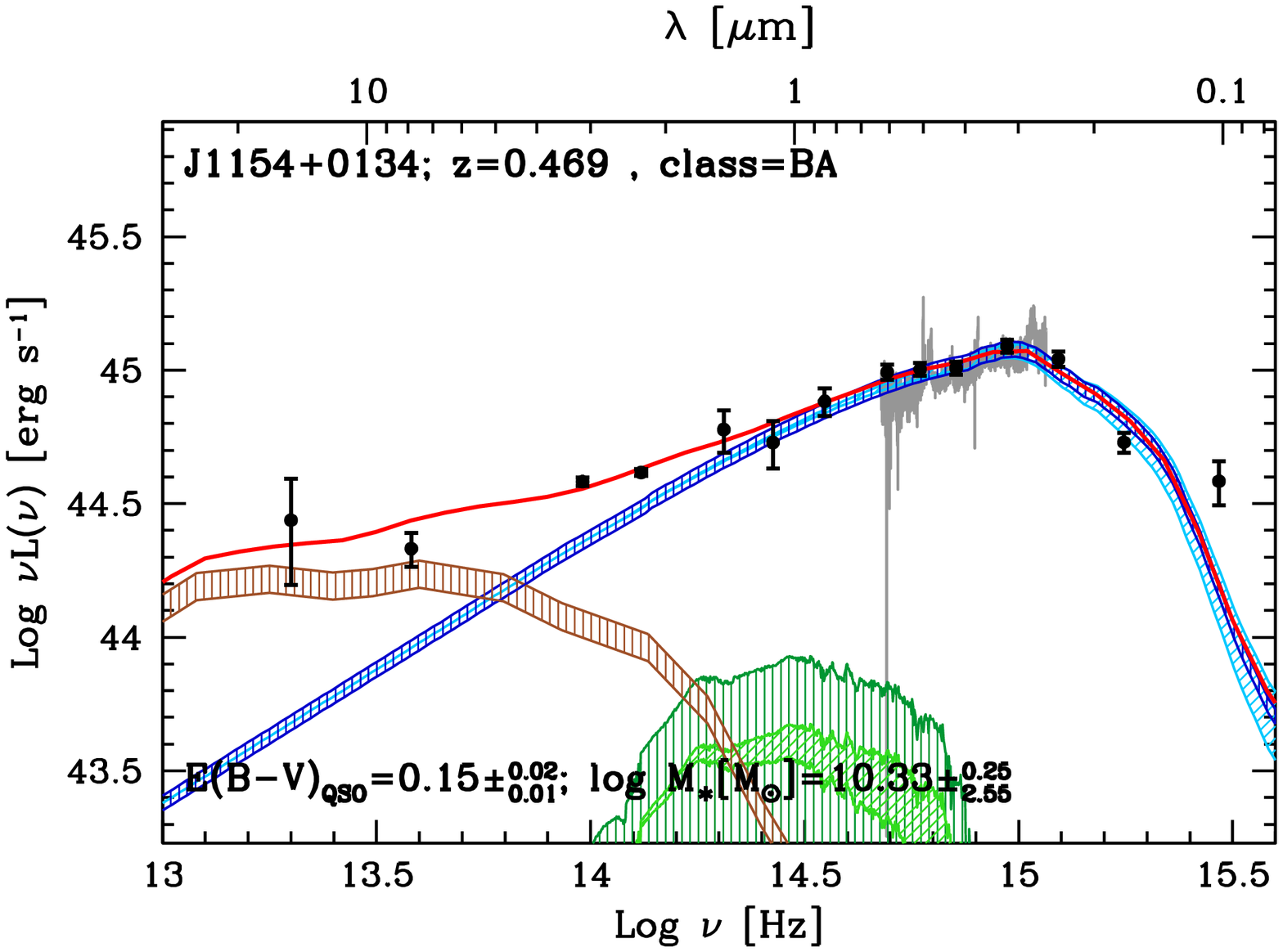}
        }
        \subfigure{
            \includegraphics[width=0.4\textwidth]{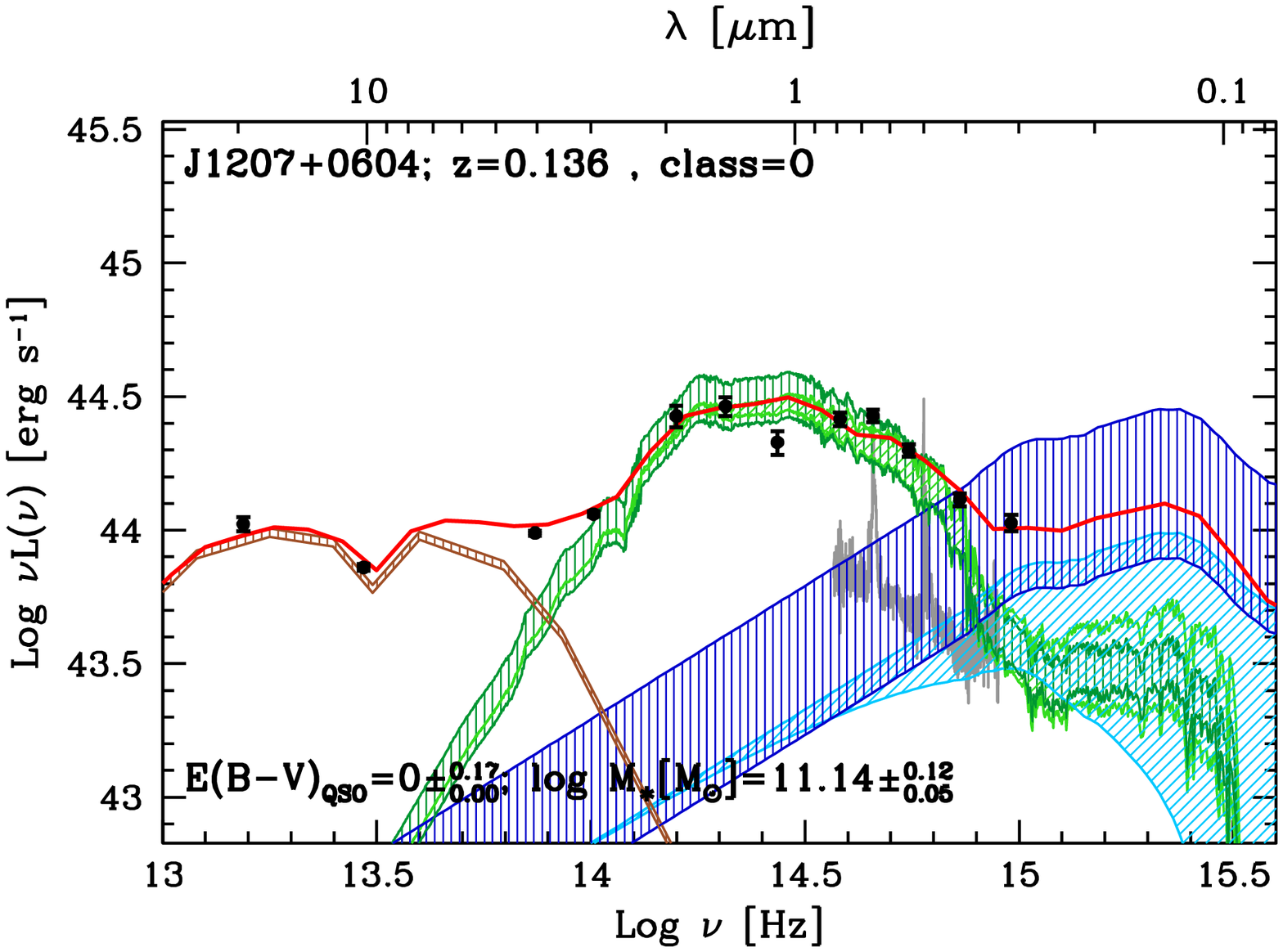}
        }\\
        \vspace{-2.2cm}
        \subfigure{
            \includegraphics[width=0.4\textwidth]{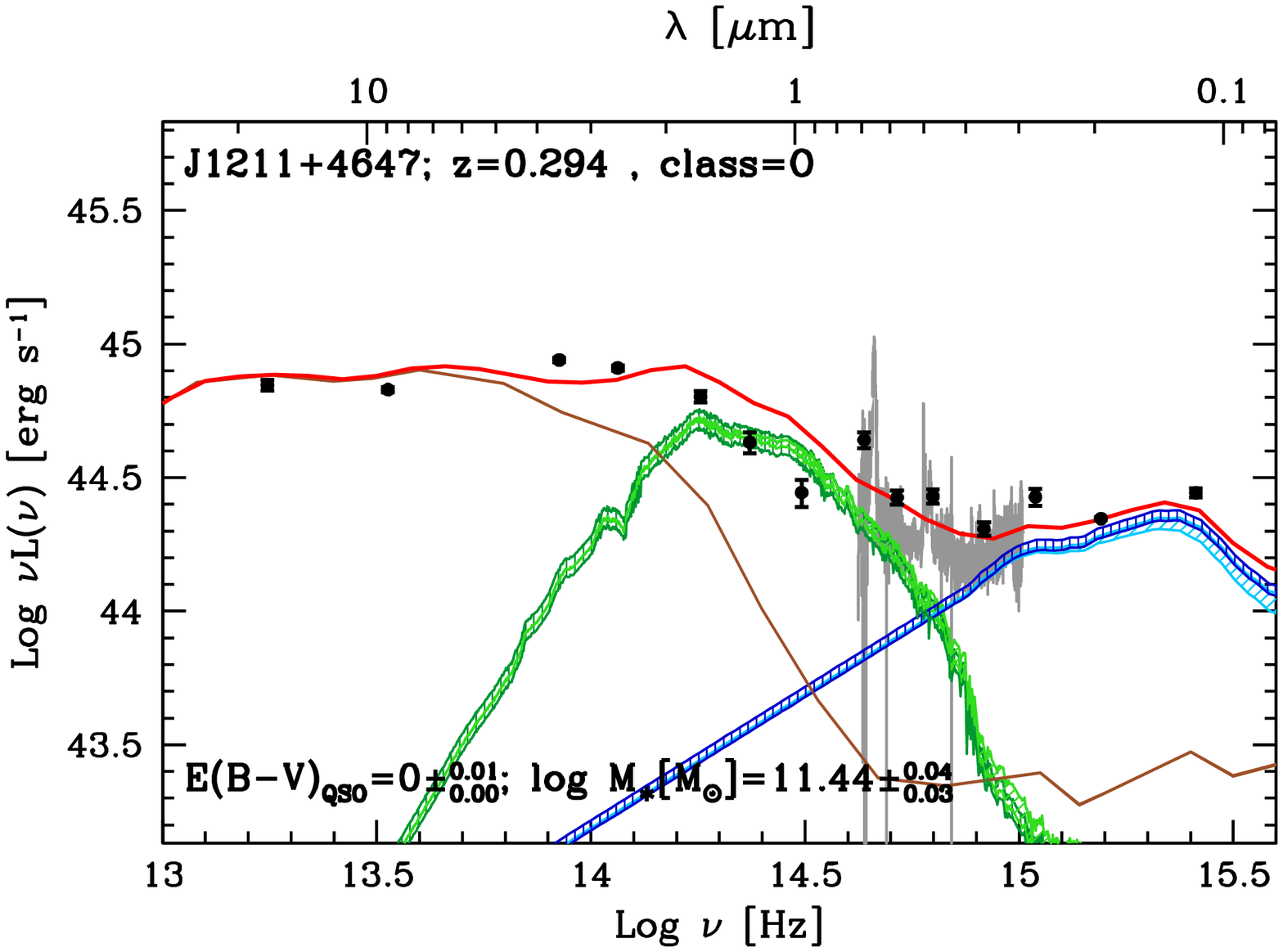}
        }
        \subfigure{
           \includegraphics[width=0.4\textwidth]{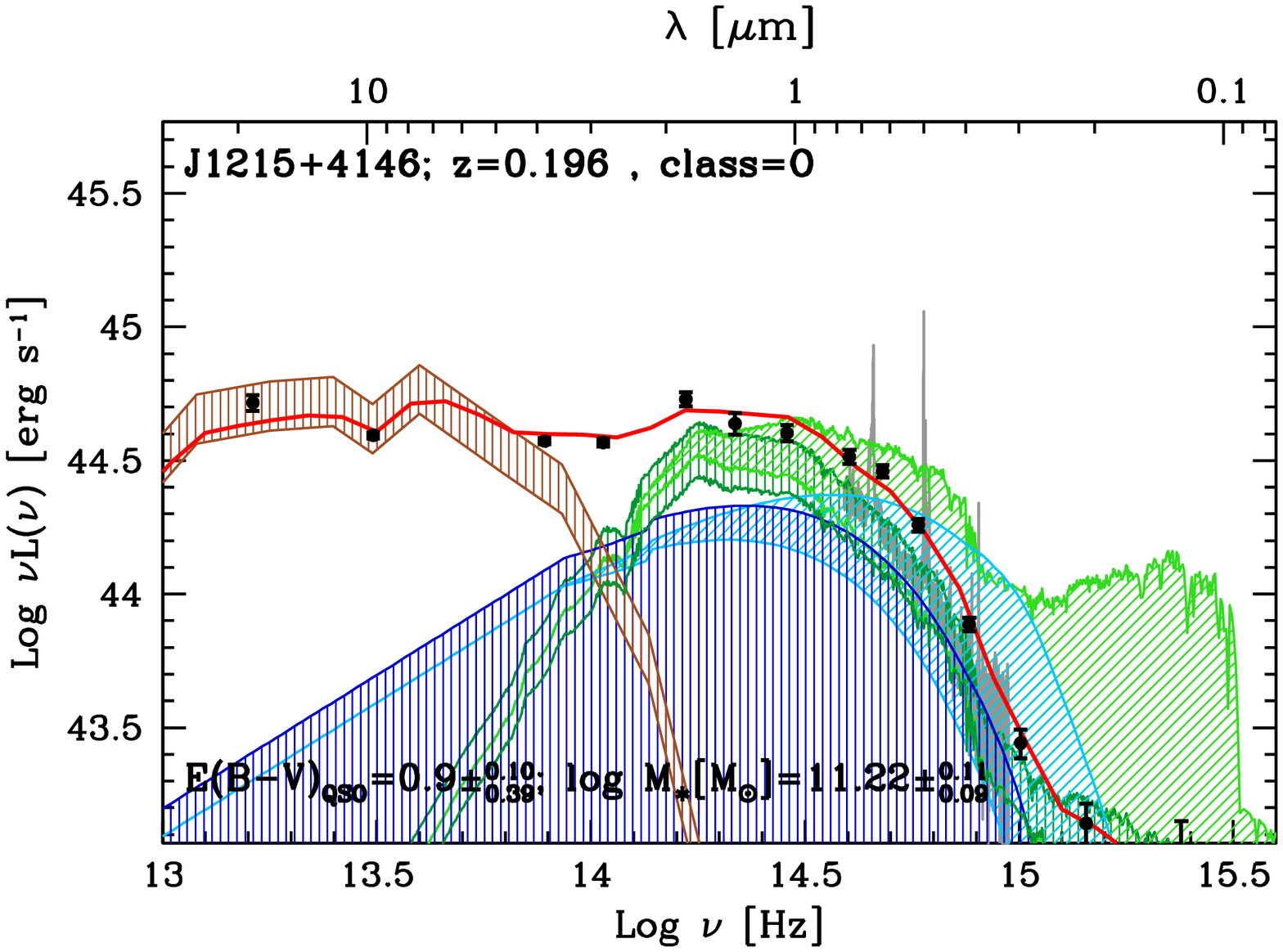}
        }\\ 
        \vspace{-2.2cm} 
        \subfigure{
            \includegraphics[width=0.4\textwidth]{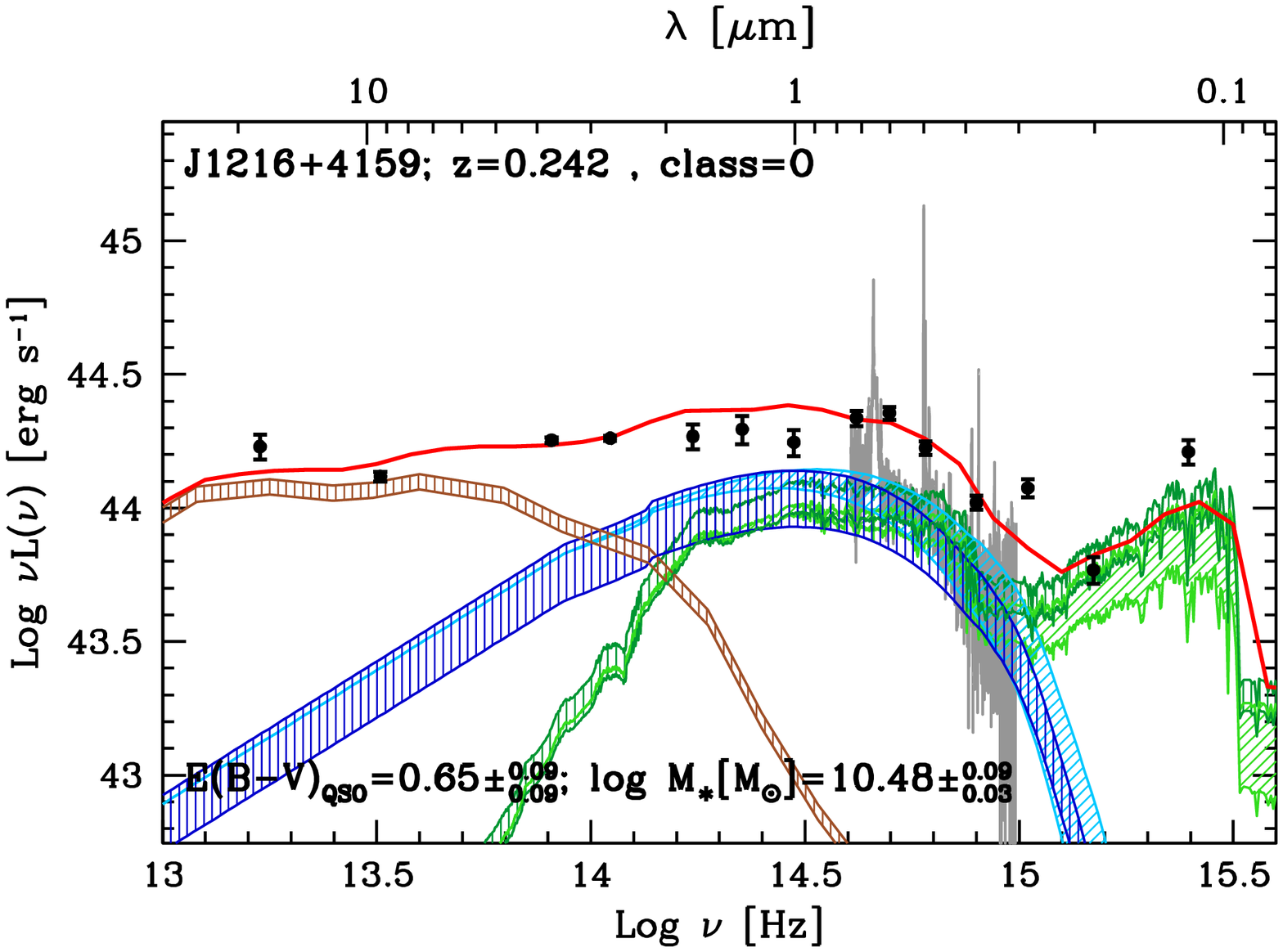}
        }
        \subfigure{
            \includegraphics[width=0.4\textwidth]{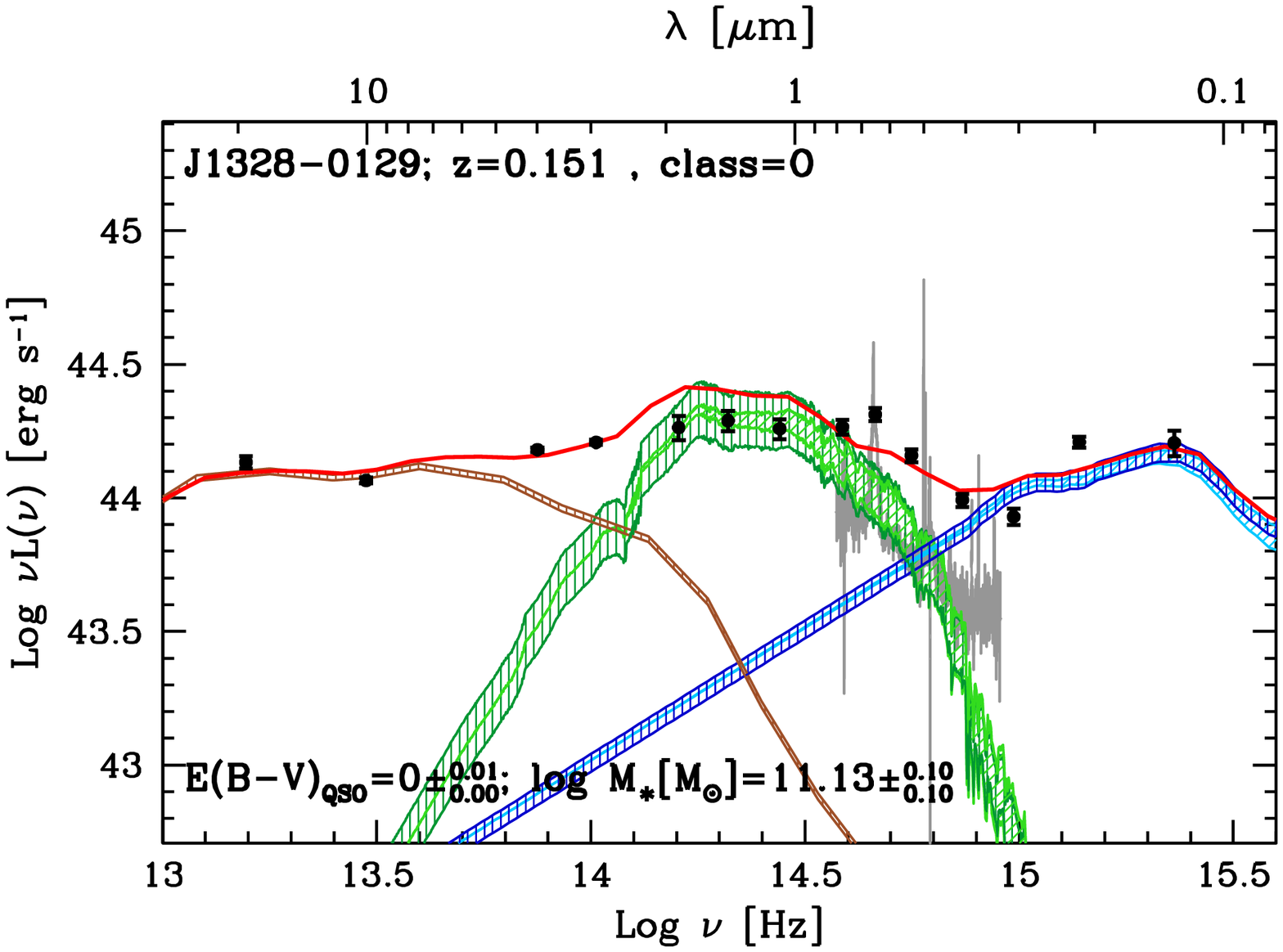}
        }\\ 
        \vspace{-2.2cm} 
        \subfigure{
            \includegraphics[width=0.4\textwidth]{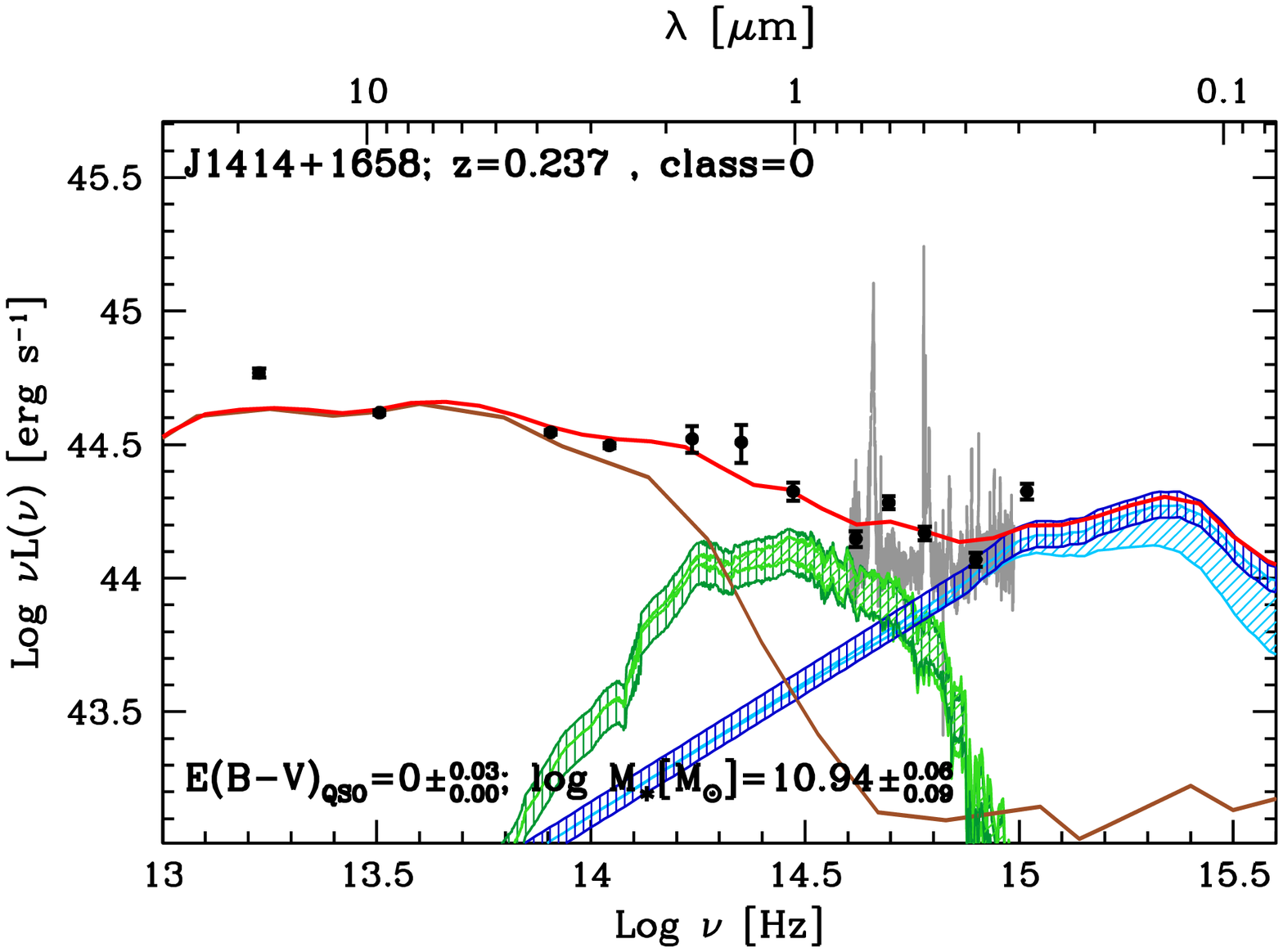}
        }
        \subfigure{
            \includegraphics[width=0.4\textwidth]{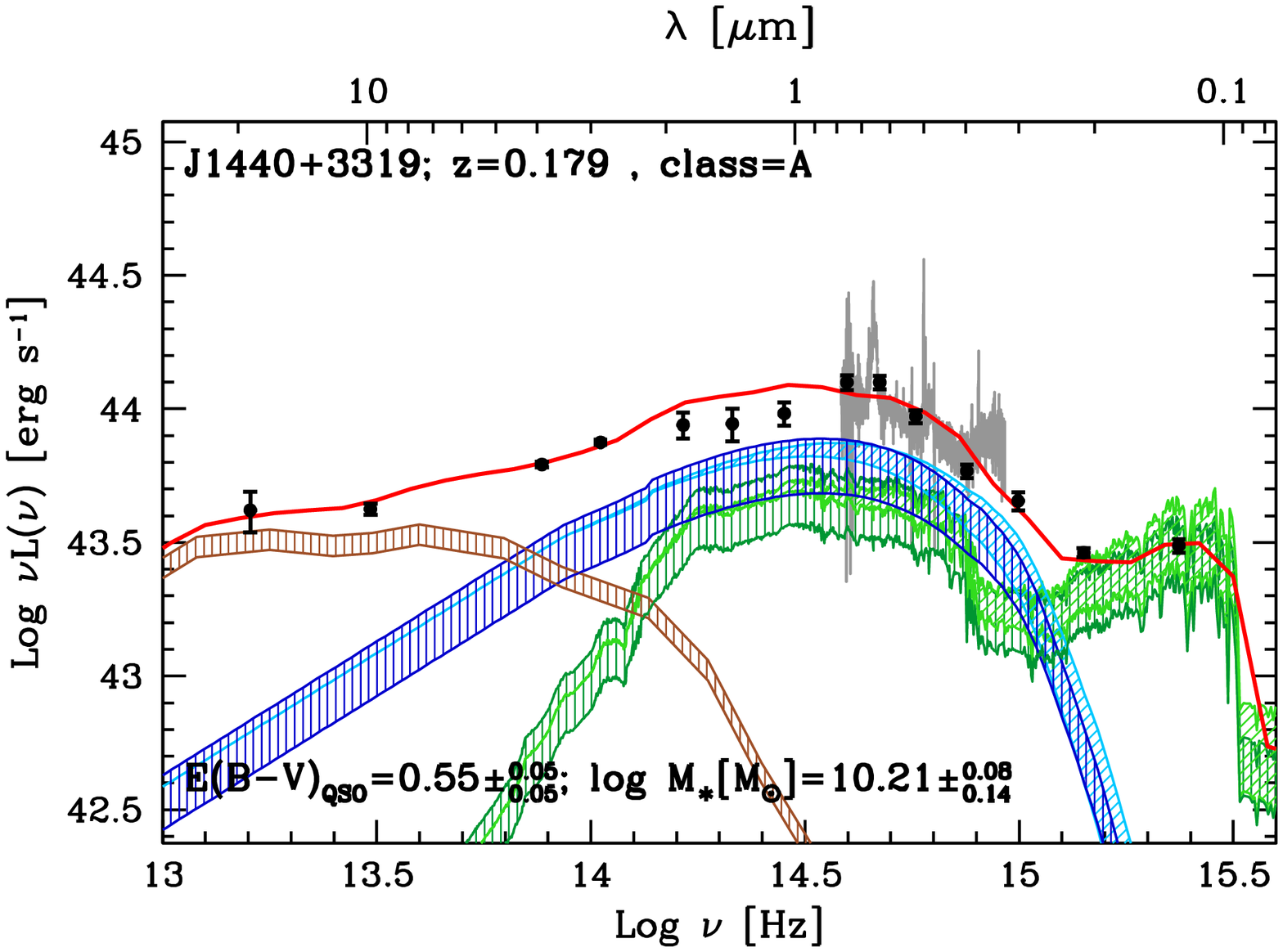}
        }
    \end{center}
    \caption[]{Continued}
\end{figure*}
\begin{figure*}
\addtocounter{figure}{-1}
     \begin{center}
        \subfigure{
            \includegraphics[width=0.4\textwidth]{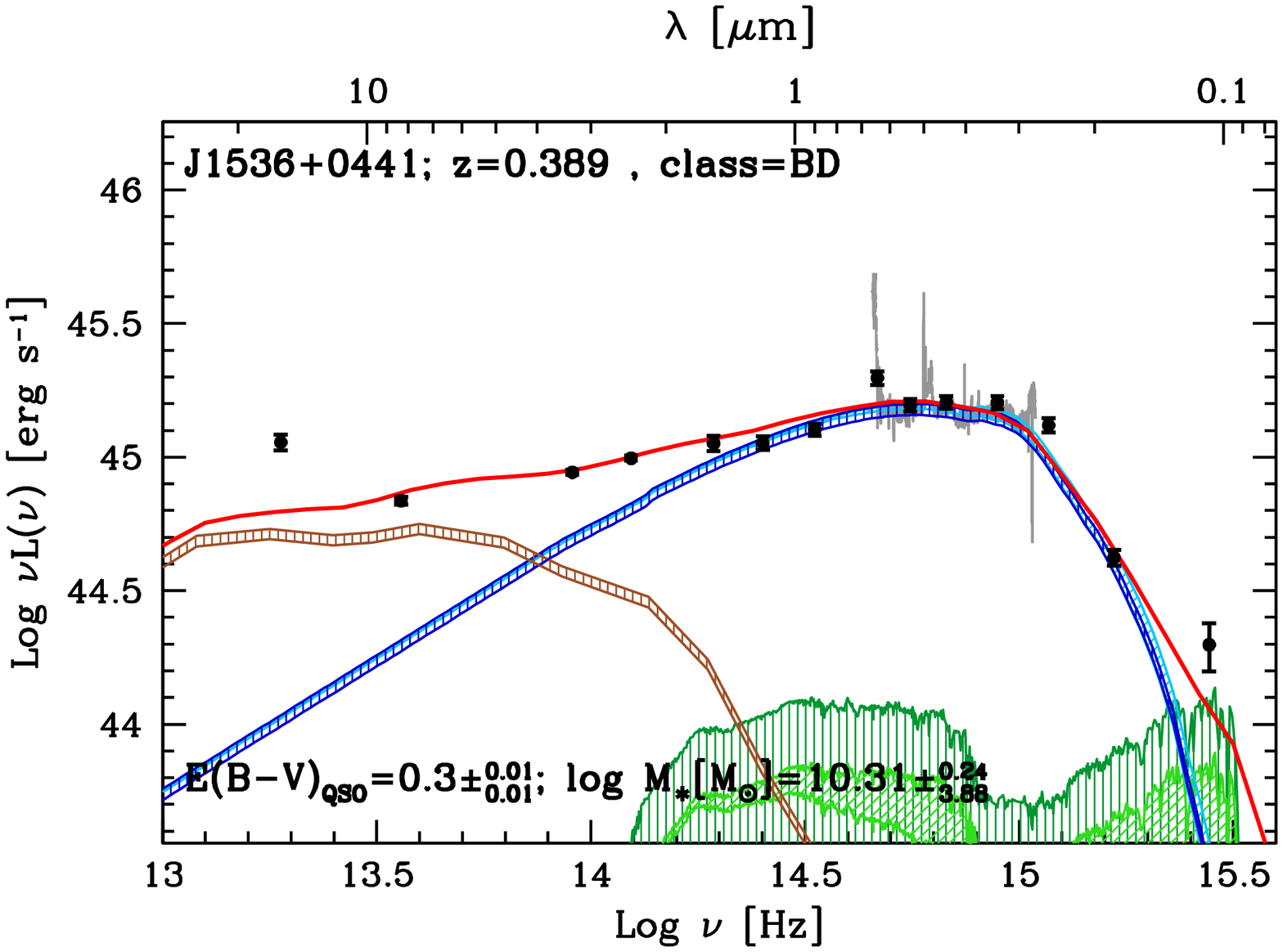}
        }
        \subfigure{
            \includegraphics[width=0.4\textwidth]{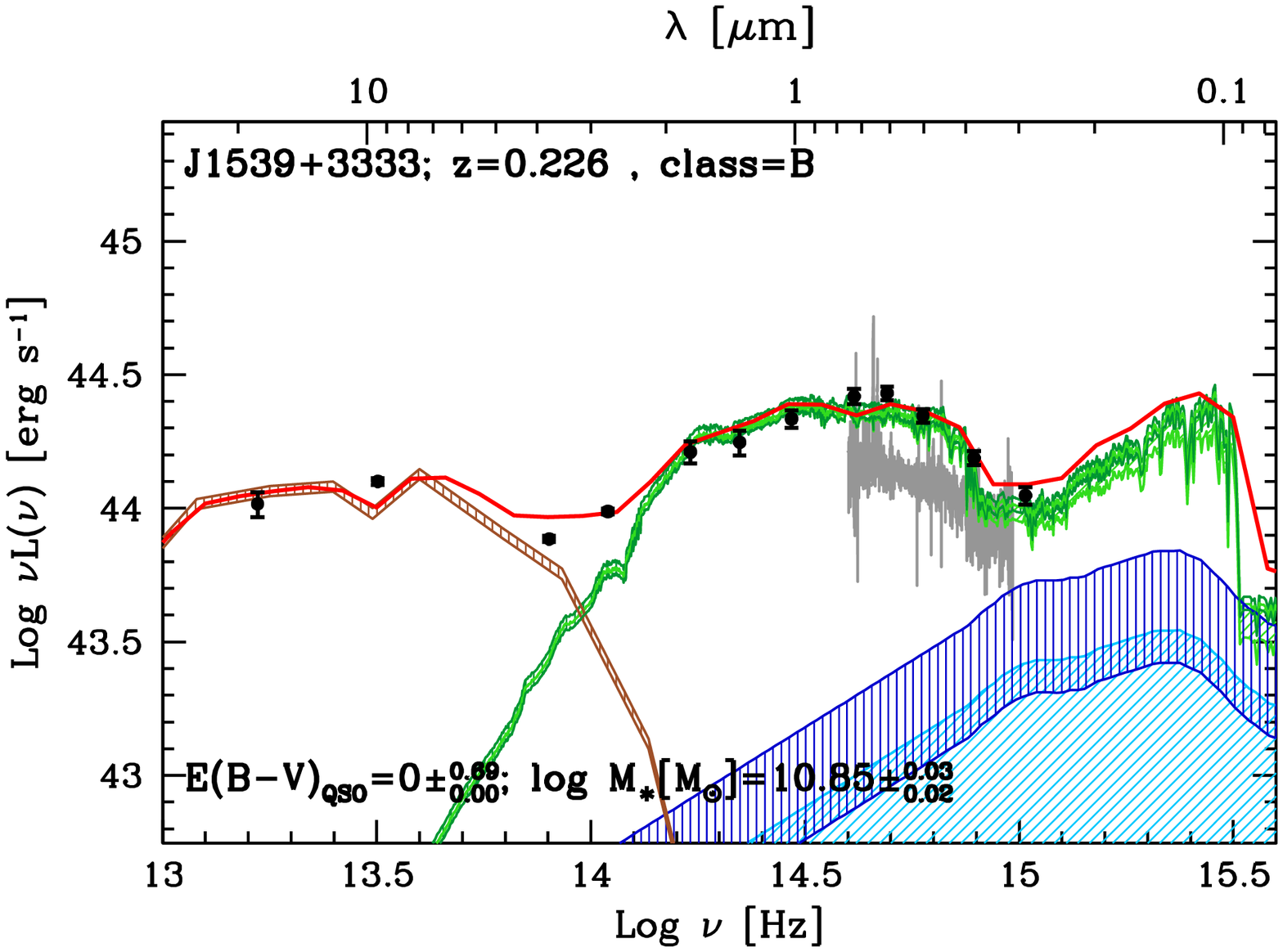}
        }\\
        \vspace{-2.2cm}
        \subfigure{
            \includegraphics[width=0.4\textwidth]{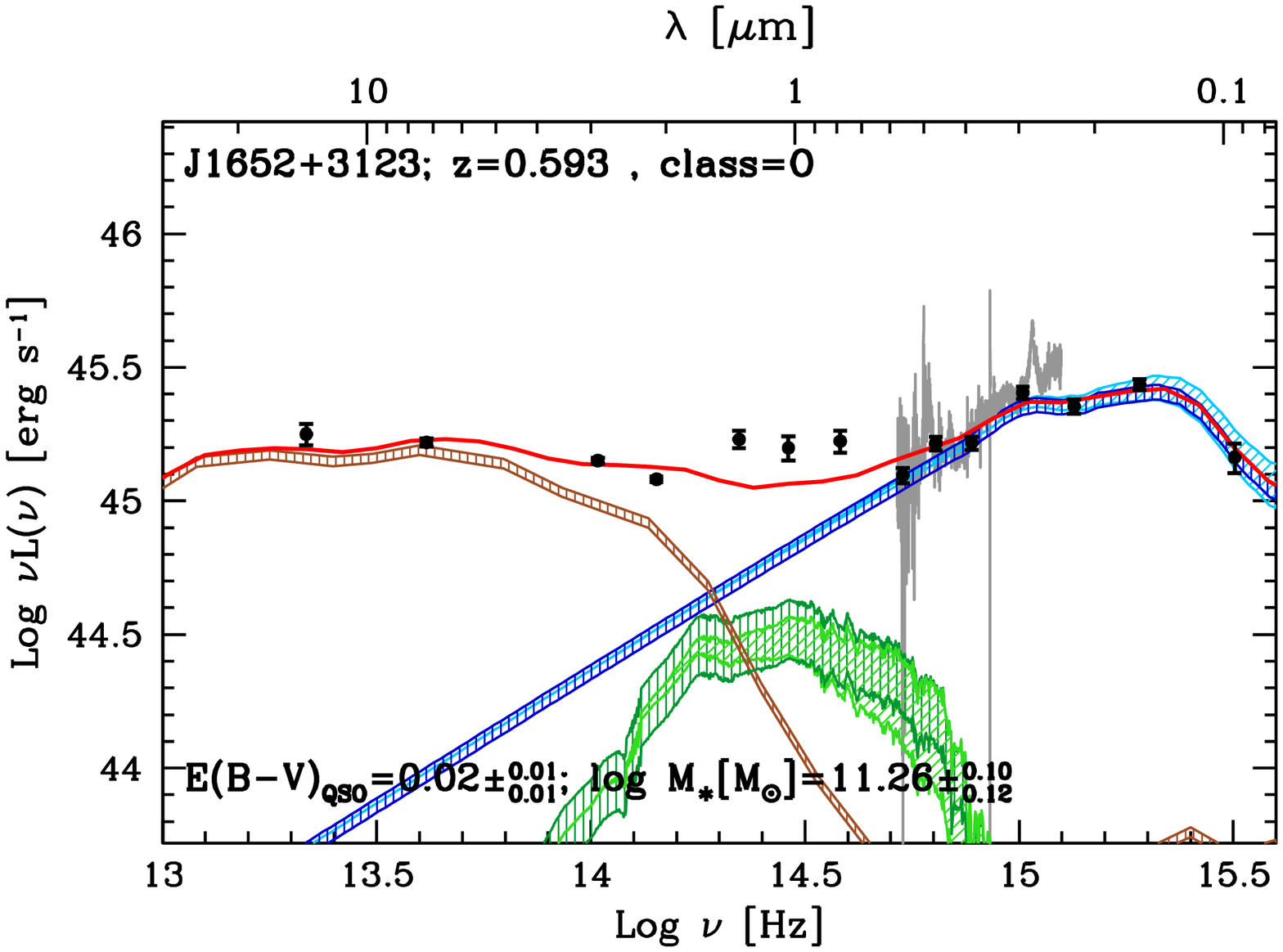}
        }
        \subfigure{
           \includegraphics[width=0.4\textwidth]{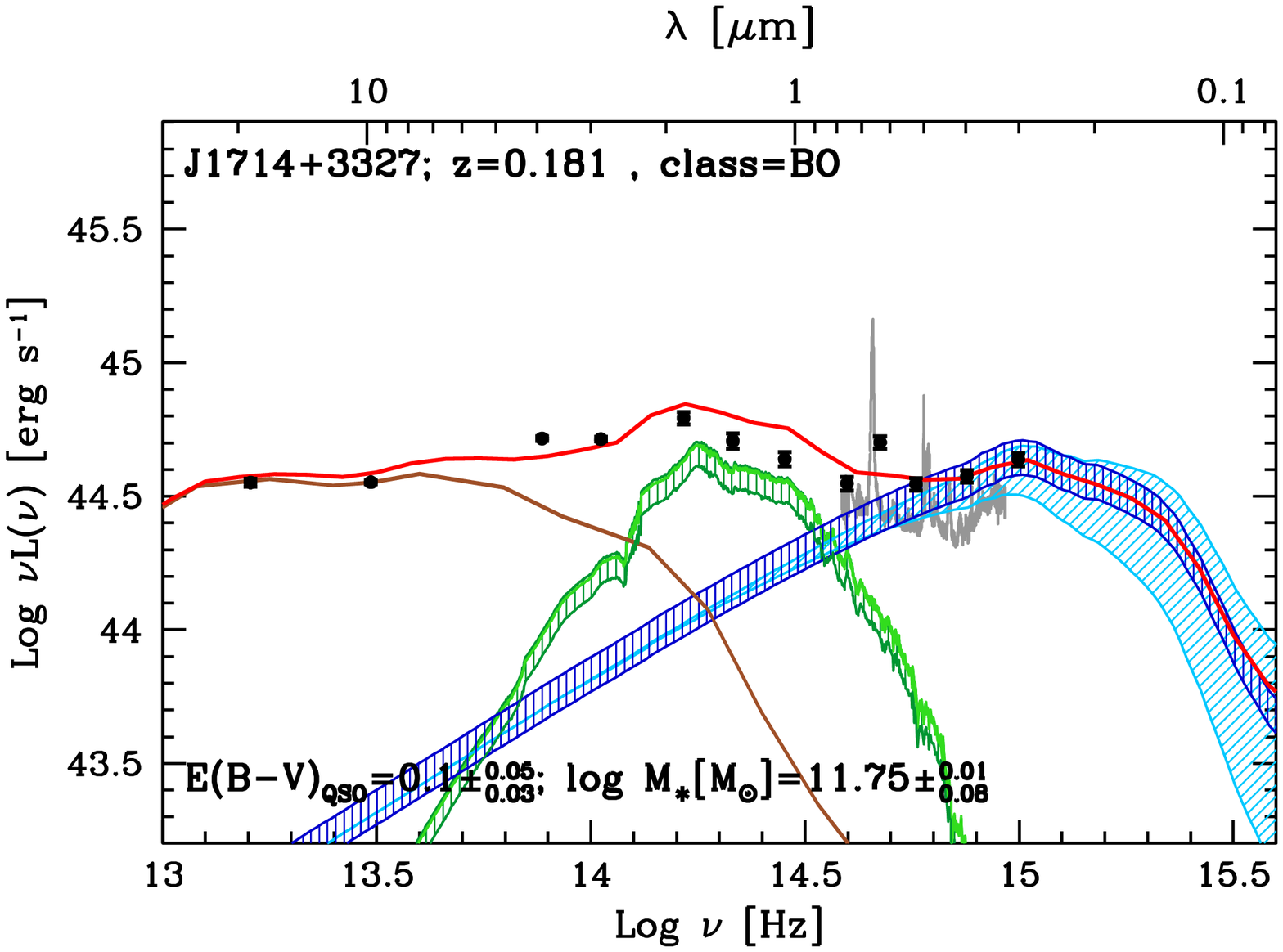}
        }
    \end{center}
    \caption[]{Continued}
\end{figure*}
\section{Model fitting}
\label{Model fitting}
The SED-fitting code utilised in the present paper is the one already presented in \citet[hereafter L13]{2013ApJ...777...86L}. 
The various components (i.e. hot-dust from a dusty torus, host galaxy, and accretion disc emission) of the rest-frame SEDs are fitted by using a standard $\chi^2$ minimization procedure (see \S~3 in L13 for details about the model fitting). The code is based on a large set of templates widely used in the literature. 
Details about the SED decomposition procedure are given in Appendix \ref{Details on the SED-fitting procedure}. 
\par
The infrared luminosity $\Lir$ and the accretion disc luminosity $\Ldisc$ are
obtained by integrating the torus template between 1$\mu$m and 1000 $\mu$m,
and the disc template between 1$\mu$m and 0.003$\mu$m (i.e., the last point of
the BBB template; \citealt{2006ApJS..166..470R}), respectively.  \par Infrared
luminosities need to be corrected for the anisotropy factor defined as the
ratio of the luminosity of face-on versus edge-on AGN, which is a function of
the column density $\NH$ \citep{2007A&A...468..603P}. This correction is
negligible for QSOs fitted with a torus template having $\NH$ lower than
$10^{22}$ cm$^{-2}$ (72\% of the total sample). Nine objects have best-fit
torus with $10^{22} < \NH $ [cm$^{-2}$] $< 10^{23}$. The anisotropy correction
for these sources is at any rate modest, of the order 1.3
(\citealt{2007A&A...468..603P,2010MNRAS.402.1081V,2011A&A...534A.110L,2012MNRAS.425..623L},
2013).  \par We point out that WISE bands might be contaminated by emission
features due to the stochastic heating of polycyclic aromatic hydrocarbon
(PAH) molecules or carbon grains (e.g., \citealt{2006A&A...453..969F}), which
are related with massive star-forming regions at much lower dust temperature
($T<100$ K).  However, PAH features are almost absent in AGN, given that these
molecules are destroyed by the extreme optical-UV and soft X--ray radiation in
AGN \citep{1991MNRAS.248..606R,2004A&A...414..123S}.
Therefore, on average our estimates of the infrared ``torus" luminosities are mainly dominated by the AGN emission.
\par
Intrinsic disc luminosities are estimated by de-reddening the BBB template according to the assumed reddening law, and employing the reddening value related with the BBB template found as the best solution.

\section{Properties of the SEDs}
\label{Properties of the SEDs}
In Figure~\ref{bhbseds} we present the best-fit decomposition for each object in our sample. 
In order to further test our fits, we have compared the best-fit templates with the optical spectra from SDSS (grey line in Figure~\ref{bhbseds}). Note that spectra are not renormalized to match the SDSS photometry and they are not corrected for aperture.
Overall, optical spectra are in good agreement with the results from the model fitting procedure. 
In general, our model fitting shows that almost all QSO SEDs are contaminated by host galaxy emission around 1~$\mu$m and/or reddening in the optical--UV. 

\begin{figure}
 \centering\includegraphics[width=9cm,clip]{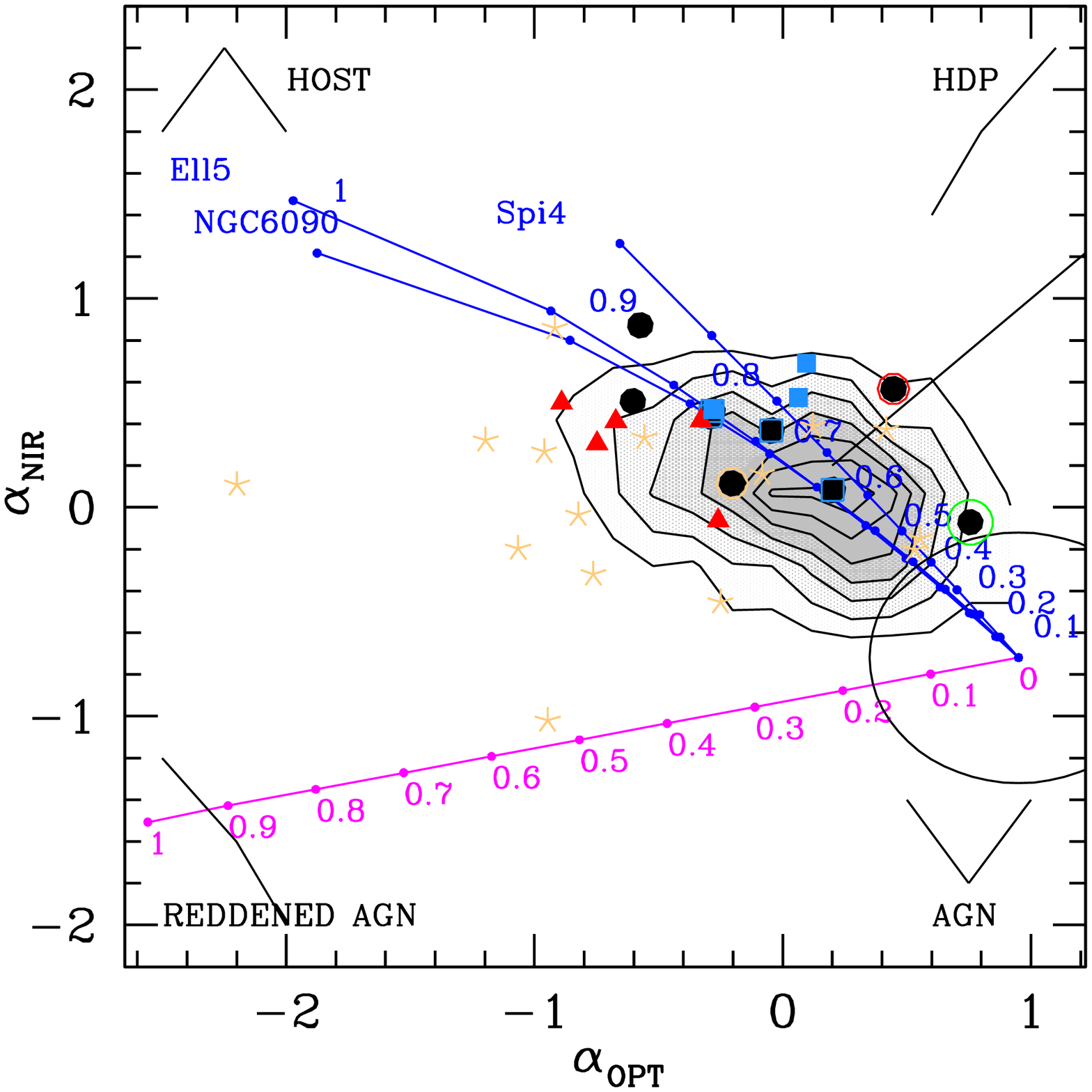}
 \caption{The mixing diagram of the quasar sample. Different regions of the plot correspond to different SED shapes, as shown by the black lines in the four corners of the plot. The black circle shows the $1\sigma$ dispersion of the quasar samples of E94 as defined by \citet{2013MNRAS.434.3104H}. 
 The blue tracks show composite QSO+host galaxy templates for three different host galaxy types \citep{polletta07}, while the numbers beside these curves represent the galaxy fraction. As QSO template, we refer here to the mean SED by E94.
The purple track shows the reddening vector applied to the E94 radio-quiet mean SED. The numbers under the reddening vector show the $\ebvq$ values. The straight black solid line shows the $\aopt=\anir$ line. Points are the sources from \citet{2011ApJ...738...20T} coded on the basis of the shape of Balmer lines: B- BHB candidates (\rev{black filled circles}); D- DPEs (\rev{blue filled squares}); A- sources with Asymmetric line profiles (\rev{red filled triangles}); O- Others (\rev{orange stars}). 
\rev{BHB candidates were the classification scheme is not univocal are marked with an additional symbol: BD- BHB/DPEs candidates (blue open squares); BA - BHB/Asymmetric (red open circle); BO - BHB/Other (orange open circle).}
The green circle marks the recoil BH candidate J0927+2943 \citep{2008ApJ...678L..81K}. We overplot the contours for the test sample of about 1000 QSOs in SDSS (see \S~\ref{Comparison with a test sample}).}
 \label{md}
\end{figure}

In order to distinguish among galaxy-dominated, QSO-dominated, and reddening dominated SEDs, we have considered the mixing diagram presented by \citet{2013MNRAS.434.3104H}.
The mixing diagram axes are the $1-3\mu$m SED slope ($\anir$) versus the $0.3-1\mu$m slope ($\aopt$). These ranges lie on either side of the 1~$\mu$m dip of the rest frame SED.
Figure~\ref{md} shows the mixing diagram our QSO sample.
The slopes $\anir$ and $\aopt$ are estimated from the best-fit SED (red line in Fig.~\ref{bhbseds}).
Different regions of the plot correspond to different SED shapes, as shown by the black lines in the four corners of the plot. 
Galaxy SEDs lie in the top left corner in this diagram. As soon as the galaxy contribution decreases, objects move along the blue tracks further down to the QSO-dominated region (bottom-left corner).
HDP QSOs and are located in the top right region of the mixing diagram.
\par Overall, the sample shows a large SED variety with moderately host galaxy
contamination and extinction.  There is no clear division among different
classes, but SEDs of BHB candidates and DPEs lie along the QSO-host galaxy
tracks, with modest intrinsic reddening, in agreement with the bulk of the
population in SDSS.  Objects classified as ``asymmetric" and most of the
``others" cover a wider range of host galaxy and intrinsic reddening
contamination. A summary of the general properties our quasar sample is given
in Table~\ref{tbl-2}. The disc luminosities reported in Table~\ref{tbl-2} have
been de-reddened considering the best-fit $\ebvq$ and the Prevot et al.
  (1984) reddening law.  Details on the SED general properties for the
different classes are given in Appendix~\ref{Notes on the SED general
  properties for the different classes}.

\begin{center}
\begin{landscape}
\begin{table}
\caption{General properties of our quasar sample. \label{tbl-2}} 
\begin{tabular}{@{}c c c c c c c c c c | c l}
 Obj.name$^{\mathrm{a}}$ & $z_{\rm NL}^{\mathrm{b}}$ & $\aopt^{\mathrm{c}}$ & $\anir^{\mathrm{d}}$ & $\Log\Lir^{\mathrm{e}}$ & $\Log\Ldiscder^{\mathrm{f}}$ & $\Log\Ldiscr^{\mathrm{g}}$ & $\ebvq^{\mathrm{h}}$ & $\Log\Lirmu^{\mathrm{i}}$ & $\size^{\mathrm{j}}$ & Class$^{\mathrm{k}}$ & $\Log M_\ast^{\mathrm{l}}$ \\ [0.5ex]
\hline\noalign{\smallskip}
    J0012-1022   &   0.228   &   -0.33   &    0.42   &   44.97   &   45.05   &   45.05   &    0.00$_{-0.00}^{+0.00}$   &   44.48    &   9.95   &       A    &  11.45$_{-0.04}^{+0.05}$   \\
    J0155-0857   &   0.165   &   -0.08   &    0.16   &   44.89   &   45.05   &   45.05   &    0.00$_{-0.00}^{+0.00}$   &   44.41    &   9.09   &       O    &  11.64$_{-0.11}^{+0.03}$   \\
    J0221+0101   &   0.354   &    0.12   &    0.39   &   44.75   &   45.38   &   44.79   &    0.15$_{-0.02}^{+0.02}$   &   44.26    &   7.71   &       O    &  10.70$_{-0.21}^{+0.18}$   \\
    J0829+2728   &   0.321   &    0.42   &    0.37   &   44.68   &   45.05   &   45.05   &    0.00$_{-0.00}^{+0.01}$   &   44.19    &   7.12   &       O    &  10.63$_{-0.02}^{+0.21}$   \\
    J0918+3156   &   0.452   &   -0.82   &   -0.03   &   45.74   &   46.10   &   45.14   &    0.35$_{-0.03}^{+0.05}$   &   45.25    &  24.05   &       O    &  12.10$_{-1.07}^{+0.03}$   \\
    J0919+1108   &   0.369   &    0.53   &   -0.20   &   45.51   &   45.68   &   45.62   &    0.01$_{-0.01}^{+0.01}$   &   45.02    &  18.49   &       O    &  11.28$_{-0.09}^{+0.19}$   \\
    J0921+3835   &   0.187   &   -0.75   &    0.31   &   44.96   &   45.45   &   44.49   &    0.35$_{-0.02}^{+0.05}$   &   44.48    &   9.91   &       A    &  11.18$_{-0.30}^{+0.07}$   \\
    J0927+2943   &   0.713   &    0.76   &   -0.07   &   45.54   &   45.89   &   45.89   &    0.00$_{-0.00}^{+0.00}$   &   45.05    &  19.07   &       B    &  11.40$_{-1.54}^{+0.05}$   \\
    J0931+3204   &   0.226   &   -1.20   &    0.32   &   44.88   &   45.86   &   44.54   &    0.65$_{-0.04}^{+0.07}$   &   44.40    &   8.99   &       O    &  11.01$_{-0.29}^{+0.06}$   \\
    J0932+0318   &   0.420   &    0.20   &    0.08   &   45.06   &   45.45   &   45.00   &    0.10$_{-0.01}^{+0.03}$   &   44.57    &  11.03   &      BD    &  11.17$_{-0.39}^{+0.11}$   \\
    J0936+5331   &   0.228   &   -0.26   &   -0.06   &   45.20   &   44.95   &   44.95   &    0.00$_{-0.00}^{+0.01}$   &   44.71    &  12.97   &       A    &  11.24$_{-0.03}^{+0.04}$   \\
    J0942+0900   &   0.213   &   -0.28   &    0.47   &   44.91   &   45.21   &   44.42   &    0.25$_{-0.05}^{+0.08}$   &   44.46    &   9.64   &       D    &  10.81$_{-0.01}^{+0.04}$   \\
    J0946+0139   &   0.220   &   -0.67   &    0.41   &   44.97   &   45.39   &   44.50   &    0.30$_{-0.02}^{+0.04}$   &   44.51    &  10.26   &       A    &  11.25$_{-0.15}^{+0.10}$   \\
    J1000+2233   &   0.419   &   -0.28   &    0.46   &   44.94   &   45.39   &   44.69   &    0.20$_{-0.06}^{+0.06}$   &   44.45    &   9.62   &      BD    &  11.09$_{-0.07}^{+0.11}$   \\
    J1010+3725   &   0.282   &   -0.95   &   -1.02   &   45.99   &   44.94   &   44.35   &    0.15$_{-0.02}^{+0.05}$   &   45.51    &  32.48   &       O    &  11.62$_{-0.08}^{+0.07}$   \\
    J1012+2613   &   0.378   &   -0.29   &    0.44   &   45.04   &   45.11   &   44.90   &    0.04$_{-0.01}^{+0.01}$   &   44.56    &  10.82   &      BD    &  11.56$_{-0.20}^{+0.11}$   \\
    J1027+6050   &   0.332   &    0.06   &    0.53   &   44.88   &   45.47   &   45.25   &    0.04$_{-0.02}^{+0.01}$   &   44.40    &   8.98   &       D    &  11.44$_{-0.02}^{+0.13}$   \\
    J1050+3456   &   0.272   &   -0.60   &    0.51   &   44.19   &   45.22   &   44.12   &    0.45$_{-0.06}^{+0.03}$   &   43.70    &   4.04   &       B    &  10.30$_{-0.06}^{+0.13}$   \\
    J1105+0414   &   0.436   &    0.10   &    0.69   &   45.37   &   45.76   &   45.31   &    0.10$_{-0.01}^{+0.01}$   &   44.89    &  15.94   &       D    &  11.40$_{-0.02}^{+0.04}$   \\
    J1117+6741   &   0.248   &   -1.07   &   -0.20   &   45.13   &   45.46   &   44.30   &    0.50$_{-0.05}^{+0.02}$   &   44.64    &  11.94   &       O    &  10.80$_{-2.15}^{+0.10}$   \\
    J1154+0134   &   0.469   &    0.45   &    0.57   &   44.71   &   45.87   &   45.28   &    0.15$_{-0.01}^{+0.02}$   &   44.22    &   7.35   &      BA    &  10.33$_{-2.55}^{+0.25}$   \\
    J1207+0604   &   0.136   &   -0.92   &    0.86   &   44.44   &   44.32   &   44.32   &    0.00$_{-0.00}^{+0.17}$   &   43.96    &   5.46   &       O    &  11.14$_{-0.05}^{+0.12}$   \\
    J1211+4647   &   0.294   &   -0.76   &   -0.32   &   45.35   &   44.68   &   44.68   &    0.00$_{-0.00}^{+0.01}$   &   44.86    &  15.33   &       O    &  11.44$_{-0.03}^{+0.04}$   \\
    J1215+4146   &   0.196   &   -2.20   &    0.11   &   45.10   &   45.58   &   44.03   &    0.90$_{-0.39}^{+0.10}$   &   44.64    &  11.95   &       O    &  11.22$_{-0.09}^{+0.11}$   \\
    J1216+4159   &   0.242   &   -0.96   &    0.27   &   44.56   &   45.38   &   44.06   &    0.65$_{-0.09}^{+0.09}$   &   44.07    &   6.16   &       O    &  10.48$_{-0.03}^{+0.09}$   \\
    J1328-0129   &   0.151   &   -0.56   &    0.33   &   44.56   &   44.51   &   44.51   &    0.00$_{-0.00}^{+0.01}$   &   44.07    &   6.17   &       O    &  11.13$_{-0.10}^{+0.10}$   \\
    J1414+1658   &   0.237   &   -0.25   &   -0.45   &   45.10   &   44.60   &   44.60   &    0.00$_{-0.00}^{+0.03}$   &   44.61    &  11.51   &       O    &  10.94$_{-0.09}^{+0.06}$   \\
    J1440+3319   &   0.179   &   -0.89   &    0.50   &   43.99   &   45.07   &   43.86   &    0.55$_{-0.05}^{+0.05}$   &   43.50    &   3.21   &       A    &  10.21$_{-0.14}^{+0.08}$   \\
    J1536+0441   &   0.389   &   -0.04   &    0.37   &   45.18   &   46.23   &   45.35   &    0.30$_{-0.01}^{+0.01}$   &   44.69    &  12.68   &      BD    &  10.31$_{-3.88}^{+0.24}$   \\
    J1539+3333   &   0.226   &   -0.57   &    0.87   &   44.53   &   43.87   &   43.87   &    0.00$_{-0.00}^{+0.69}$   &   44.08    &   6.21   &       B    &  10.85$_{-0.02}^{+0.03}$   \\
    J1652+3123   &   0.593   &    0.55   &   -0.15   &   45.65   &   45.86   &   45.74   &    0.02$_{-0.01}^{+0.01}$   &   45.16    &  21.69   &       O    &  11.26$_{-0.12}^{+0.10}$   \\
    J1714+3327   &   0.181   &   -0.20   &    0.12   &   45.03   &   45.31   &   44.86   &    0.10$_{-0.03}^{+0.05}$   &   44.54    &  10.63   &      BO    &  11.75$_{-0.08}^{+0.01}$   \\
\hline\hline
\end{tabular}

\flushleft\begin{list}{}
 \item[${\mathrm{a}}$]{ Quasar name.}
 \item[][${\mathrm{b}}$]{ Redshift of narrow lines.}
 \item[][${\mathrm{c}}$]{ Optical slope estimated from the best-fit SED in the wavelength range $0.3-1~\mu$m.}
 \item[][${\mathrm{d}}$]{ Near-infrared slope estimated from the best-fit SED in the wavelength range $1-3~\mu$m.}
 \item[][${\mathrm{e}}$]{ Logarithm of the infrared luminosity (in erg s$^{-1}$) integrated from the best-fit torus template between 1-1000 $\mu$m.}
 \item[][${\mathrm{f}}$]{ Logarithm of the de-reddened disc luminosity (in erg s$^{-1}$) integrated from the best-fit BBB template from 1$\mu$m to $\Log \nu=17$ Hz.}
 \item[][${\mathrm{g}}$]{ Logarithm of the reddened disc luminosity (in erg s$^{-1}$) integrated from the best-fit BBB template from 1$\mu$m to $\Log \nu=17$ Hz.}
 \item[][${\mathrm{h}}$]{ Best-fit AGN reddening value.}
 \item[][${\mathrm{i}}$]{ Logarithm of the 12$~\mu$m luminosity (in erg s$^{-1}$) estimated from the best-fit torus template.}
 \item[][${\mathrm{j}}$]{ The approximate size of the emitter (in pc) at 12$~\mu$m from Eq.~(\ref{sizelum}).}
 \item[][${\mathrm{k}}$]{ Object classification based on the shape of the Balmer lines: B- BHB candidates; D- DPEs; A- sources with Asymmetric line profiles; O- Others. \rev{The classification scheme is not univocal and different classes are not mutually exclusive (e.g., BHB candidates may have spectral feature similar to the DPEs, see Paper I for details): BD- BHB/DPEs candidates; BA - BHB/Asymmetric; BO - BHB/Other.}}
 \item[][${\mathrm{l}}$]{ Best-fit stellar mass (in $M_\odot$) from BC03 model using a Chabrier IMF.}
\end{list}
\end{table}
\end{landscape}
\end{center}

\section{Hot-dust-poor candidates}
\label{Hot-dust-poor candidates}
Almost all objects in our sample show emission from warm/hot dust which is a
clear indication of optical-UV AGN luminosity reprocessed by a dusty torus.  A
HDP quasar should have small mid-infrared to disc luminosity ratios
($R=\Lir/\Ldisc$) \footnote{$R$ should be defined as the mid-infrared to bolometric luminosity ratio, where the 
bolometric luminosity is estimated in the optical--UV/X--ray wavelength range. 
X--ray data are not available for the majority of the sources in our 
sample, hence are not considered in the present analysis. 
Since the X--ray emission is of the order of $\sim10-20\%$ of the bolometric budget, the $R$ values may be overestimated by a similar factor.}. Following
\citet{2011ApJ...733..108H}, HDP quasars should have R values between
$\sim2\%$ to 30\%, well below the 75\% predicted by the unified model.

\par
We have thus considered, for the following analysis, the $R$ values where the $\Ldisc$ estimates are corrected for both host galaxy contamination and disc reddening, and
in order to classify a quasar HDP, we have searched for those objects with low mid-infrared to disc luminosity ratios. 
\par
 Only one source, J1154+0134 (class=BA) shows significantly lower hot dust
  luminosity compared with what expected from its optical-UV emission,
  by about one order of magnitude assuming an average $R$ of $0.3-0.6$ (e.g. \citealt{2008ApJ...679..140T,2012arXiv1212.4245M}; L13).
According to the position of J1154+0134 relative to the equal slope in the $\aopt-\anir$ plot, the infrared emission should be the continuation of the disc emission to longer wavelengths ($2-3~\mu$m, see \citealt{2013MNRAS.434.3104H}.). The flattening of the $W3$ and $W4$ bands at $\lambda>7\mu$m indicates still little mid-infrared emission.
\rev{The $R$ values of J1154+0134 is 0.07 considering the de-reddened $\Ldisc$ and 0.27 if we use the reddened $\Ldisc$ value, which are consistent with the expected $R$ values for class II HDP\footnote{Class I, II, and III HDP are sources lying below, on, and above the equal slope line in the mixing diagram, respectively (see Section~2 by H10).} estimated by H10. Assuming a conservative error on $R$ of the order of 5\% (see L13), the difference between the value predicted by unified models and the one of J1154+0134 is significant at $6.5\sigma$ level (for $R=0.27$).}

\par
In conclusion, all quasars in our sample (aside from J1154+0134) show mid-infrared emission, and therefore it is unlikely that our objects have experienced a recoil event (in a scenario of a single black hole displaced from the center of the galaxy).


\subsection{The case of J1154+0134}
\begin{figure}
 \centering\includegraphics[width=8cm,clip]{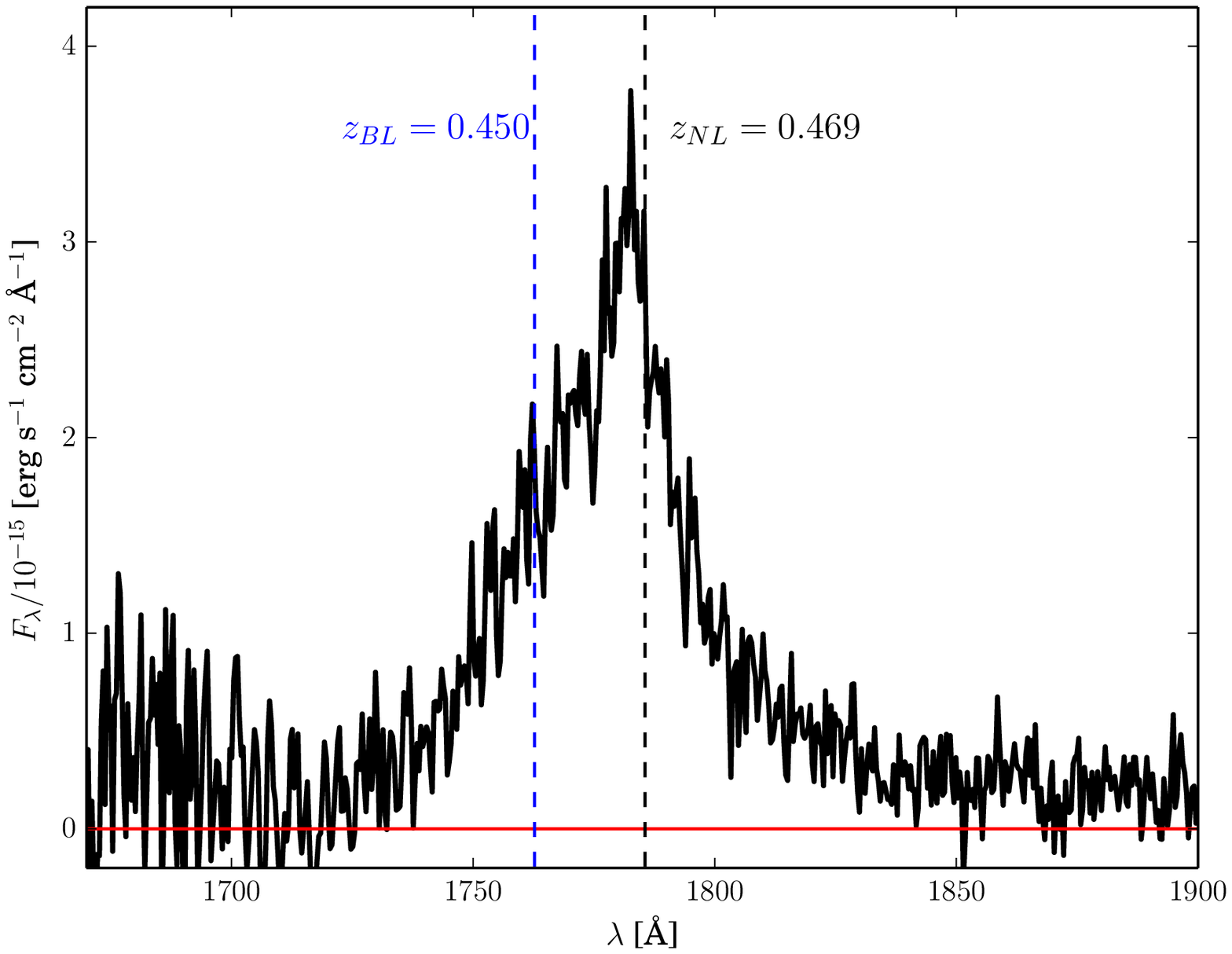}
 \caption{{\it HST}/COS ultraviolet spectrum of J1154+0134 displaying \ion{Ly}{$\alpha$} emission. The red solid line denotes the zero level, while the two vertical dashed lines mark the broad line (blue) and narrow line (black) redshift.}
 \label{1154}
\end{figure}

\rev{
J1154+0134 has been newly selected as BHB candidate by both \citet{2011ApJ...738...20T} and  \citet{2012ApJS..201...23E}. The optical spectrum of this object and the evolution of the broad line profiles are discussed in Paper I (see also \citealt{2012ApJS..201...23E}).
From the spectral point of view, this source presents \ion{H}{$\beta$} and \ion{Mg}{ii} lines with similar profiles, blueshifted peaks ($\sim3500$ km s$^{-1}$), red asymmetries, and do not show any significant evolution of the shift of broad line peaks between the SDSS spectrum and our follow-up observations. 
J1154+0134 has been also targeted with the Cosmic Origins Spectrograph (COS) on the Hubble space Telescope (HST) in May 2011 with the grating NUV/G230L (central wavelength 2950\AA~, covering the windows 1649--2051\AA~ and 2749--3151\AA) to observed the \ion{Ly}{$\alpha$} line (see Figure~\ref{1154}).  
The line shows blue asymmetries with the peak of \ion{Ly}{$\alpha$} consistent with the redshift from narrow lines. This spectrum has low signal-to-noise and no other lines are present.
The UV observation hints at a DPE nature of J1154+0134, but more detailed analysis and discussion of the UV spectrum will be given in Eracleous et al. 2014, in preparation.

J1154+0134 is not detected by ROSAT, but it is clearly detected in the Faint Images of the Radio
Sky at Twenty cm (FIRST; \citealt{1995ApJ...450..559B}) survey with a measured peak flux density of 1.54 mJy.
If J1154+0134 is a recoiling SMBH, high-resolution radio and HST observations would provide precise measurement of its location. 
}

\begin{table*}
\caption{RASS properties of our quasar sample. \label{tbl-5}} 
\begin{tabular}{@{}l c c c c c c}
 Obj.name$^{\mathrm{a}}$ & Count rate$^{\mathrm{b}}$  & Exposure$^{\mathrm{c}}$  & $N_{H}$$^{\mathrm{d}}$    &  F$_{[0.5-2]\rm keV}$$^{\mathrm{e}}$ & L$_{[0.5-2]\rm keV}$$^{\mathrm{f}}$ \\ [0.5ex]
                                           &     s$^{-1}$  & s               &  cm$^{-2}$ &   erg s$^{-1}$cm$^{-2}$  & erg s$^{-1}$ \\ [0.5ex]
\hline\noalign{\smallskip}
1RXS J015530.4-085720 &   0.1336$\pm$0.0269  &   222    &   2.14$\times10^{20}$ &       9.43$\times10^{-13}\pm$     1.90$\times10^{-13}$   &     7.08$\times10^{43}$  \\
1RXS J091834.7+315632 &   0.0655$\pm$0.0162  &   364    &   1.62$\times10^{20}$ &       3.99$\times10^{-13}\pm$     9.86$\times10^{-14}$   &     3.01$\times10^{44}$  \\
1RXS J091930.5+110842 &   0.0749$\pm$0.0191  &   323    &   2.78$\times10^{20}$ &      6.08$\times10^{-13}\pm$     1.55$\times10^{-13}$   &     2.84$\times10^{44}$  \\
1RXH J092713.2+294334 &   0.0145$\pm$0.0020  &   9056   &   1.75$\times10^{20}$ &     3.22$\times10^{-13}\pm$     4.45$\times10^{-14}$   &     7.32$\times10^{44}$  \\
1RXH J092712.8+294344 &   0.0066$\pm$0.0013  &   17796  &   1.75$\times10^{20}$ &   1.48$\times10^{-13}\pm$     2.77$\times10^{-14}$   &     3.36$\times10^{44}$  \\
1RXH J094214.9+090024 &   0.0056$\pm$0.0012  &   5148   &   2.83$\times10^{20}$ &    1.42$\times10^{-13}\pm$     3.04$\times10^{-14}$   &     1.88$\times10^{43}$  \\
1RXS J100022.3+223327 &   0.0476$\pm$0.0138  &   335    &   2.57$\times10^{20}$ &       3.70$\times10^{-13}\pm$     1.08$\times10^{-13}$   &     2.33$\times10^{44}$  \\
1RXS J101226.6+261335 &   0.0454$\pm$0.0128  &   379    &   2.46$\times10^{20}$ &       3.45$\times10^{-13}\pm$     9.73$\times10^{-14}$   &     1.71$\times10^{44}$  \\
1RXS J102736.8+605025 &   0.0320$\pm$0.0095  &   631    &   7.00$\times10^{19}$ &      1.34$\times10^{-13}\pm$     3.96$\times10^{-14}$   &     4.88$\times10^{43}$  \\
1RXS J110539.2+041449 &   0.0220$\pm$0.0087  &   381    &   4.89$\times10^{20}$ &     2.37$\times10^{-13}\pm$     9.39$\times10^{-14}$   &     1.64$\times10^{44}$  \\
1RXS J141441.9+165812 &   0.1424$\pm$0.0218  &   412    &   1.19$\times10^{20}$ &      7.45$\times10^{-13}\pm$     1.14$\times10^{-13}$   &     1.26$\times10^{44}$  \\
1RXS J153635.6+044118 &   0.0312$\pm$0.0102  &   473    &   4.19$\times10^{20}$ &    3.13$\times10^{-13}\pm$     1.02$\times10^{-13}$   &     1.66$\times10^{44}$  \\
1RXS J165256.4+312346 &   0.0205$\pm$0.0076  &   647    &   2.56$\times10^{20}$ &     1.60$\times10^{-13}\pm$     5.88$\times10^{-14}$   &     2.31$\times10^{44}$  \\
1RXS J171448.2+332737 &   0.0400$\pm$0.0086  &   749    &   2.90$\times10^{20}$ &      3.32$\times10^{-13}\pm$     7.15$\times10^{-14}$   &     3.06$\times10^{43}$  \\
 \hline\hline
\end{tabular}

\flushleft\begin{list}{}
 \item[${\mathrm{a}}$]{ ROSAT All-Sky Survey Catalogue source name.}
 \item[][${\mathrm{b}}$]{ Count rates in the ROSAT band (0.1--2.4 keV).}
 \item[][${\mathrm{c}}$]{ Exposure time.}
 \item[][${\mathrm{d}}$]{ Galactic column density \citep{2005A&A...440..775K}.}
 \item[][${\mathrm{e}}$]{ X--ray fluxes at 0.5--2 keV calculated from the count rates in the ROSAT band by employing a power law spectrum with a photon index $\Gamma=2.0$ and corrected for Galactic absorption \citep{2005A&A...440..775K}.}
 \item[][${\mathrm{f}}$]{ Rest frame luminosity in the 0.5--2 keV band (estimated from the unabsorbed X-ray flux).}
\end{list}
\end{table*}

\section{X--ray detection}
\label{X--ray detection}

\rev{

In the last decade several authors (e.g. \citealt{2009MNRAS.398.1392L,2012MNRAS.420..860S,2012MNRAS.420..705T,2012ApJ...761...90G}, and references therein) have modeled the observable signatures related to annular gaps or larger central cavities in the accretion disc around a SMBH caused by the presence of secondary BH orbiting in a binary system. 
The SED of such systems is predicted to show standard emission in the optical bands with a peculiar dimming 
in the UV and possibly in the X-rays, depending on the gap extent and the amount of material present in this region.
In order to observe a decrease in the disc emission at UV or at shorter wavelength, the 
binary separations ($a$) considered in these works are $a\leq0.01$ pc for BH masses of $10^8-10^9\msun$.
The peculiar AGN targeted in our analysis are characterized by the presence of displaced BLs that in the binary 
scenario are associated with the Broad Line Region (BLR) bound to the single accreting BHs. 
Consistently the binary separations have to be $a \geq R_{BLR}$. Taking into account the range of optical luminosities computed for our targets $a\geq0.1$ pc \citep{2013ApJ...767..149B}. 
Hence we do not expect our objects to be unusually under luminous in the UV/X-rays. 
However due to the large uncertainties in the theoretical modeling and the complete lack of observations of accreting sub-pc BHBs, we decided to check also the X-ray properties of our objects.  
We found 11 objects serendipitously detected in the ROSAT All-Sky Survey (RASS), while 2 sources have ROSAT/HRI images. 
The X--ray fluxes at 0.5--2 keV were calculated from the count rates in the ROSAT band (0.1--2.4 keV) by employing a power law spectrum with no intrinsic absorption and a photon index $\Gamma=2.0$ modified by Galactic absorption \citep{2005A&A...440..775K}. Table~\ref{tbl-5} lists the X-ray properties of the detected RASS objects in our sample.
Among these 13 quasars, two have additional spectral information from XMM-{\it Newton} (J0155-0857) and {\it Chandra} (J0927+2943). 

J0155-0857 was also found serendipitously in an XMM-{\it Newton} observation 
(nominal exposure time of $\sim$13~ks) in June 2007. Although the observation 
shows a high level of background, the source is bright enough to allow an 
X-ray spectral analysis. A total of $\sim$2000/1500/1300 source net counts 
were collected in the pn/mos1/mos2 (0.3--7~keV band) using circular extraction regions of 
radius 30/25/20 arcsec, respectively. 
Since the source is located partly on a CCD gap in the pn camera, we focused 
our attention to the mos1 and mos2 spectral data. A power-law with photon 
index $\Gamma=1.94\pm0.06$ provides a good fit to the data, with no apparent 
need for additional spectral components. This is confirmed by the pn data. 
The observed 0.5--2~keV flux is 
$\sim7.0\times10^{-13}$~erg~s$^{-1}$~cm$^{-2}$, consistent with the ROSAT 
observation. 

J0927+2943 was targeted by {\it Chandra} with ACIS-S in February 2009; 
the nominal exposure of 26.7 ks provided $\sim$60 net counts in the 
0.5--7 keV band. 
The spectrum is fitted with a power law of $\Gamma=0.1$, which is unusual for 
AGN, suggesting a significant level of obscuration. In this case, assuming a 
more canonical photon index of 1.8, we obtain 
$N_{H}=7.7(+8.4-5.1)\times10^{22}$ cm$^{-2}$. The observed flux in the 
0.5--2 keV band (adopting the fitted $\Gamma$) is 
$3.6\times10^{-15}$ erg s$^{-1}$ cm$^{-2}$, much lower than the one obtained 
from the ROSAT observation. Long-term X--ray variability may be responsible 
for the observed difference in the X-ray flux. 

J1328-0129 has not been detected by ROSAT, but was serendipitously observed 
by {\it Chandra} with ACIS-I in August 2012, with a nominal exposure of 5 ks; 
the number of source net counts in the 0.5--7~keV bad is $\sim$830.
The spectrum is fitted with a power law of $\Gamma=1.2\pm0.1$, which may 
suggest some level of obscuration. 
Fitting the data with a power law with $Gamma=1.8$ (i.e., typical for 
AGN emission) modified by absorption, we obtain 
$N_{H}=5.5(+1.2-1.1)\times10^{21}$ cm$^{-2}$. 
The observed flux in the 0.5--2 keV band is 
$5.3\times10^{-13}$ erg s$^{-1}$ cm$^{-2}$.

Among the detected objects, all soft luminosities are, on average, typical of bright AGN/quasars (aside from the {\it Chandra} observation of J0927+2943).

Sources without a ROSAT detection might be either intrinsically faint, X--ray obscured, and/or highly variable. 
Bolometric luminosities of the detected objects are, on average, brighter than the undetected ones, as expected. Given that the X-ray luminosity scales with the bolometric luminosity by a factor of about 30 for optically selected QSO \citep{marconi04}, we expect the undetected AGN to have X--ray luminosities of the order of $10^{43}$ erg s$^{-1}$ or lower\footnote{The ROSAT flux limit is $5\times10^{-13}$ erg s$^{-1}$ cm$^{-2}$ for a mean effective exposure time of 400 sec, but the range of ROSAT exposure times
is large, and the sensitivity limit is different from field to field (\citealt{1999A&A...349..389V}).}. Deep X-ray observations with {\it Chandra} and XMM-{\it Newton} are needed in order to establish if undetected RASS objects are intrinsically faint or X--ray obscured.
}

\section{Additional Properties}

\subsection{Comparison with a test sample}
\label{Comparison with a test sample}
Are the SEDs of quasars with spectra showing high velocity shifts between broad and narrow lines peculiar with respect to {\it typical} QSOs? To answer this question we have selected an appropriate control sample of \rev{Type-1} QSOs from the one described by \citealt{2013ApJS..206....4K} (K13 hereafter). The main sample in K13 contains 119,652 QSOs from both SDSS-DR7 (103,895 objects with $i<19.1$ for $z<3$ and $i<20.2$ for $z>3$) and the Two Degree Filed QSO Redshift survey (2QZ; 15,757 with $b_J<20.85$ for $z<3$; \citealt{2004MNRAS.349.1397C}) with $0.064< z < 5.46$. Mid-IR photometry is coming from {\it Spitzer} and WISE, near-infrared data from 2MASS and UKIDSS, optical data from SDSS, and UV data from GALEX.
\par
From the K13 sample we have selected all objects with the same redshift range as the one explored in the present analysis (17,666 objects with $0.136\leq z\leq 0.713$). Well-constrained $\Lir$ measurements require a detection at $\sim10\mu$m, we thus considered a sub-sample of sources detected in W4 (i.e., the W4 band is redshifted at $\sim$13$\mu$m for our maximum redshift of 0.713), bringing the final sample size to 13,634 QSOs.
The redshift distribution of this sample is very different than the one covered by our quasars. The median redshift of the typical QSO sample is 0.474, while the median redshift of our sample is 0.272.
We have thus further cut the sample of typical QSOs by extracting at
  random $\sim1000$ objects following the redshift distribution of our sample. A Kolmogorov-Smirnov (K-S) test confirms that the two redshift distributions are drawn from the same parent population (see Table~\ref{tbl-1}).
We then run our SED-fitting code on this control sample and we estimate intrinsic reddening, torus and disc luminosities for all objects.

\begin{table}
\caption{Kolmogorov-Smirnov test and their significance from $t-$Student test to determine if our sample (32 objects) and the control sample ($\sim1000$ objects) are significantly different for a given parameter of interest. \label{tbl-1}} 
\centering
\begin{tabular}{@{}l l c |l}
 Parameter$^{\mathrm{a}}$ & K-S$^{\mathrm{b}}$ & $t-$Student $(\sigma)$ \\ [0.5ex]
\hline\noalign{\smallskip}
  $z$              &  $D=0.12$ with prob=0.72  &   0.25  \\  
  $\ebvq$       &  $D=0.30$ with prob=$5.0\times10^{-3}$  & 2.70 \\  
  $\Ldiscr$     &  $D=0.10$ with prob=0.88  & 0.21  \\  
  $\Ldiscder$ &  $D=0.37$ with prob=$2.5\times10^{-4}$  &   3.00     \\  
  $\size$        &  $D=0.32$ with prob=$2.7\times10^{-3}$  &  2.20  \\  
\hline\hline
\end{tabular}

\flushleft\begin{list}{}
 \item[${\mathrm{a}}$]{ Parameter of interest in the comparison of our sample with the control sample.}
 \item[][${\mathrm{b}}$]{ Kolmogorov-Smirnov test and two-sided probability that the two samples are drawn from the same parent population.}
\end{list}
\end{table}
\subsection{Intrinsic reddening distribution}
\begin{figure}
 \centering\includegraphics[width=9cm,clip]{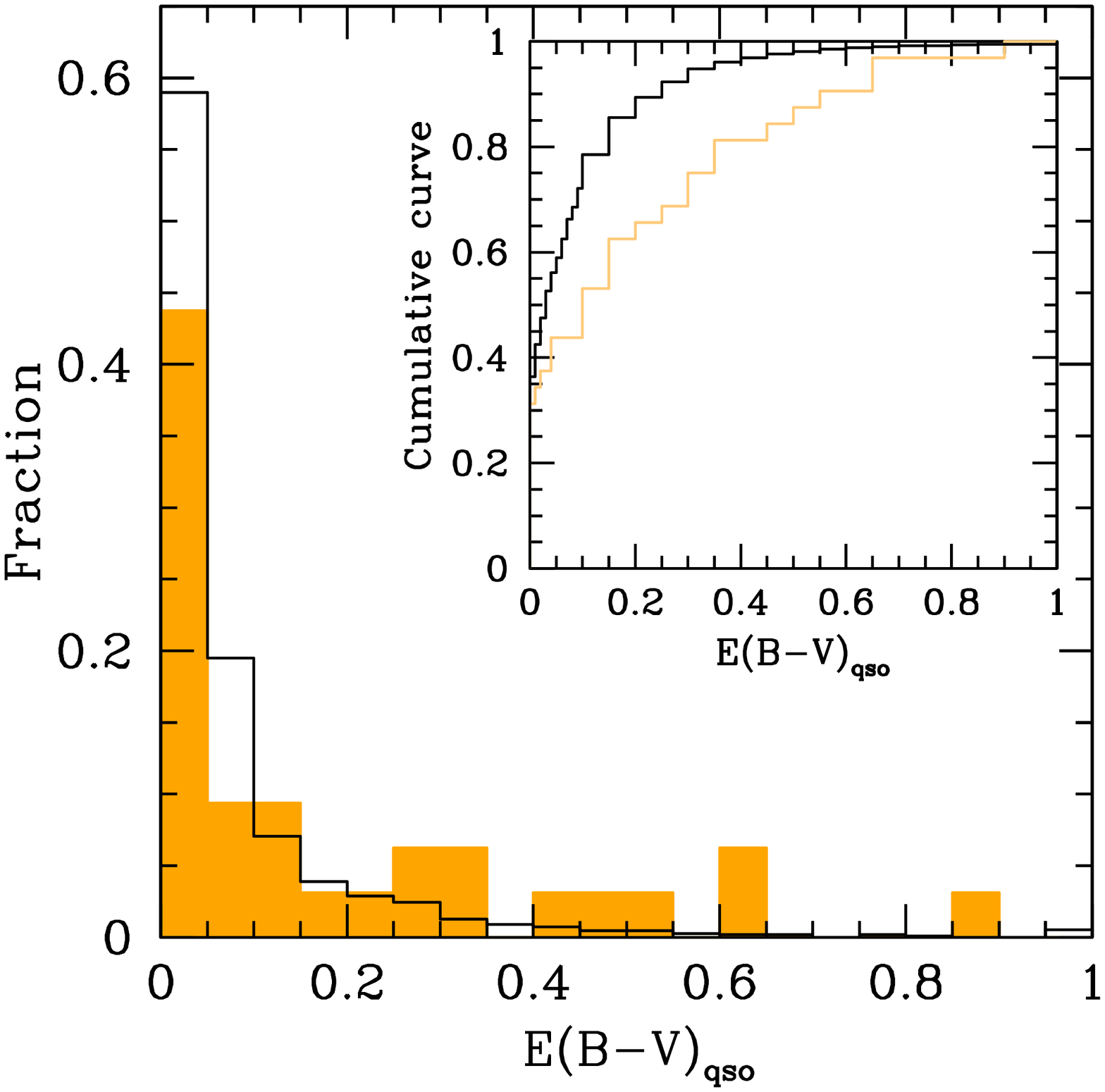}
 \caption{Disc reddening distribution for our sample ({\it orange filled histogram}) and for the control sample of typical QSOs ({\it black open histogram}). The two histograms are normalized to the total number of sources in the two samples. The internal panel shows the cumulative distributions of $\ebvq$ for our sample (orange line) the control sample (black line).}
 \label{ebvqcomp}
\end{figure}

Figure~\ref{ebvqcomp} shows a comparison of the disc reddening distribution between our sample and the control sample. On average, our sample presents a higher level of intrinsic obscuration with respect to the match sample. The mean disc reddening for our sample is $\langle \ebvq\rangle\simeq0.20$ with a standard dispersion of 0.24 (the median $\ebvq$ is 0.1 with 16$^{\rm th}$ and the 84$^{\rm th}$ percentile at 0 and 0.45, respectively), while the control sample has $\langle \ebvq\rangle\simeq0.08$ with $\sigma=0.14$ (median $\langle \ebvq\rangle\simeq0.03$ with 16$^{\rm th}$ and the 84$^{\rm th}$ percentile at 0 and 0.15, respectively). 
A K-S test rules out the hypothesis that the two samples have, on average, similar $\ebvq$ values.
However, the difference in the reddening distributions is mainly dominated by ``others" and "asymmetric", and it is no longer statistically significant (1.4$\sigma$ level) when we restrict to the objects classified as BHB candidates (9 sources classified as B, BD, BO, and BA). The AGN reddening value for each object is listed in Table~\ref{tbl-2}, while Table~\ref{tbl-4} contains the average reddening values for each class.

\begin{table}
\caption{Average reddening values for the single classes. \label{tbl-4}} 
\centering
\begin{tabular}{ccccc}
 $\langle\ebvq\rangle$ & $\sigma$ & median & class & $N_{\rm obj}$ \\ [0.5ex]
\hline\noalign{\smallskip}
0.15  & 0.14 & 0.10 & B & 9$^{\mathrm{a}}$ \\
0.13  & 0.10 & 0.10 & D & 3 \\
0.24  & 0.21 & 0.30 & A & 5 \\
0.26  & 0.30 & 0.02 & O & 15 \\
\hline
 0.20 & 0.24 & 0.1 & Tot & 32 \\
 0.08 & 0.14 & 0.03 & Ref & 1100 \\
\hline\hline
\end{tabular}

\flushleft\begin{list}{}
 \item[${\mathrm{a}}$]{ This sample contains objects with class B, BD, BA, and BO.}
\end{list}
\end{table}

\begin{figure}
  \centering
  {\includegraphics[width=0.45\textwidth]{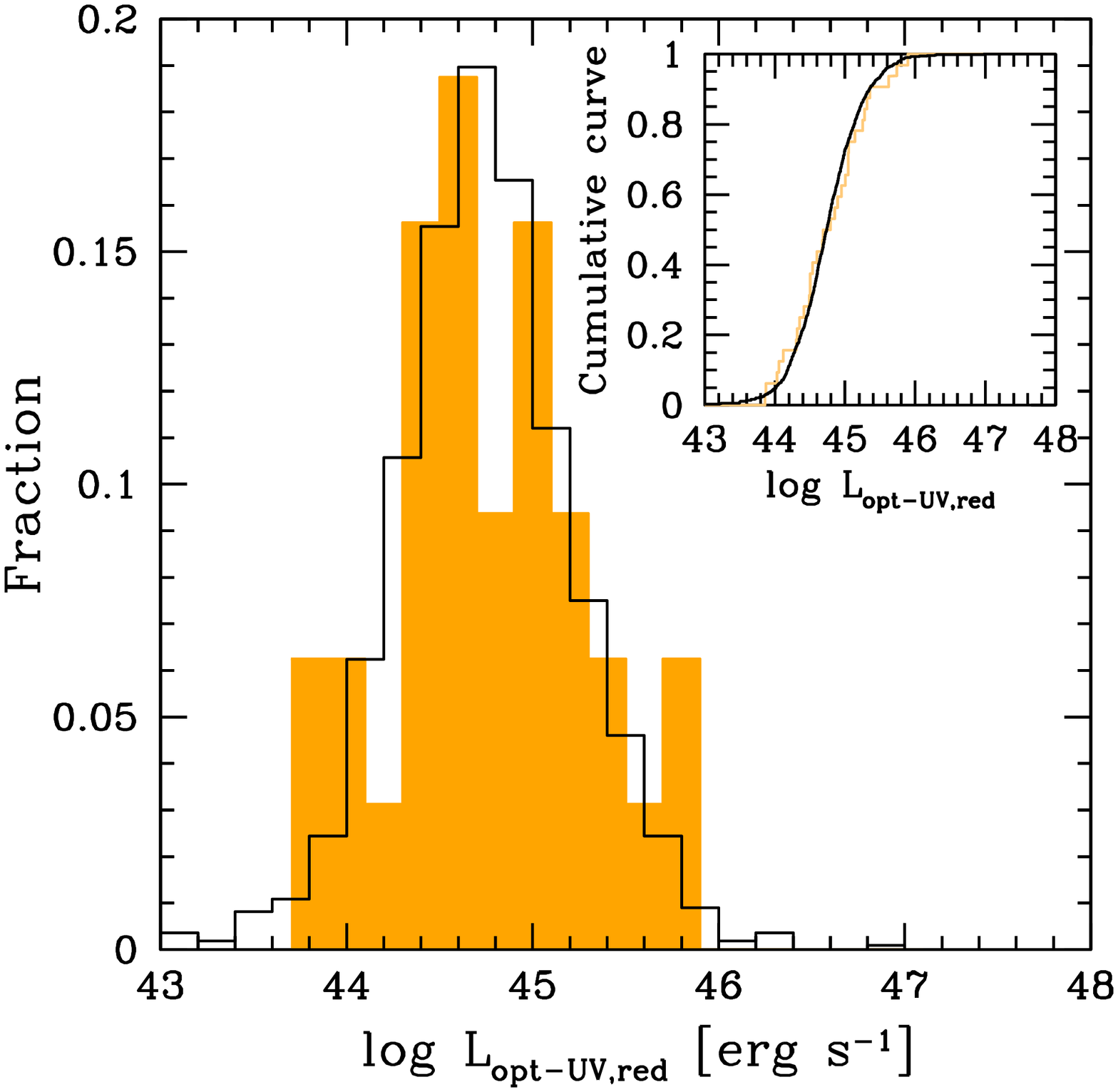}}
  {\includegraphics[width=0.45\textwidth]{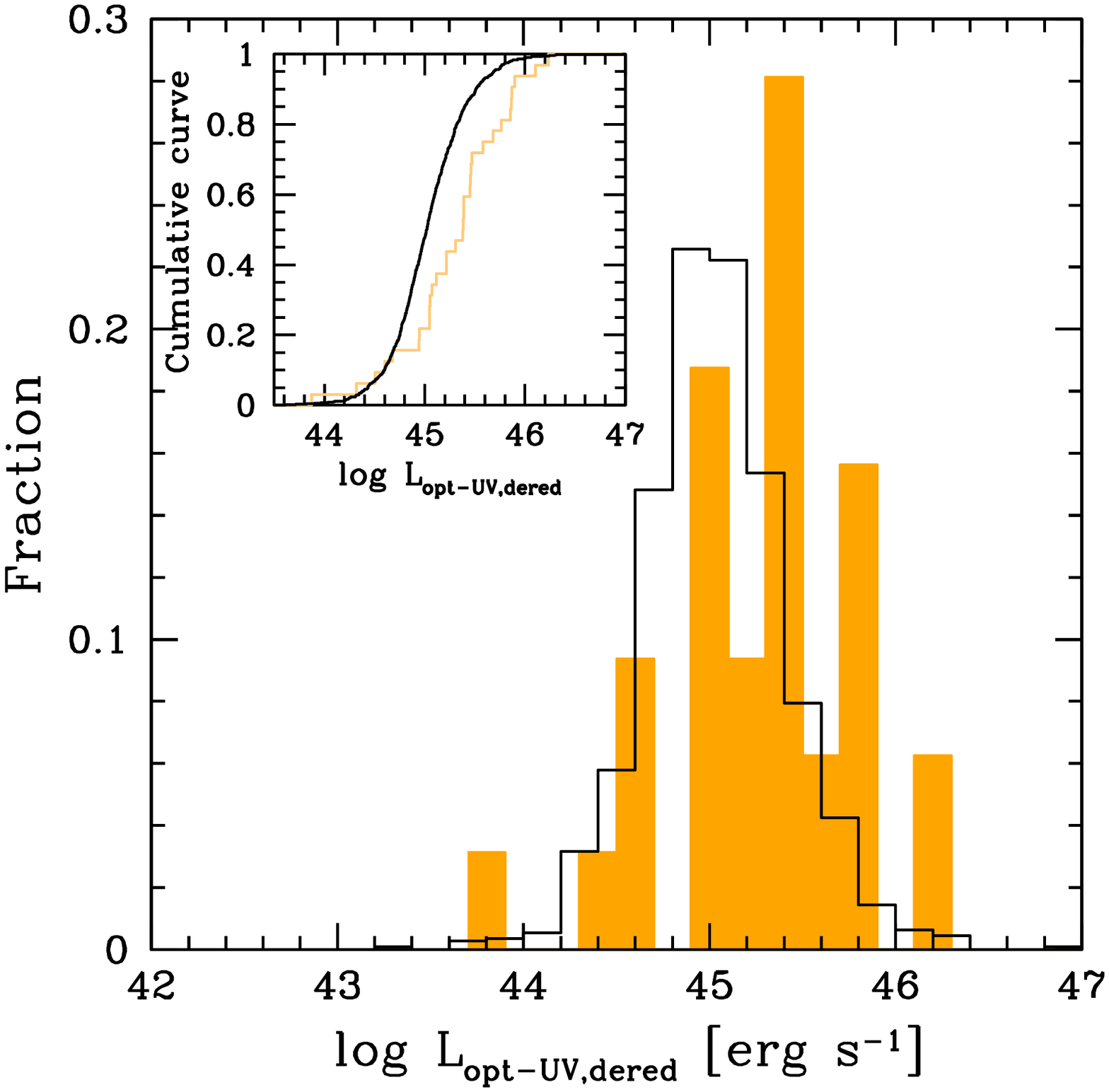}}                        
  \caption{Disc luminosity distribution for our sample ({\it orange filled histogram}) and for the control sample of typical QSOs ({\it black open histogram}). The two histograms are normalized to the total number of sources in the two samples.  The internal panels show the cumulative distributions for our sample (orange line) the control sample (black line). {\it Top panel: } Best-fit disc luminosity before applying the reddening correction (host galaxy contamination removed). {\it Bottom panel: } Best-fit disc luminosity after applying the reddening correction (host galaxy contamination removed).}
  \label{lcomp}
\end{figure}
\subsection{Luminosity distributions}
\label{Luminosity distributions}

We have further checked if the control and our sample are probing different luminosities in the same redshift bin. In the top panel of Figure~\ref{lcomp} we present a comparison of the best-fit disc luminosities between the control and our sample. These luminosities are host galaxy subtracted but with disc reddening left in. 
The mean disc luminosity for our sample is $\langle \Log\Ldisc\rangle\simeq44.77$ erg s$^{-1}$ with a standard dispersion of 0.53 (the median $\Log\Ldisc$ is 44.79 erg s$^{-1}$), while the control sample has a mean of 44.75 erg s$^{-1}$ with $\sigma=0.48$ (median $\langle \Log\Ldisc\rangle\simeq44.74$ erg s$^{-1}$). The two distributions are found to be consistent, as verified via a K-S test.
\par
We have also computed the host-galaxy luminosity ($\Lhost$) from the best-fit galaxy template for each quasar in our sample and in the control sample over the same wavelength range of $\Ldisc$. 
We found that the ratio between $\Lhost$ and $\Ldisc$ is more than 0.7 for $\sim$31\% of the quasars in our sample, while for the control sample is about 10\%. This means that the host galaxy is generally brighter in our sample for a given $\Ldisc$.
For these objects the host galaxy fraction might be considered an upper limit given that the best-fit host galaxy solution is degenerate with their reddened BBB as shown in Fig.~\ref{bhbseds} (the shaded areas correspond to the higher and lower uncertainty in the normalization).
In fact, the shape of a highly reddened BBB is similar to the one of the galaxy and therefore the finding of higher host galaxy contamination in our sample than in the control sample might be simply biased by the reddening distribution. Further details on the SED fits for these 10 objects are given in Appendix~\ref{Notes on the SED general properties:  GALEX data}.

Given that our sample have the tendency to be more reddened than the control sample, once we have corrected for the intrinsic $\ebvq$, the disc luminosities in the BHB sample are actually higher than the ones in the control sample.
In the bottom panel of Figure~\ref{lcomp} a comparison of the intrinsic disc luminosities (host galaxy subtracted and corrected for reddening) for the two samples is presented.
It is apparent that after applying the reddening correction the disc luminosities of our sample are, on average, higher than the one in the control sample.
The mean disc luminosity for our sample is $\langle \Log\Ldisc\rangle\simeq45.30$ erg s$^{-1}$ with a standard dispersion of 0.52 (the median $\Log\Ldisc$ is 45.38 erg s$^{-1}$), while the control sample has a mean of 45.03 erg s$^{-1}$ with $\sigma=0.38$ (median $\langle \Log\Ldisc\rangle\simeq45.02$ erg s$^{-1}$).
If we perform again a K-S test, we rule out the hypothesis that the two samples have similar means.
\par
\begin{figure}
 \centering\includegraphics[width=9cm,clip]{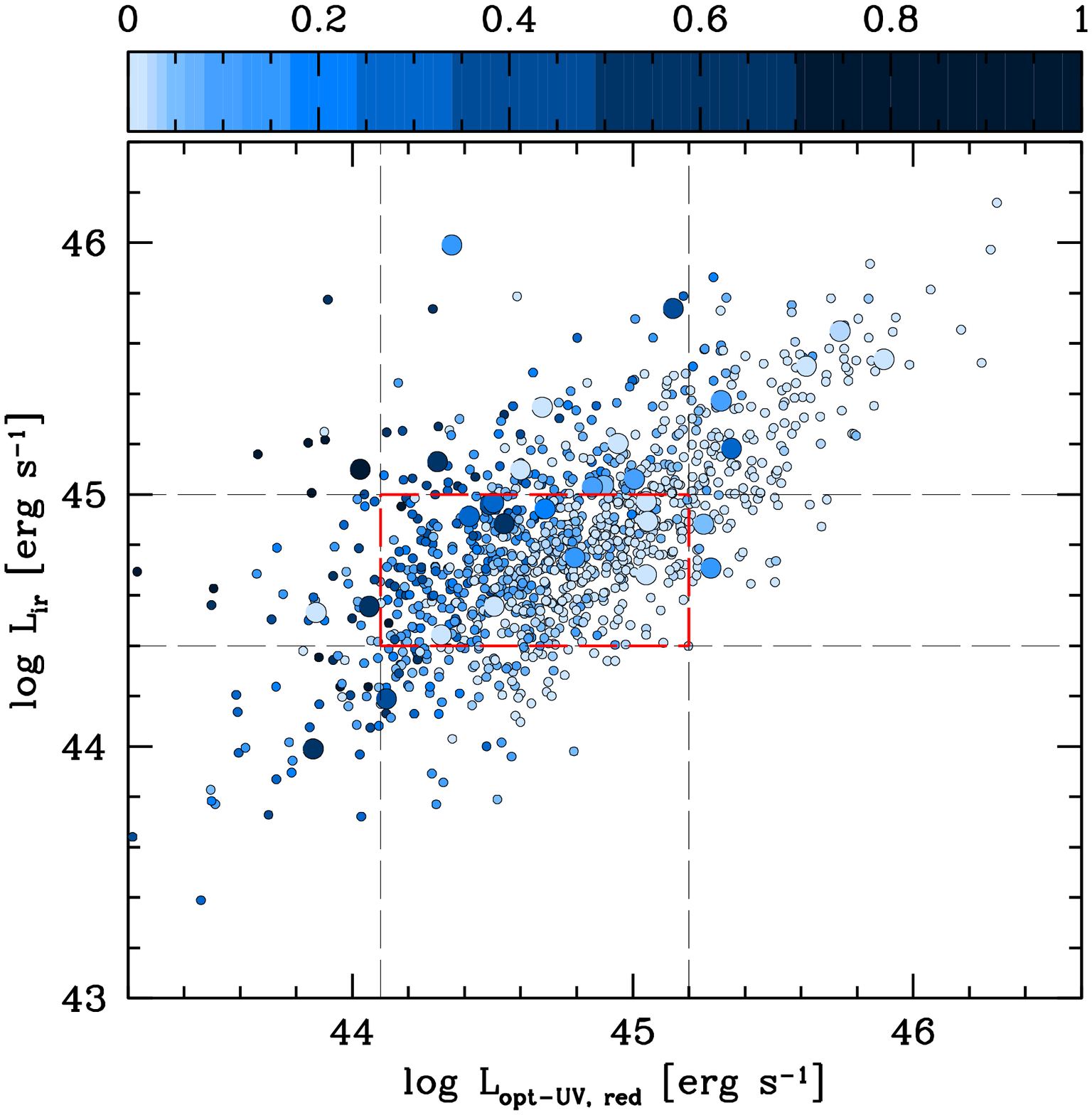}
 \caption{Infrared ``torus" luminosities as a function of disc ``accretion" luminosities (not corrected for reddening) for our sample (big circles) and for the control sample (small circles). The bluescale indicates the AGN reddening for each object: light blue represent low reddening (white means $\ebvq=0$) and dark blue high reddening values (the darkest circles have $\ebvq=1$). \rev{The red dashed lines encompass the region where the two samples have similar average $\Lir$ and $\Ldiscr$ values.}}
 \label{lirlopt}
\end{figure}
We further tested whether the difference in the average reddening values between our sample and the control may be explained by the small sample differences in luminosity.  \rev{Figure~\ref{lirlopt} shows $\Lir$ as a function of $\Ldiscr$ for our sample and the control objects. We have then selected a region in this plot where our sample and the control one have similar $\Lir$ and $\Ldiscr$ values. 
The red dashed lines encompass the region where the average values for $\Lir$ and $\Ldiscr$ are similar within $\sim1.6\sigma$.} 
The average values for $\Lir$, $\Ldiscr$, and $\ebvq$ are reported in Table~\ref{tbl-3}.  
We find that there is no significant difference (1.7$\sigma$ level) between the mean reddening values in the two samples at matched $\Lir$ and $\Ldiscr$, although we note that the $\Lir$ and $\Ldiscr$ intervals are quite narrow. 
\begin{table}
\caption{Values of $\Lir$, $\Ldiscr$, and $\ebvq$ for the control and our sample inside the region encompasses by the red dashed lines in Fig.~\ref{lirlopt} ($44.10\leq\log\Ldiscr\leq 45.20$ and $44.4\leq\log\Lir\leq 45.00$). \rev{This region has been selected in order to have luminosity-matched samples (within $\sim1.6\sigma$).} \label{tbl-3}} 
\centering
\begin{tabular}{@{}l c c c l}
 Parameter & median & mean & $\sigma$ \\ [0.5ex]
\hline\noalign{\smallskip}
 &  our sample (N=11) & \\
\hline\noalign{\smallskip}
  $\log \Ldiscr$     & 44.54 & 44.67 & 0.27  \\  
  $\log \Lir$          & 44.89 & 44.81 & 0.18  \\  
  $\ebvq$             & 0.15 & 0.17 & 0.21  \\   
\hline {\smallskip}
 &  control sample (N=596) & \\
\hline\noalign{\smallskip}
  $\log \Ldiscr$     & 44.69 & 44.68 & 0.27  \\
  $\log \Lir$          & 44.72 & 44.72 & 0.16 \\  
  $\ebvq$             & 0.03   & 0.07  & 0.10   \\    
\hline\hline
\end{tabular}
\end{table}

\par
In summary, the SEDs of our sample are, on average, {\it 1)} characterized by moderately higher reddening values than the ones estimated in a control sample of typical QSOs, and {\it 2)} the host galaxy contamination is higher in our sample than in the control sample. The latter result might be due to degeneracies between host galaxy and BBB in the case of high AGN reddening values, while the first is valid only in the case of redshift matched samples.
Nonetheless, winds can also contribute in modifying the shape of the SEDs providing obscuration in the optical-UV wavelength. 
Broad absorption line troughs in the bluest part of the \ion{C}{iv} $\lambda1550$ and/or of the  \ion{Mg}{ii} line could be used as an indicator of the possible presence of outflows. The \ion{C}{iv} line is not covered by our SDSS spectra, while for the \ion{Mg}{ii} line we have sufficient coverage only for objects with redshift higher than $\sim$0.42 (7 objects, but two of those have $z\simeq0.42$ where the bluest part is barely covered).
However, there is no clear evidence of BAL troughs in the bluest part of this line. 

\subsection{Torus sizes and geometry}
\label{Torus sizes}

\begin{figure}
 \centering\includegraphics[width=9cm,clip]{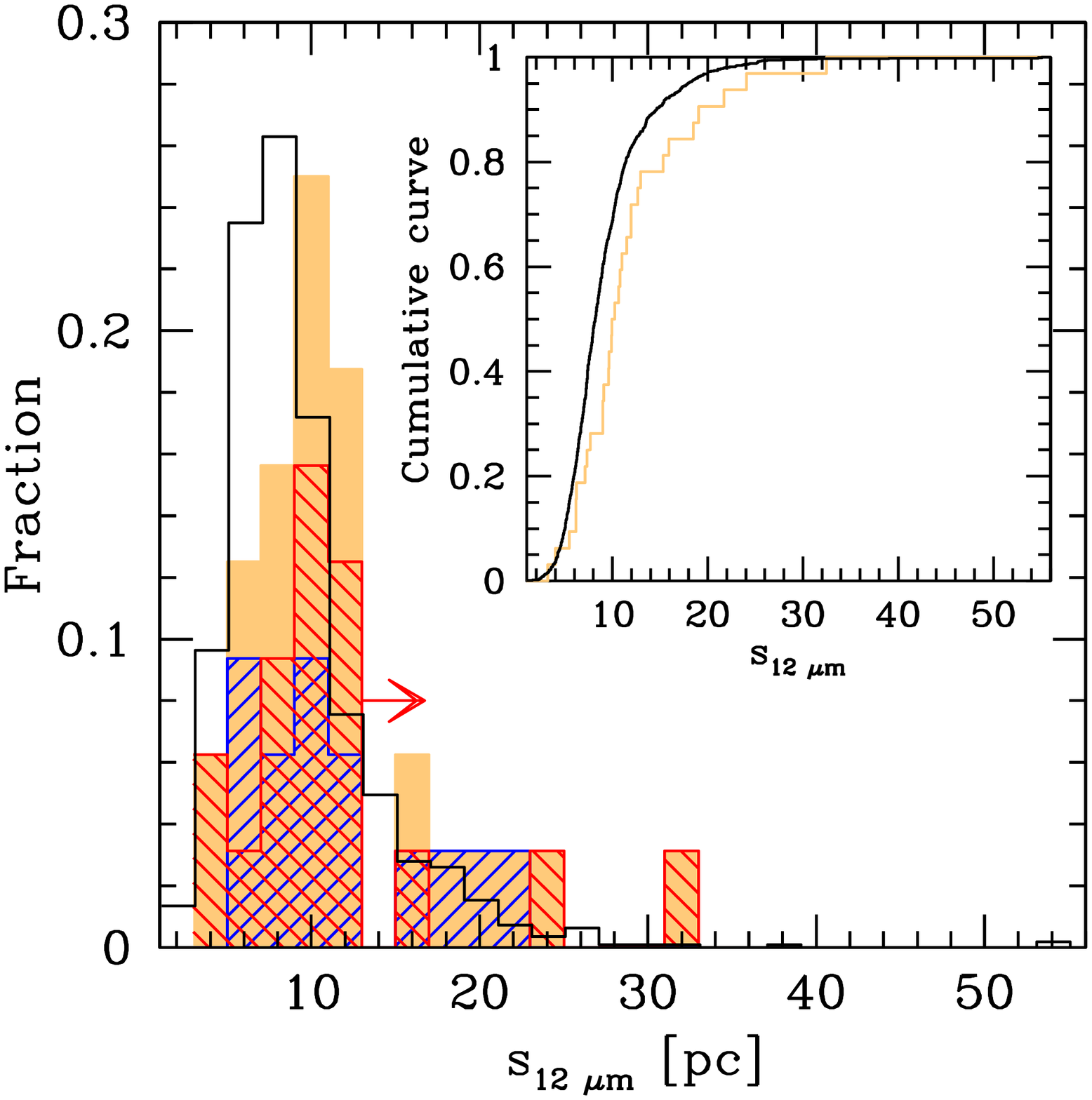}
 \caption{Distribution of the sizes (in parsec) of the 12$\mu$m emitter for our sample ({\it orange filled histogram}) and for the control sample of typical QSOs ({\it black open histogram}). The two histograms are normalized to the total number of sources in the two samples. The red hatched histogram is the $\size$ distribution of our sample with $\ebvq\geq0.1$ (18 objects), while the blue hatched histogram is for our sample with $\ebvq<0.1$ (14 objects).}
 \label{sizecomp}
\end{figure}

Based on Eq.~(\ref{sizelum}), we estimated the size of the $12~\mu$m emitting region in our sample and in the control sample. Results are shown in Figure~\ref{sizecomp}.
The mean $\size$ for our sample is $\langle \size\rangle\simeq11.6$ pc with a
standard dispersion of 6.2 (the median $\size$ is 10.3 pc), while the control
sample has a mean of 9.2 with $\sigma=4.7$ (median $\size \simeq 8.2$ pc).
The two samples are different at $\sim$2$\sigma$ level from a K-S test.
\par
If we assume that the torus is optically thick to its own radiation (e.g., \citealt{1994MNRAS.268..235G}; L13), the infrared emission along an equatorial line-of-sight (i.e., aligned with the torus) is smaller than along a polar line-of-sight.
Sources with high reddening values (i.e., 18 out of 32 objects with $\ebvq\geq0.1$) are expected to have, on average, lower $\Lir$ values, and thus lower $\size$ values. Yet, the $\size$ values for our sample are marginally higher than the ones of the control sample.
This is consistent with the larger intrinsic brightness (once we have
corrected for reddening) in the optical \citep{2006ApJ...639...46S,2009ApJ...705..298M} of our sample with respect to the
control sample.
In this scenario, inclination
effects are not playing a major role in the optical AGN obscuration (otherwise
we should observe lower $\Lir$ and thus lower $\size$ values for an equatorial
line of sight) under the assumption that the torus is optically
thick. Inclination can be still a possible interpretation of the observed AGN
obscuration if the torus is actually relatively optically thin to its own
infrared radiation (this is consistent with the findings of L13, but see also
\citealt{2012A&A...548A..45D}).  Another possibility is that our sample
presents small scale structure which is {\it intrinsically different} than the
one of the control sample. More complex discs than the circular
  Keplerian disc, such as a warped disc, thought to occur around rotating
  black holes \citep{1975ApJ...195L..65B}, may give rise to both obscuration
  and a wide range of line profiles (e.g. double-horned profiles;
  \citealt{1999A&A...348...71B,2009MNRAS.397.1510Z}), but this is just a
speculation and the current data do not yet place significant constraints on
disc structure.

\rev{We can use the velocity difference between the broad and narrow
emission lines together with the torus sizes to test the recoiling
scenario. Through equation 2 we estimate the minimum BH mass $M_{\rm
min}$ required to keep the torus bound to a recoiling hole. To be as
conservative as possible, we will assume that the candidate recoiling
BH is actually moving exactly along the line of sight. For 28 out of
the 32 objects in our sample BH masses $M_{\rm Shen}$ as been
estimated by \cite{2011ApJS..194...45S} through the single epoch method
\citep[e.g.][]{vestergaard06}. The results of this test is that for 22
objects out of 28 $0.0001<M_{\rm shen}/M_{\rm min}<0.2$, i.e. these
objects should show a significant reduction of the IR emission.  For 5
objects $0.2<M_{\rm Shen}/M_{\rm min}<1$, and 1 object (J1652+3123)
has $M_{\rm shen}/M_{\rm min}\approx 2$. J1154+0134, the only object
showing a reduced IR flux, is not among these 6.

We note that $1)$ all the BH masses of these last 6 objects are
inconsistent with the galaxy masses we estimate from the SED
fitting. Even assuming that these objects are extreme outliers of the
BH-host galaxy relations, $2)$ for two of them (J1027+6050 and J1105+0414)
$M_{\rm min}>10^{11} \msun$ is required to keep the tori
bound\footnote{Masses between $10^{10}$ and $10^{11}$ $\msun$ are required
for a recoiling MBH to keep its torus for the 4 objects (J1207+0604,
J1328-0129, J1440+3319 and J1539+3333) without mass estimates as
well.}. 
The unrealistic estimate of the BH masses for these two
objects is caused by their double peaked emitter nature, and single
epoch measurements should not be attempted for the elements of this
peculiar class.  
The remaining 4 objects have $M_{\rm shen}/M_{\rm min}\sim 1$ 
because of the very small velocity shifts measured, $\simeq1000$ km s$^{-1}$ 
for all of them. 
None of these four objects have bell-shaped broad lines 
(i.e. they have not being classified as binary
candidates), hence the simplest explanation is that these are standard
AGN with slightly asymmetric lines.

We stress that in this test we neglected any projection effect. In the
scenario of a recoiling MBH with a 3-D velocity inclined by 45 degrees
with respect to the line of sight $M_{\rm shen}/M_{\rm min} < 1$ for
the whole sample.

}

\section{Summary and conclusions}
We have presented a homogeneous and comprehensive study of the broad-band properties of 32 quasars in the sample identified by \citet{2011ApJ...738...20T}. These objects have been selected to have large velocity shifts ($>$1000 km s$^{-1}$) between NLs and BLs. According to the line profiles, source are subsequently divided in four classes: {\it 1)} fairly bell-shaped, strongly shifted BLs identify good BHB candidates; {\it 2)} BLs with evidence of double-horned profiles are classified as DPEs; {\it 3)} objects with Asymmetric lines; {\it 4)} other sources with complex profiles, or having lines with relatively small shifts ($\sim$1000 km s$^{-1}$).
\par
One possible interpretation that might explain the BLs displacement is a recoil induced by the anisotropic emission of gravitational waves in the final merge of SMBHs pairs.
This phenomenon would leave a characteristic imprint in the SED of these objects (i.e. a mid-infrared deficit).
This analysis aims to investigate the multi-wavelength SEDs of such peculiar QSOs, in particular in the context 
of the recoiling scenario depicted above. 

 
To achieve this goal, we have employed the SED-fitting code already presented in
Lusso et al. (2013), which models simultaneously three components of the AGN SED
i.e. hot-dust from the torus, optical-UV emission from the evolving stellar population, and optical-UV from the accretion disc.  
\par
Our main findings are summarized in the following:
\begin{description}

\item[(1)] All quasars analyzed here present a significant amount of 12~$\mu$m
  emission. The SED of hot-dust poor quasars should be
  a further indication of displaced SMBH from the center of a galaxy, maybe
  attributed to an undergoing recoil event. We do not find any evidence
    of such objects in the presented sample aside from J1154+0134, whose IR
    emission is a factor $\sim 10$ lower than what expected from its UV-optical
    emission. Further follow-ups (e.g. high resolution imaging) is required to
    test \rev{the nature (recoil vs DPE)} of J1154+0134.

\item[(2)] We find that our sample covers a wide range in terms of SED shapes with host galaxy contamination higher with respect to the QSO control sample (modulo degeneracies between the galaxy and a highly reddened BBB). Nevertheless, SEDs of our sample show, on average, moderately higher levels of intrinsic disc obscuration than the control sample (if we match solely on redshift). 


\item[(3)] Disc luminosity estimates (before applying the reddening correction but host galaxy subtracted) for our sample and for the control one are not significantly different. 
Given the higher disc reddening contamination in our sample, then such objects should be intrinsically brighter than the QSO control sample. 

\item[(4)] There is a tendency to have higher average sizes of the $12~\mu$m emitter for our sample than the one for the control sample. This is consistent with the larger intrinsic (once we have corrected for reddening) brightness in the optical of our sample with respect to the control sample. 

\item[(5)] If we assume that the torus is optically thick to its own
  radiation, the infrared emission observed from a line-of-sight aligned with
  the torus should be smaller than one along a polar direction.  Sources with
  high reddening values (i.e., $\ebvq\geq0.1$) are expected to have lower
  $\Lir$ values, and thus lower $\size$ values. However, we find that the
  $\size$ values for our sample are, on average, somewhat higher than the ones
  of the control sample. This indicates that inclination effects are not
  playing a major role in the optical AGN obscuration.
The source of
  obscuration might be due to intrinsic differences in the small scale
  structure of the quasars in our sample.

\end{description}

As a note of caution we also point out that the difference in the
average AGN reddening is no longer significant when we select a sub-sample in
both our data-set and the control sample with similar $\Ldiscr$ and $\Lir$.
Moreover objects classified BHB and DPEs in our sample have no significant
difference in the intrinsic reddening even when compared to the larger
control sample.

\section*{Acknowledgements}
\label{sec_acknow}
The authors acknowledge the anonymous reviewer who provided many useful suggestions for improving
the paper.
EL gratefully thanks Kate Rubin for useful discussions on galaxy outflows and Cristian Vignali for having reduced the Chandra and the XMM--Newton spectra of J0927+2943, J1328-0129, and J0155-0857, and for illuminating discussion on the X-ray properties of these sources.
EL also thank Joseph F. Hennawi, Gianni Zamorani, Andrea Comastri, and the members of the ENIGMA group\footnote{http://www.mpia-hd.mpg.de/ENIGMA/} at the Max Planck Institute for Astronomy (MPIA) for helpful discussions.
MD and CM are grateful for the hospitality of the Max Planck Institute for Astronomy (MPIA).
Support for MF was provided in part by NASA through Hubble Fellowship grant HF-51305.01-A awarded by the Space Telescope Science Institute, which is operated by the Association of Universities for Research in Astronomy, Inc., for NASA, under contract NAS 5-26555.

\bibliographystyle{mn2e}
\bibliography{bibl}

\appendix
\label{appendix}

\section{Details on the SED-fitting models and procedure}
\label{Details on the SED-fitting procedure}
Our SED-fitting code models simultaneously three components in the quasar SED i.e. hot-dust from the torus, emission from the evolving stellar population, and emission from the accretion disc.  
The nuclear hot-dust SED templates are taken from \citet{2004MNRAS.355..973S}. They were constructed from a large sample of Seyfert galaxies selected from the literature for which clear signatures of non-stellar nuclear emission were detected in the near-IR and mid-IR, and also using the radiative transfer code GRASIL \citep{1998ApJ...509..103S}. 
The infrared SEDs are divided into 4 intervals of absorption: $\NH<10^{22}$ cm$^{-2}$ for Seyfert 1, $10^{22}<\NH<10^{23}$ cm$^{-2}$, $10^{23}<\NH<10^{24}$ cm$^{-2}$, and $\NH>10^{24}$ cm$^{-2}$ for Seyfert 2. 
Since all the objects in our sample are broad-line QSOs, the latter case is neglected in our analysis.

We employed a set of 18 galaxy templates built from the \citet{2003MNRAS.344.1000B} spectral synthesis models, built at Solar metallicity and with a Chabrier IMF (\citealt{2003ApJ...586L.133C}).
To make the galaxy templates representative of the entire luminous-AGN-hosting galaxy population, a set of models was selected with a range of star-formation histories;
the models combine 6 exponentially decaying star formation histories (SFHs) with characteristic times ranging from $\tau = 0.1$ to $3$\,Gyr,
with 3 total stellar-population ages of 6, 9, and 11\,Gyr.
For context, early-type galaxies (characterized by a small amount of ongoing star formation) are represented by the models with characteristic times $\tau<1$\,Gyr and old ages, whereas more actively star forming galaxies are represented by models with larger characteristic times $\tau$ and smaller ages.
An additional hard constraint applied to the fits is that, for each source, the only templates considered for fitting are those with total ages smaller than the age of the Universe (in our fiducial cosmological model) at the redshift of the source.
Each template is reddened according to the \citet{2000ApJ...533..682C} reddening law, with 11 permitted $E(B-V)_{\rm gal}$ values range between 0 and 0.5 with a step-size of $0.05$.

The BBB template representative of the accretion disc emission is taken from \citet{2006ApJS..166..470R}. 
The near-infrared bump is neglected since we have already covered the mid-infrared region of the SED with the hot-dust templates. 
This template is reddened according to the \citet{prevot84} reddening law for the Small Magellanic Clouds (SMC, which seems to be appropriate for Type-1 AGN, \citealt{2004AJ....128.1112H,salvato09}). 
The AGN-reddening ($\ebvq$) ranges between 0 and 1 with a variable step ($\Delta \ebvq=0.01$ for $\ebvq$ between 0 and 0.1, and $\Delta \ebvq=0.05$ for $\ebvq$ between 0.1 and 1.0) for a total of 29 templates.
\par
The total number of free parameters in each of the linear fits is eight: three amplitudes corresponding to the three components involved in the fit, two parameters concerning the shape of the galaxy (i.e. $\tau$ and age), one parameter concerning the shape of the torus component, the AGN and the galaxy reddening.
\par
We obtain approximate uncertainty estimates on model parameters by considering the variation in best-fit $\chi^2$ as a function of template property.
In detail, the uncertainty estimate on each parameter $x$ is produced by considering each alternative component combination, drawing a parabola in the space of $x$ and $\chi^2$ with a minimum at the best-fit $x$ and $\chi^2$ but also passing through the alternative component combination's $x$ and $\chi^2$, and finding the $x$ value where the parabola crosses $\Delta\chi^2=1$.
From among these crossings (one per component combination), we use the most extreme values to determine the uncertainty estimates.
\par
The resulting SED fits are presented in Figure~\ref{bhbseds}.

\section{Notes on the SED general properties for the different classes}
\label{Notes on the SED general properties for the different classes}

\subsection{Binary BH candidates \& DPEs}
BHB candidates (9 objects with class=B, plotted with filled circles in Fig.~\ref{md}) and DPEs  (3 objects with class=D, marked with filled squares in Fig.~\ref{md}) are almost all characterized by high host galaxy contamination around $1~\mu$m and AGN reddening lower than 0.5. 
J1050+3456 presents the highest reddening value in these two classes (i.e., $\ebvq=0.45$ from the best-fit, $\ebvq\sim0.4$ from the mixing diagram), while J0927+2943 ($\aopt=0.76$ and $\anir=-0.07$, marked with the green circle in Fig.~\ref{md}) has rather standard AGN SED with little host galaxy contamination and no AGN reddening.
\par
J1539+3333 is the only object that has been fitted with a galaxy template with very little contribution from the accretion disc. The optical spectrum of this source shows strongly shifted broad lines, no clear indication of  [\ion{Ne}{v}] is observed at any redshift which suggests modest or no AGN contribution in the optical-UV or strong extinction. 
The best-fit SED of this source can be interpreted as the one of an obscured AGN, however it is unclear whether $W3$ and $W4$ are powered by PAH feature as well. For this objects, {\it Herschel} data are necessary in order to disentangle between the torus or PAHs scenario.

\subsection{Asymmetric line profiles}
Sources with lines in the optical spectra showing a symmetric base, centered at the redshift of the narrow lines, but an asymmetric core with a shifted peak are called {\it asymmetric} (class=A, 5 objects, see Paper I), and they are plotted with red triangles in Fig.~\ref{md}. These AGN present both high host galaxy contamination ($\sim70-90\%$), and reddening ($0.3\lesssim\ebvq\lesssim0.5$).

\subsection{Others}
QSOs with complex line profiles or lines with small ($v_{\rm peak} \sim1000$
km s$^{-1}$) velocity shifts are marked as {\it others} (15 objects, marked as
stars in Fig.~\ref{md}).  These sources cover a wide range of host galaxy
contaminations and reddening.  \par For example, the best-fit of J1010+3725
($\aopt=-0.95$ and $\anir=-1.02$) shows conspicuous infrared emission compared
with the BBB, with an SED similar to what has been observed in starburst/ultra
luminous infrared galaxies (ULIRGs, e.g., Arp220, NGC6240,
\citealt{1997A&A...325L..21K}).  The AGN reddening value estimated from the
mixing diagram ($\ebvq\sim0.5$) is about 3.3 times higher than the one from
the SED-fit ($\ebvq=0.15$). 
\ion{H}{$\beta$} only, we cannot use the \ion{H}{$\alpha$}/\ion{H}{$\beta$}
ratio.  The high \ion{H}{$\alpha$}/\ion{H}{$\beta$} ratio observed in broad
lines (see Paper I) may be a further hint of high extinction. Note however
that the \ion{H}{$\alpha$}/\ion{H}{$\beta$} ratio in the broad-line regions
can be much larger than the case B recombination value of 3.0 usually adopted
for narrow-line region (\citealt{1984PASP...96..393G}) due to effects of
high-density and optical depth, and thus high
\ion{H}{$\alpha$}/\ion{H}{$\beta$} ratios could not entirely be due to
reddening of the broad line region \citep{1982MNRAS.198..589N}.

The spectrum of this AGN shows broadened [\ion{O}{iii}] (FWHM=1140 km s$^{-1}$), with various components at different velocities, likely caused by massive gaseous outflows driven either by star formation or AGN winds. Such a feature is not observed in other lower ionization lines (e.g., [\ion{O}{ii}]).
\par
J1215+4146 ($\aopt=-2.20$ and $\anir=0.11$) presents the highest broad
\ion{H}{$\alpha$}/\ion{H}{$\beta$} ratio in the sample
(\ion{H}{$\alpha$}/\ion{H}{$\beta$}$\sim20$, see Fig.~3 in Paper I), which
might imply severe reddening (\citealt{2008MNRAS.383..581D}), consistent with
both SED best-fit and mixing diagram estimates ($\ebvq=0.90$). 
Note however that the same uncertainties discussed above for J1010+3725 apply here as well.

\section{Notes on the SED general properties:  GALEX data}
\label{Notes on the SED general properties:  GALEX data}
GALEX data for the 10 sources with high host-galaxy
contamination\footnote{J0931+3204, J0942+0900, J1000+2233, J1050+3456,
  J1105+0414, J1207+0604, J1215+4146, J1216+4159, J1440+3319, and
  J1539+3333. The latter does not have any GALEX detection.} described in
\S~\ref{Luminosity distributions} are usually too bright to be well fitted
with the BBB template only and, in fact, a galaxy component in the optical--UV
is required. An additional constraint on the contribution of the AGN to
  the UV emission could come from flux variability. As
  discussed in Section~\ref{Photometry}, only J0221+0101, J0829+2728, and
  J1440+3319 show significant variability (by a factor $\sim 2-3$ in
  fluxes). The SED fitting of J0221+0101 and J0829+2728 requires a significant
  to dominant contribution of the AGN light to the UV spectrum, as it is
  expected for a varying object. The case of J1440+3319 is more subtle. Our
  SED analysis suggests that most of the UV emission comes from a young
  stellar population in the host, and such an origin would not justify the
  temporal behavior of the quasar. To check whether the spectrum of J1440+3319
  as well as of the others quasars with strong host contamination in the UV
  can be fitted with a BBB only we forced the analysis to assume low reddening
  values (i.e. we have imposed $\ebvq\leq 0.1$).  In four objects
(J1050+3456, J1215+4146, J1216+4159, and J1440+3319) out of ten we have
significant changes, in all other cases the mid-infrared to disc luminosity
ratio is higher than one, which is not physical (dust cannot reprocessed more
optical--UV luminosity than what it is actually emitted from the disc).  For
J1215+4146 the disc component disappears and the photometry is well fitted
solely by passive stellar population. We consider this fit unlikely, given
that we have independent indications that this object is a red Type-1 and the
continuum should come from the AGN.  J1050+0414 and J1216+4159 show better
fits in the UV, nonetheless with high host galaxy contamination in the
near-IR/optical but, J1216+4159 presents mid-infrared to disc luminosity ratio
close to one.  Forcing low reddening value, the SDSS and GALEX data of
J1440+3319 are well fitted by relatively young stellar population with an
almost negligible disc component. Also in this case, the mid-infrared to disc
luminosity ratio is higher than one, making the interpretation of the
J1440+3319 unclear. We stress however that, even if we consider these fits
plausible, the difference between the reddening distribution of our and the
control sample is still moderately significant at 2$\sigma$ (instead of
2.7$\sigma$).

\end{document}